\documentclass[useAMS,usenatbib,usegraphicx]{mn2e}

\newcommand{\lya}{{\rm Ly}\alpha}
\newcommand{\lyb}{{\rm Ly}\beta}
\newcommand{\hkpc}{h^{-1}{\rm kpc}}
\newcommand{\hmpc}{h^{-1}{\rm Mpc}}

\newcommand{\kms}{\;{\rm km}\,{\rm s}^{-1}}
\newcommand{\cms}{\;{\rm cm}^{-2}}
\newcommand{\cmc}{\;{\rm cm}^{-3}}
\newcommand{\cmsst}{\;{\rm cm}^2 {\rm s}^{-3}}
\newcommand{\Zsolar}{\;{\rm Z}_{\odot}}
\newcommand{\msolar}{\;{\rm M}_{\odot}}
\newcommand{\logrm}{{\rm log}}

\newcommand{\vw}{{v_{\rm wind}}}

\newcommand{\msol}{M_{\odot}}

\newcommand{\gad}{{\sc Gadget-2}}
\newcommand{\CII}{{\hbox{C\,{\sc ii}}}}
\newcommand{\CIII}{{\hbox{C\,{\sc iii}}}}
\newcommand{\CIV}{\hbox{C\,{\sc iv}}}
\newcommand{\SiII}{{\hbox{Si\,{\sc ii}}}}

\newcommand{\SiIV}{\hbox{Si\,{\sc iv}}}

\newcommand{\OVI}{\hbox{O\,{\sc vi}}}

\newcommand{\HI}{{\hbox{H\,{\sc i}}}}
\newcommand{\HII}{{\hbox{H\,{\sc ii}}}}
\newcommand{\HeII}{{\hbox{He\,{\sc ii}}}}
\newcommand{\NeVIII}{{\hbox{Ne\,{\sc viii}}}}
\newcommand{\NaI}{{\hbox{Na\,{\sc i}}}}

\title[The Nature and Origin of Low-Redshift $\OVI$ Absorbers]{The Nature and Origin of Low-Redshift $\OVI$ Absorbers}

\author[B. D. Oppenheimer \& R. Dav\'e]{Benjamin D. Oppenheimer$^1$,
Romeel Dav\'e$^1$ \\$^{1}$Astronomy Department, University of Arizona,
Tucson, AZ 85721}

\begin{document}

\pagerange{\pageref{firstpage}--\pageref{lastpage}} \pubyear{2008}

\maketitle

\label{firstpage}

\begin{abstract}

  The $\OVI$ ion observed in quasar absorption line spectra is the
  most accessible tracer of the cosmic metal distribution in the low
  redshift ($z<0.5$) intergalactic medium (IGM).  We explore the
  nature and origin of $\OVI$ absorbers using cosmological
  hydrodynamic simulations including galactic outflows with a range of
  strengths.  We consider the effects of ionization background
  variations, non-equilibrium ionization and cooling, uniform
  metallicity, and small-scale (sub-resolution) turbulence.  Our main
  results are (1) IGM $\OVI$ is predominantly photo-ionized with
  $T\approx 10^{4.2\pm0.2}$ K.  A key reason for this is that $\OVI$
  absorbers preferentially trace over-enriched (by $\sim\times 5$)
  regions of the IGM at a given density, which enhances metal-line
  cooling such that absorbers can cool to photo-ionized temperatures
  within a Hubble time.  As such, $\OVI$ is not a good tracer of the
  Warm-Hot Intergalactic Medium.  (2) The predicted $\OVI$ properties
  fit observables if and only if sub-resolution turbulence is added,
  regardless of any other model variations.  The required turbulence
  increases with $\OVI$ absorber strength.  Stronger absorbers arise
  from more recent outflows, so qualitatively this can be understood
  if IGM turbulence dissipates on the order of a Hubble time.  The
  amount of turbulence is consistent with other examples of turbulence
  observed in the IGM and galactic halos.  (3) Metals traced by $\OVI$
  and $\HI$ do not trace exactly the same baryons, but reside in the
  same large-scale structure.  Our simulations reproduce observed
  alignment statistics between $\OVI$ and $\HI$, yet aligned absorbers
  typically have $\OVI$ arising from cooler gas, and for stronger
  absorbers lower densities, than $\HI$.  Owing to peculiar velocities
  dominating the line structure, coincident absorption often arises
  from spatially distinct gas.  (4) Photo-ionized $\OVI$ traces gas in
  a variety of environments, and is not directly associated with the
  nearest galaxy, though is typically nearest to $\sim 0.1L_*$
  galaxies.  Weaker $\OVI$ components trace some of the oldest cosmic
  metals.  (5) Very strong absorbers ($EW\ga 100$m\AA) are more likely
  to be collisionally ionized, tracing more recent enrichment ($\la
  2$~Gyr) within or near galactic halos.

\end{abstract}


\section{Introduction}

The exploration of metals in the low redshift ($z<0.5$) intergalactic
medium (IGM) has taken a large leap forward with the advent of
high-resolution space-based ultraviolet (UV) spectroscopy with the
{\it Hubble Space Telescope Imaging Spectrograph} (STIS) and {\it Far
Ultraviolet Spectroscopic Explore} (FUSE).  The strongest and most
common metal transition seen in quasar absorption line spectra is
the $\OVI$ doublet at 1032,1038\AA.  Many recent papers
\citep[e.g.][]{tri00, sav02, pro04, ric04, sem04a, leh06, coo08}
have examined $\OVI$ absorbers along single sight lines, and have
fit ionization models to individual systems with associated $\lya$
and lower ionization metal transitions.  Three recent studies (Tripp
et al. 2008, hereafter T08; Danforth \& Shull 2008, hereafter DS08;
and Thom \& Chen 2008a, 2008b) have compiled samples of $\OVI$,
providing the largest statistical surveys to date.

Understanding $\OVI$ in the low-$z$ Universe is extremely important,
because it may hold the key to locating a significant reservoir of
cosmic baryons and metals \citep[e.g.][]{tri00b}.  The inventory of
observed baryons \citep[e.g.][]{fuk98} falls well short of the
predicted cosmological values from the cosmic microwave background
\citep{hin08}, leading to the well-known ``missing baryons problem.''
Cosmological simulations \citep{cen99,dav99,dav01a} suggest that a
large fraction of baryons today ($>30\%$) reside in a hard-to-observe
warm-hot intergalactic medium (WHIM), with temperatures of
$T=10^{5}-10^{7}$~K.  The WHIM results primarily from shocks created
during large-scale structure formation.  More recent simulations
including feedback \citep[hereafter OD08]{cen06a,opp08} find that
galactic winds may increase the fraction of cosmic baryons in the WHIM
to 50\% or more.

A possibly related problem is that only $\frac{1}{3}$ of metals have
been accounted for observationally~\citep{fuk04}, judging from the
mismatch between the amount of metals nucleosynthesized and ejected by
observed stars and those observed in various cosmic baryonic
components.  $\OVI$ potentially represents the Holy Grail of all
things missing in the low-$z$ Universe, because its collisionally
ionized equilibrium (CIE) maximum temperature, $10^{5.45}$ K, provides
a unique and relatively easily accessible tracer of WHIM metals and
baryons.  The incidence of $\OVI$ with broad $\lya$ absorbers (BLAs)
thought to trace gas at $T>10^{5}$ K \citep[e.g.][]{ric04} supports
this notion, while $\OVI$ seen with very broad $\HI$ having line
widths $>100\;\kms$ may indicate even hotter $\OVI$ at $T\sim10^6$ K
\citep[e.g.][]{dan06}.

Early investigations into $\OVI$ using cosmological simulations
\citep{cen01,fan01,che03} predicted that stronger $\OVI$ absorbers
tend to be collisionally ionized while weaker ones tend to be
photo-ionized, with the cross-over equivalent width of $\sim
30-50$~m\AA.  \citet{cen06b} added non-equilibrium ionization and
used higher resolution simulation, finding similar behavior compared
to their earlier work.  The above simulations generally focused on
fitting only the observed cumulative equivalent width ($EW$)
distribution.  Recent observations now provide new challenges for
simulations to fit a wider range of low-$z$ $\OVI$ observables.

Recent surveys have renewed confusion about the nature of $\OVI$
absorbers.  Ionization models for absorbers showing aligned $\OVI$,
$\HI$, and $\CIII$ are usually forced to invoke multi-phase gas, with
both $\OVI$ in CIE and lower ionization species at photo-ionized
temperatures \citep{pro04, dan06, coo08}.  Yet T08 find that the
majority of aligned $\HI-\OVI$ absorbers can also be explained by
photo-ionization alone.  The ``blind'' searches of $\OVI$ without
$\HI$ by \citet{tho08a} indicate $\ga 95\%$ of $\OVI$ is associated
with $\HI$.  These studies suggest that $\OVI$ traces baryons at least
partially accounted for by $\HI$ absorbers, and is not necessarily a
good tracer of the WHIM.  Studies of galaxy-absorber correlations find
a wide variety of environments for $\OVI$ absorbers including voids,
filaments, galaxies, and groups
\citep[e.g.][]{tri01,tum05b,pro06,tri06}.  Furthermore, it is not
clear how the $\OVI$ in intermediate- and high-velocity clouds (IVCs
and HVCs) associated with the Milky Way (MW) halo
\citep[e.g.][]{sem03,sav03,fox05} relate to $\OVI$ observed in quasar
absorption line spectra.  A unified model was proposed by
\citet{hec02} where $\OVI$ absorbers are all radiative cooling flows
passing through the coronal temperature regime, with a modification by
\citet{fur05} for long cooling times in the IGM.  However,
observations are inconclusive as to whether this scenario also applies
to IGM $\OVI$ \citep[e.g][]{dan06,leh06}.  Hence there remains much
uncertainty in regards to which cosmic gas and metal phases $\OVI$
absorbers actually trace.

In this paper we explore $\OVI$ and $\HI$ absorbers in a range of
physical models using our modified version of the cosmological
hydrodynamics code \gad.  We pay close attention to three key
observables confirmed in multiple studies: the cumulative $EW$
distribution, $\OVI$ linewidths ($b$-parameters) as a function of
$\OVI$ column density, and the alignment of $\OVI$ with $\HI$.  Our
two main goals are (1) to see how self-consistent metal enrichment via
galactic superwinds reproduces the $\OVI$ observations, including an
honest evaluation of the short-comings of our simulations, and (2) to
understand the physical conditions and environments of $\OVI$
absorbers.  This study combines state-of-the-art modeling with the
most recent $\OVI$ data to further understand the nature of IGM
$\OVI$, in light of the much anticipated installation of the {\it
Cosmic Origins Spectrograph} (COS) and the re-activation of STIS on
{\it Hubble}.

The structure of the paper flows as follows.  In \S2, we introduce
our simulations all run to $z=0$ with three different galactic
outflow models, plus other post-run input physics variations.  Then,
in \S3 we see how the various models fit the $\lya$ forest and our
three $\OVI$ observables, including a test of resolution convergence.
The crux of this paper lies in the following sections where we
dissect our simulated absorber population, acknowledging the
imperfections in our modeling while assessing what we believe is
physically significant.  \S4 advocates that IGM $\OVI$ absorbers
are primarily photo-ionized, with an in-depth analysis of the
association with $\HI$, and forwards an explanation for the non-thermal
component of $\OVI$ line profiles as arising from small-scale
turbulence.  \S5 examines the origin of $\OVI$ absorbers with an
emphasis on environment.  We discuss the minority population of
collisionally ionized $\OVI$ here, and examine tell-tale signs of
such WHIM absorbers with an eye towards COS.  We summarize in \S6.

Throughout this paper, we adopt \citet{asp05} solar abundances, as
most low-$z$ $\OVI$ observations use this or a similar value.  This
contrasts to our use of \citet{and89} values in previous publications
(Oppenheimer \& Dav\'e 2006, hereafter OD06; Dav\'e \& Oppenheimer
2007; OD08), and results in a significant decrease in solar oxygen
mass fraction (0.00541 versus 0.00962).  The Type II supernovae (SNe)
yields we use as simulation inputs are from \citet{lim05}, and are
independent of solar values.  For ease of discussion, we classify
$\OVI$ absorbers into three categories according to their column
density: weak ($N(\OVI)<10^{13.5} \cms$), intermediate
($N(\OVI)=10^{13.5-14.5} \cms$), and strong ($N(\OVI)=10^{14.5-15.0}
\cms$).  We often split the intermediate absorbers into two bins,
since this comprises the majority of observed $\OVI$.

\section{Simulations} \label{sec:sim}

We use cosmological simulations run to $z=0$ using our modified
version of the N-body + Smoothed Particle Hydrodynamics code
\gad~\citep{spr05} in order to explore the nature of $\OVI$ in the
low-$z$ Universe.  Our simulations directly account for cosmic metal
enrichment via enriched galactic outflows, where outflow parameters
are tied to galaxy properties following that observed in local
starbursts.  Our most successful model follows scalings for
momentum-driven winds; this fairly uniquely matches a wide range of
observations such as early IGM enrichment (OD06), the galaxy
mass-metallicity relations~\citep{fin08}, and the enrichment and
entropy levels seen in intragroup gas~\citep{dav08b}.  We introduce
the simulations first, then describe variations applied to our favored
outflow model, and finally describe post-run variations to the input
physics.

\subsection{Model Runs} \label{sec:modelruns}

Our simulations adopt cosmological parameters based on 5-year WMAP
results~\citep{hin08}.  The parameters are $\Omega_{0} = 0.25$,
$\Omega_{\Lambda} = 0.75$, $\Omega_b = 0.044$, $H_0 = 70$ km s$^{-1}$
Mpc$^{-1}$, $\sigma_8 = 0.83$, and $n = 0.95$; we refer to this as the
$d$-series.  The value of $\sigma_8$ is slightly higher than favored
from WMAP data alone, but it agrees better with combined results from
Type Ia SNe and baryon acoustic oscillation data.  Our general naming
convention is d[{\it boxsize}]n[{\it particles/side}][{\it wind model}].

We use the momentum-driven wind model based on the analytic derivation by
\citet{mur05} and explored previously in OD06 and OD08.  Briefly, wind
velocity ($\vw$) scales linearly with the galaxy velocity dispersion
($\sigma$), and mass loading factor ($\eta$) scales inversely 
with $\sigma$.  The mass loading factor represents the mass loss rate
in outflows in units of the galaxy star formation rate.  We use the
following relations for $\vw$ and $\eta$:
 \begin{eqnarray}
  \vw &=& 3\sigma \sqrt{f_L-1}, \label{eqn: windspeed} \\
  \eta &=& {\sigma_0\over \sigma} \label{eqn: massload},
 \end{eqnarray} 
where $f_L$ is the luminosity factor in units of the galactic Eddington
luminosity (i.e. the critical luminosity necessary to expel gas from
the galaxy potential), and $\sigma_0$ is the normalization of the mass
loading factor.  A minor difference from OD06 and OD08 is that we no
longer impose an upper limit on $\vw$ due to SN energy limitations.

\begin{table*}
\caption{Simulations}
\begin{tabular}{lcccccc}
\hline
Name$^{a}$ &
$L^{b}$ &
$\epsilon^{c}$ &
$m_{\rm SPH}^{d}$ &
Wind Model & 
$\sigma_0^{e}$ &
$\Omega_{*}/\Omega_{b}(z=0)^{f}$
\\
\hline
\multicolumn {6}{c}{} \\
d32n256vzw150   & 32 & 2.5  &  34.6 & Momentum-driven & 150 &  0.102 \\ 
d16n128vzw150   & 16 & 1.5  &  34.6 & `` & 150 &  0.144 \\
d16n256vzw150   & 16 & 1.25  &  4.31 & `` & 150 &  0.144 \\
d32n256hzw075    & 32 & 2.5  &  34.6 & Hybrid Momentum-SNe & 75 & 0.061 \\
d32n256lzw400   & 32 & 2.5 &  34.6 & Low-$\vw$ Momentum  & 400 & 0.116 \\
\hline
\end{tabular}
\\
\parbox{15cm}{
$^a${\it vzw} suffix refers to the momentum-driven winds. {\it
hzw} suffix refers to momentum-driven winds with an added boost from
SNe. {\it lzw} suffix refers to momentum-driven winds launched with
51\% as much kinetic energy relative to {\it vzw}.  All models are run to
$z=0$\\
$^b$Box length of cubic volume, in comoving $\hmpc$.\\
$^c$Equivalent Plummer gravitational softening length, in comoving
 $\hkpc$.\\
$^d$Masses quoted in units of $10^6M_\odot$.\\
$^e$Normalization for the momentum-driven wind mass-loading factor
where $\eta = \sigma_0/\sigma$, units in $\kms$.\\
$^f$Fraction of baryons in stars.  
\\
}
\label{table:sims}
\end{table*}

\begin{figure*}
\includegraphics[scale=0.50]{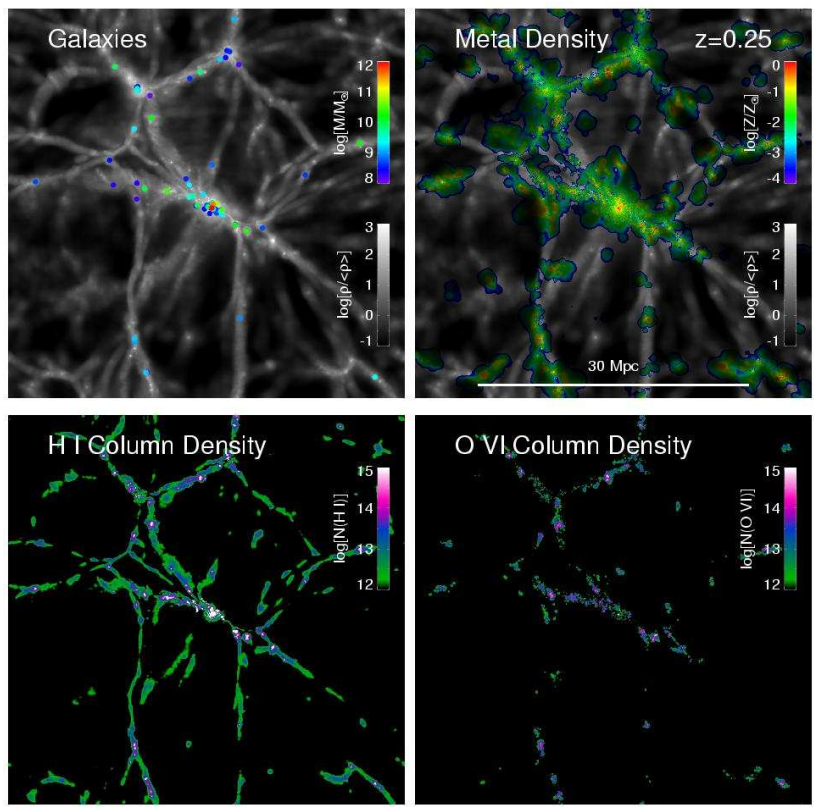}
\caption[]{A 25 $\kms$ slice of IGM spanning the 32 comoving $\hmpc$s
  of the d32n256vzw150 simulation at $z=0.25$. {\it Upper left:} The
  locations of galaxies are shown as colored points, where the color
  corresponds to stellar mass, overlaid on the gas overdensity
  represented in greyscale.  {\it Upper right:} Color indicates the
  enrichment level; 90\% of the IGM remains unenriched as the
  superwinds are unable to enrich voids.  {\it Bottom panels:} $\HI$
  (left) and $\OVI$ (right) column densities in this 25~$\kms$ slice,
  assuming a uniform \citet{haa01} ionization background. $\OVI$ is
  not as extended as $\HI$, partially due to the small volume filling
  factor of metals, however the strongest $\OVI$ absorbers are often
  found associated with sub-$M^*$ galaxies, unlike $\HI$, which has
  its strongest absorption around the large group in the center.  The
  vast majority of $\OVI$ in our simulations is photo-ionized tracing
  overdensities of 10-200.}
\label{fig:fullslice}
\end{figure*}

The model runs, detailed in Table \ref{table:sims}, have [{\it wind
model}] suffixes denoting a variation of the momentum-driven wind
model run.  Our previously used favored model, {\it vzw}, contains a
spread in $f_{L}=1.05-2.00$, includes a metallicity dependence for
$f_{L}$ owing to more UV photons output by lower-metallicity stellar
populations, and adds an extra boost to get out of the galaxy
potential well simulating continuous acceleration by UV photons.

Figure \ref{fig:fullslice} illustrates a 25 $\kms$ slice spanning the full
extent of the d32n256vzw150 simulation box at $z=0.25$.  The location of
galaxies (upper left) is well correlated with the enriched IGM (upper
right).  The $\lya$ forest traces baryons in filaments (bottom left)
down to the lowest observable column densities.  The $\OVI$ absorbers
(bottom right) mainly trace the diffuse IGM enriched via feedback,
showing a more grainy structure indicative of inhomogeneous enrichment.
Much of this paper focuses on quantifying the relationship between these
various observational and physical components of the IGM.

\begin{figure}
\includegraphics[scale=0.80]{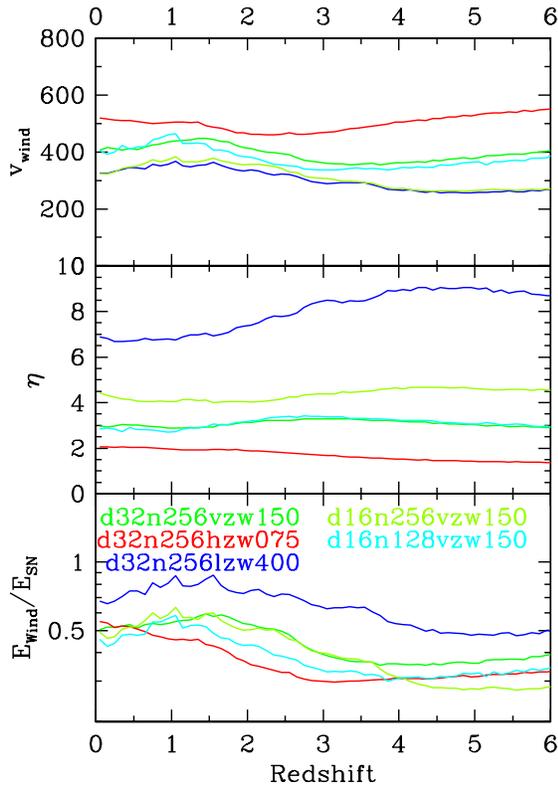}
\caption[]{Wind properties as a function of redshift for our three
wind models, plus one higher resolution simulation (d16n256vzw150),
and a simulation to test box size convergence (d16n128vzw150).  In
order to create a reasonable baryonic fraction in stars at $z=0$,
weaker winds (top panel) must have a higher mass loading factor
(middle panel) resulting in greater energy input per unit star
formation as shown by the $E_{wind}/E_{SN}$ ratio (lower panel).  The
opposite applies for the {\it lzw} model.  $E_{SN}$ is set to
$1.3\times10^{49}$ ergs $\msolar^{-1}$.  }
\label{fig:wind}
\end{figure}

We run 16 and 32 $\hmpc$ (comoving) sized boxes at the same mass and
spatial resolution (with different numbers of particles) of this wind
model.  The baryon fraction in stars is identical, while the wind
speed and mass loading shown in Figure \ref{fig:wind} are in good
agreements from $z=6\rightarrow 0$, despite a volume difference of
8$\times$ (compare green vs.  cyan lines).  $E_{wind}/E_{SN}$
decreases slightly in the smaller volume, because fewer massive
galaxies having higher wind energy efficiencies form at early times.

We also run a $16 \hmpc$ box length, $2\times256^3$ particle
simulation to $z=0$ (d16n256vzw150) to explore resolution convergence.
This run with its smaller time stepping is the most computationally
expensive run, using over 30,000 CPU hours on Harpertown Intel Xeon
processors.  The wind properties in Figure \ref{fig:wind} indicate
lower $\vw$'s and higher $\eta$'s due to increased contributions from
small galaxies below the resolution limit of the d32n256vzw150 box.
The stellar baryon fraction is 4\% higher in this simulations, mainly
due to late time star formation.  We suspect this is due to the
non-cosmological scale of this box, which cannot form as many groups
and clusters at late times where star formation is inhibited, as well
as the fact that it probes further down the relatively steep mass
function of star-forming galaxies.

In this paper we explore two other variations of the momentum-driven wind
model that bookend, at high and low wind speeds, physically plausible
winds that are able to enrich the IGM as observed.  In addition to varying
$\vw$, we must also vary $\sigma_0$ to match the observed baryon fraction
in stars today of $\sim 6-10\%$~\citep{col01}.

The high-$\vw$ case is denoted {\it hzw} in which a portion of the
energy from Type II SNe is used to increase $\vw$, while keeping the
momentum-driven wind derived $\eta$ the same.  We use $1.9\times10^{48}$
erg $\msolar^{-1}$, or one-seventh of the total supernova energy from
a Chabrier IMF, assuming each star between 8-100 $\msol$ explodes
with $10^{51}$ ergs.  Additionally, we add $1.9\times10^{48}$ erg
$\msolar^{-1}$ of SNe energy to the thermal energy of each wind particle,
and allow adiabatic expansion (i.e. no cooling) while the wind particle
is hydrodynamically decoupled.

The {\it hzw} model is designed to explore the case where we push
feedback to extreme physical limits with the specific purpose to see if
such strong feedback can make IGM metals both more widely distributed
and hotter in the low-$z$ Universe.  In order to obtain a reasonable
stellar density by low-$z$ we set $\sigma_0=75 \kms$, which results in
6.1\% of baryons in stars at $z=0$.  Figure \ref{fig:wind} shows the
wind parameters as a function of redshift.  Ironically, the lower $\eta$
of the {\it hzw} model reduces the efficiency of winds as indicated by
the lowest $E_{wind}/E_{SN}$ ratio of all three wind models explored.

To explore the low-$\vw$ case, we run the {\it lzw} model in which the
wind velocity is reduced by 50\% relative to the {\it vzw} model.  In
order to prevent the formation of too many stars by $z=0$, $\sigma_0$
is set to $400 \kms$, resulting in $\Omega_* /\Omega_b = 0.103$ at
$z=0$.  The {\it lzw} model generates the most wind energy per
baryonic mass ($E_{wind}/E_{SN}$), because of the high mass loading
factor.



\subsection{Post-Run Physics Variations}

A variety of post-run treatments are performed on the model runs,
using modifications within our quasar absorption line spectral
generator, {\tt specexbin}.  Some of these variations are used to
explore the consequences of unlikely but intuitive scenarios (i.e.  a
uniform metallicity distribution), while others are meant to account
for physics not included in our simulations (i.e. non-equilibrium
ionization).  We list these variations below, and denote such model
variations throughout the text with an additional suffix at the end of
the model name.

{\bf Z01}: This model variation uses a uniform metallicity of $0.1
\Zsolar$ for every SPH particle instead of the metallicity from the
simulation.  This unrealistic model presents an exercise to see how
$\OVI$ observations vary if metals are uniformly distributed.  $0.1
\Zsolar$ is the fiducial value that was needed in early simulation
studies to match the equivalent width distribution of low-$z$ $\OVI$
absorbers~\citep{cen01,fan01,che03}.  It is also similar to the
average $z=0$ oxygen content of the d32n256vzw150 simulation, $0.13
\Zsolar$.  We do not alter the gas densities or temperatures, so the
ionization conditions remain the same.

{\bf ibkgd}: By default, we use a spatially uniform ionization
background given of \citet{haa01} background, {\it with} a
contribution from quasars {\it and} star forming (SF) galaxies (10\%
escape fraction) included.  We use the normalization taken directly
from \citet{haa01} with no adjustment, since as we will show this
reproduces $z\sim 0$ $\lya$ forest statistics quite well \citep[see
also][]{pas08}.  The hydrogen photo-ionization rate is $\Gamma =
1.84\times 10^{-13} s^{-1}$ at $z=0.25$.  To explore variations in
photo-ionization, we consider two other uniform ionization
backgrounds:

\noindent 1. {\it ibkgd-Q1} uses intensity from the \citet{haa01}
background, minus the contribution from SF galaxies.  This mainly
affects $\HI$ properties as SF galaxies contribute $\sim\frac{2}{3}$
of the ionization at the Lyman limit, therefore the absorption from
the $\lya$ forest will grow.  The ionization intensity at the $\OVI$
ionization potential, 8.4~Rydbergs, remains mostly unaffected by
subtracting star forming galaxies.  This background may be applicable
in regions where the mean free paths of $\HI$ ionizing photons are
shorter than $\OVI$ ionizing photons, e.g. denser regions in and
around galactic haloes.

\noindent 2. {\it ibkgd-Q3} uses triple the quasar-only \citet{haa01}
background, which better reproduces the global statistics of the
$\lya$ forest.  There are now more photons at 8.4~Rydberg, which leads
to photo-ionized $\OVI$ at higher overdensities.  We explore this
alternative since little is known about the true ionization field at
these high energies.

{\bf noneq}: Our default case assumes ionization equilibrium for
all species, which we calculate using CLOUDY~\citep{fer98}.  To test
sensitivity to this assumption for collisionally ionized $\OVI$, we apply
the non-equilibrium ionization tables of \citet{gna07} as a function of
temperature and metallicity rather than our CLOUDY-generated equilibrium
tables.  \citet{gna07} calculated non-equilibrium ionization fractions
as a function of metallicity assuming collisional ionization only.
High-metallicity shocked gas cools more rapidly than it can recombine,
resulting in higher ionization states than expected at lower temperatures
when assuming CIE.  This has implications for $\OVI$ at densities where it
is collisionally ionized, because ionization fractions of a few percent
for $\OVI$ can exist at well below its CIE peak at $10^{5.45}$ K if the
gas is $\Zsolar$ or above.  There does not exist a non-equilibrium treatment for
photo-ionized gas as of yet, since this depends additionally on density;
however these ionization fractions are more likely to be regulated by
the intensity of the ionization background.  Ideally it would be best
to directly include non-equilibrium ionization as in \citet{cen06b}, but
we leave that for future work, and consider this test as bracketing the
plausible impact of non-equilibrium ionization.

To more fully explore the impact of non-equilibrium ionization, we also
run a d16n128hzw075 simulation with the \citet{gna07} cooling tables as
a function of metallicity.  The metal line cooling peaks for individual
ions are spread out over larger temperature ranges as individual ion
coolants exist over a broader swath of temperature.  The cooling is also
of the order 2-4$\times$ less in some places than \citet{sut93}, because
ion coolants exist at lower temperatures where they collide less rapidly.
This run is denoted as d16n128hzw075-noneq.  To ensure that this volume
is not too small, we run a d16n128vzw150 simulation and find general
convergence with d32n256vzw150 for the star forming, wind, and IGM
enrichment properties; therefore this box size should be sufficient
for comparison with larger volumes.


{\bf bturb}: The most critical post-run modification we employ is the
by-hand addition of sub-resolution turbulence.  As we will show, our
simulations do not reproduce the broadest observed absorption lines,
especially those with $N(\OVI)>10^{14}$ cm$^{-2}$.  Often, where
observations see one broad component, our simulations would show a
number of narrow components in a system.  This is understandable when
one considers our simulations have a mass resolution of $\bar{m}_{\rm
SPH}\approx 3.5\times 10^{7} \msol$, and metals may exist in
structures of significantly smaller scale \citep[e.g][]{rau99, sim06,
fra07, hao07, sch07}.  Even at the best-available instrumental
resolution of STIS, thermal broadening at CIE $\OVI$ temperatures
cannot explain the bulk of the line widths in the T08 dataset.
Furthermore, the alignment with $\HI$ suggests that some non-thermal
broadening mechanism is needed.  Turbulence is often used as the
overarching explanation \citep[e.g.][]{tho08b}.  We will argue that
$\OVI$ absorbers, especially strong ones, are likely to be made up of
numerous smaller metal concentrations with a range of velocities
resulting in a component dominated by turbulent broadening.

We add turbulence residing on scales below the SPH particle resolution
(``sub-SPH turbulence") by adding a turbulent $b$-parameter as a function of
hydrogen density ($n_{\rm H}$).  With no physical guide available, we
add just enough turbulence to approximately match the observed line width
distribution as a function of column density.  Specifically, we apply
a linear fit to the STIS-only sample of T08 $b(\OVI)-N(\OVI)$ relation
between $10^{12.9}-10^{14.9} \cms$ ($b_{obs} = 21.40\times \logrm[N(\OVI)]
- 268.4 \kms$), and subtract our relation from the d32n256vzw150 lines
of sight at STIS resolution ($b_{noturb} = 1.50\times \logrm[N(\OVI)]
- 8.9 \kms$), using the equation
\begin{equation} \label{eqn:bturbform}
b_{turb}^2 = b_{obs}^2-b_{noturb}^2.  
\end{equation}
$b_{turb}$ is the turbulent broadening describing the small-scale
motions within an SPH particle for this specific simulation resolution.
We cast $b_{turb}$ in terms of $n_{\rm H}$ by employing the relation
\begin{equation} \label{eqn:nh_NOVI}
\logrm[n_{\rm H}] =  0.66\times \logrm[N(\OVI)] - 13.98~\cmc,  
\end{equation}
which is a dependence we find in \S\ref{sec:photomodel}.  Putting it
together, the relation we use in {\tt specexbin} is
\begin{equation} \label{eqn:bturb1}
b_{turb} = \sqrt{1405 \logrm[n_{\rm H}]^2 + 15674 \logrm[n_{\rm H}] + 43610}~\kms ,
\end{equation}
applied only over the range $n_{\rm H}=10^{-5.31}-10^{-4.5} \cmc$, 
describing the majority of $\OVI$ within filaments but outside galactic halos.

Since the sub-SPH turbulence from Equation \ref{eqn:bturb1}~would
extrapolate to artificially high values at higher densities, we use
two more observational constraints from the $\OVI$ surveys of MW IVCs
and HVCs to describe sub-SPH turbulence associated with galactic
halos.  \citet{sem03} finds high-velocity $\OVI$ features have
$b(\OVI)=40\pm14 \kms$, and suggest this $\OVI$ arises between the
interfaces of cool/warm clouds and the $T>10^6$ K galactic
corona/intragroup medium at $R>70$ kpc with $n<10^{-4}-10^{-5} \cmc$.
We assume that these lines are predominantly turbulently broadened, as
temperature and spatial explanations fall well short, requiring
another broadening mechanism possibly related to the motion of the gas
\citep{sem03}.  If the absorption arises from CIE $\OVI$ ($T\sim
10^{5.5}$ K) then $n_{\rm H}\approx 10^{-4.5} \cmc$ applies for this
gas assuming approximate pressure equilibrium with the galactic
corona.  Therefore $b_{turb}\sim 40 \kms$ corresponds to $n_h=10^{-4.5}
\cms$ both in Equation \ref{eqn:bturb1} and in HVCs.

\citet{sav03} find $b(\OVI)\approx 60 \kms$ in IVCs, which they find
most likely reside in the galactic thick disk with $\OVI$ densities
tracing $n_{\rm H}\sim 10^{-3} \kms$ for $\OVI$ flows.  Using a linear
interpolation with these two constraints, we find the relation 
\begin{equation} \label{eqn:bturb2}
b_{turb} = 13.93 \log(n_{\rm H}) + 101.8~\kms 
\end{equation}
for $n_{\rm H}=10^{-4.5}-10^{-3.0} \cmc$, and assume a maximum
$b_{turb}=60 \kms$ for higher densities.  Very rarely do our random
sight lines intersect densities as high as $n_{\rm H}=10^{-3} \cms$,
but such dense regions can create a strong absorption profile.

At $z=0.25$, the sub-SPH turbulent broadening is $b_{turb} = 13, 22,
40,$ and $51 \kms$ at $\rho/\bar{\rho} = 20, 32, 100,$ and 320
respectively.  Besides producing wider lines, it turns out to also
help explain various other observed $\OVI$ absorber properties, as we
show in \S\ref{sec:turbulence}.


\section{Simulated vs. observed $\OVI$ and $\HI$ absorbers} 

In this section we place constraints on our models from observations
of the low-$z$ $\lya$ and $\OVI$ forests.  We begin with the $\lya$
forest, considering column density and line width distributions, and
the evolution from $z=1.5\rightarrow0$.  We then focus on three key
$\OVI$ observables: (1) the cumulative equivalent width distribution,
(2) $b(\OVI)$ as a function of $N(\OVI)$, and (3) the $N(\OVI)$ as a
function of $N(\HI)$.  We explore how each of our models fits these
observables, and explain why models the turbulent broadening provide
the best fit.  We end the section discussing numerical resolution
convergence among our simulations.

We generate 30 continuous lines of sight from $z=1\rightarrow 0$ shot
at a variety of angles through each simulation using {\tt specexbin}.
We use a continuum-normalized spectral template with 0.02 \AA~bins
convolved with the STIS instrumental resolution (7 $\kms$), and add
Gaussian noise with $S/N=10$ per pixel.  This creates a large sample of
high quality spectra covering $\Delta z=15$, larger than observational
samples, with the intention to extract subtle differences between our
various models.  For the sake of simplicity, we generate $\HI$-only and
$\OVI$-only spectra using only the strongest transition (i.e. Ly-$\alpha$
and the 1032 \AA~line).  See \S2.5 of OD06 for a more detailed description
of {\tt specexbin}.  The AutoVP package \citep{dav97} is used to fit
all component absorption lines, yielding e.g. a sample of 475 $\OVI$
components with $N(\OVI)\ge 10^{13} \cms$ in the d32n256vzw150-bturb
model.

\subsection{$\HI$ observables} \label{sec:lya}

The observationally well-characterized $\lya$ forest provides
baseline constraints on our simulations.  A successful model of
$\OVI$ absorbers must first reproduce observations of the low-$z$
$\lya$ forest, especially given that $\OVI$ is usually associated
with $\HI$ absorption \citep{tho08a}.  The $\lya$ forest traces the
warm photo-ionized IGM (i.e. $T\sim10^4$ K), which contains $\sim30\%$
of cosmic baryons \citep{pen04} and likely traces some of the same gas
as photo-ionized $\OVI$ absorbers.  We examine briefly three key $\HI$
observables spanning a range of physical and evolution properties that
our simulations must match.

\begin{figure*}
\includegraphics[scale=0.90]{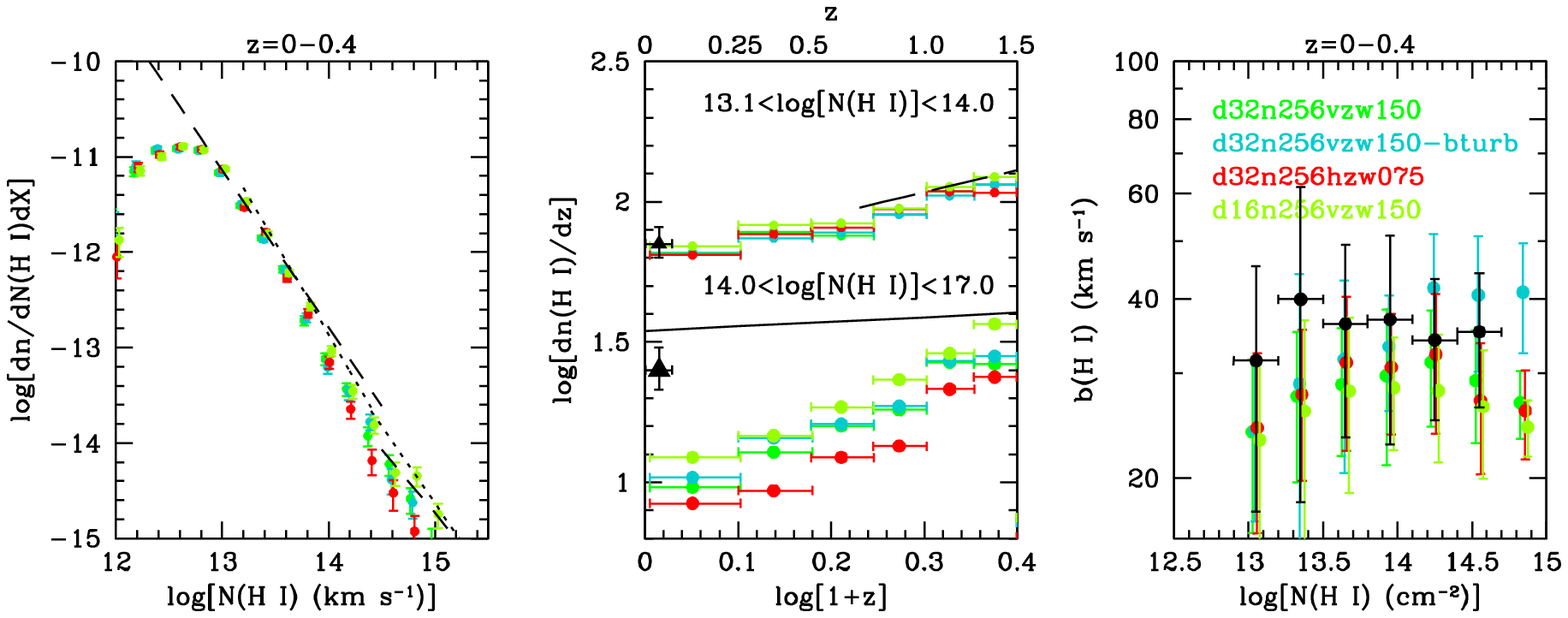}
\caption[]{Four of our models, d32n256vzw150 with and without sub-SPH
turbulence added, d32n256hzw075, and d16n256vzw150, are shown against
three general $\HI$ observables.  The left panel shows our
differential column density distribution compared to the power law
fits to data of \citet[][dotted line, $b<40\kms$ only]{leh07} and
\citet[][dashed line]{pen04}.  In the central panel, the evolution
below $z=1.5$ in two bins ($N(\HI)=10^{13.1-14.0}$ \&
$10^{14.0-17.0}$) is plotted relative to \citet[][solid line]{wey98},
\citet[][long dashed line]{jan06}, and \citet[][triangles with
large/small symbols corresponding to the strong/intermediate
bin]{pen04}.  Line width as a function of $N(\HI)$ in our simulations
is shown in the right panel compared to \citet{leh07} with one extra
bin considered for our larger simulated sample.  We fit $\lya$ for
$N(\HI)\leq10^{14.4}$ and $\lyb$ for higher column densities when
considering $b$-parameters.  Turbulence broadens the stronger $\HI$
lines noticeably.}
\label{fig:HI}
\end{figure*}

Figure \ref{fig:HI} shows the column density distribution (left),
evolution of line densities from $z=1\rightarrow 0$ (middle), and the
$N-b$ relation (right) from observations and four of our simulations,
namely d32n256vzw150, d16n256vzw150, d32n256vzw150-bturb, d32n256hzw075.
We display only these models because there exist only minor differences
in the $\lya$ forest amongst the other models.	The comparison to data
shows reasonably good agreement, except for an underestimate by at least
a factor of two of strong $\HI$ lines ($EW>240$ m\AA).	There are two
likely reasons for this: The first is that the observations compared to
have lower signal-to-noise and/or resolution than our simulated spectra,
so line blending may artificially enhance the observed equivalent widths;
and second, many of these lines arise from gas within galactic haloes,
where we may not be resolving halo substructures adequately.  We note
however that the higher resolution in the d16n256vzw150 simulation does
little to resolve this discrepancy, so either we need much higher
resolution or else line blending issues are important.

We find the low-$z$ $\lya$ forest is relatively unaffected by wind
strength; this is similar to the findings of \citet{ber06} at high-$z$,
who find galactic winds do very little to alter the main statistical
properties of the $z\sim3$ $\lya$ forest.  The only difference worth
noting is that the strong winds of the {\it hzw} reduce the frequency
of strong $\HI$ lines ($EW>240$ m\AA) by about $\sim$20\%.

We now compare our baseline {\it vzw} model with and without turbulence,
since we will later show that the d32n256vzw150-bturb model best fits
the $\OVI$ observables, and we want to quantify how sub-SPH turbulence
influences the $\lya$ forest.  Ideally, turbulence should not affect
$\lya$ observables significantly since the low-$z$ $\lya$-forest is
well-described by simulations without any added turbulence~\citep{dav99},
and the $b$-parameters are well-described primarily by thermal and Hubble
broadening \citep{dav01b}.

The left panel shows the differential $\HI$ column density distribution,
i.e. the number of absorbers per $\HI$ column density interval per
redshift pathlength $d^2n/dN(\HI)dX$, from our models compared to
power law fits by \citet[][dashed line, $z=0.002-0.069$]{pen04} and
\citet[][dotted line, $z<0.4$]{leh07}.  The different models show
little variation from each other below $N(\HI)=10^{14} \cms$, agreeing
well with the slightly steeper power law of the \citet{leh07} fit
($d^2n/dN(\HI)\propto N(\HI)^{-1.83}$) over the same redshift interval.
We use fits to $\lyb$ for $N(\HI)\ge 10^{14.4} \cms$ absorbers, but find
no statistical difference versus using only $\lya$ for the frequency
of strong lines fit with AutoVP.  We find fewer higher column density
lines than observed (as also indicated by our comparison to the $dN/dz$
of high equivalent width lines), but we note that an observer typically
climbs up the Lyman series to fit the strongest lines resulting in a
more accurate fit.  

We consider the redshift evolution of $\lya$ absorbers in the central
panel of Figure \ref{fig:HI}.  The evolution is rapid down to $z\sim
2$ but slows significantly below $z\sim 1$ \citep{bah91,imp96,wey98},
due to a decreasing ionization field strength counter-balancing
decreasing rates of recombination due to Hubble expansion
\citep{dav99}.  Matching this trend is important, because the lack of
evolution in $\HI$ may well be reflected in $\OVI$.  In this paper we
do not explore $\OVI$ evolutionary trends, leaving it for future work,
but we broadly find a comparable amount of evolution as in $\HI$.  We
count the frequency of lines, $dN/dz$, for intermediate ($240>EW\geq
60$ m\AA, $N(\HI)=10^{13.1-14.0} \cms$) and strong absorbers
($EW\geq240$ m\AA, $N(\HI)=10^{14.0-17.0} \cms$).  Our simulated
intermediate absorbers agree with the evolution from higher $z$
\citep{jan06} to the local Universe \citep{pen04}.  We do not
calibrate the factor of two mismatch in stronger absorbers, because
this more likely has to do with unresolved halo substructure and the
breakdown of the optically thin assumption of the ionization
background in such regions; this is beyond the scope of this paper,
and is not very relevant to the bulk of $\OVI$ absorbers, as it turns
out.


The rightmost panel of Figure \ref{fig:HI}~shows the $b(\HI)$-$N(\HI)$
correlation, with 1-$\sigma$ dispersions, compiled from a number of quasar
sight lines by \citet{leh07}.  Our absorbers show a slow but steady
increase until $N(\HI)=10^{14} \cms$.  Above $N(\HI)>10^{14.4} \cms$,
fitting to $\lyb$ rather than $\lya$ avoids saturation effects, and yields
systematically smaller $b$-parameters in AutoVP.  The \citet{leh07}
sample combines FUSE and STIS data, and could result in different
$b$-parameters than our uniform simulated STIS spectra (with typically
higher S/N).  The turbulent broadening starts making an impact in $\HI$
lines with $N(\HI)\ga 10^{14.0} \cms$, tracing overdensities around 30,
although it is not obvious whether the agreement with data is improved
or worsened.  A detailed comparison must wait for when the simulated
spectra are generated and analyzed identically to observations.

An important marker of WHIM gas is broad $\lya$ absorbers (BLAs).
Ideally, we would like to derive the frequency of BLAs since they
are often thought to trace the same WHIM gas as collisionally ionized
$\OVI$ \citep[e.g.][]{ric04}.  However, continuum fitting uncertainties,
spectral resolution, and $S/N$ sensitivity likely make any such comparison
fraught with systematics.  We find BLA frequencies of $dn(BLA)/dz=16$
and 12 using $S/N=20$ and 10, respectively, for $N(\HI)\ge 10^{13.2}
\cms$ in the d32n256vzw150-bturb simulation, compared to $30\pm 4$ from
\citet{leh07}.  However, 39\% of these absorbers have $N(\HI)>10^{14.0}
\cms$, while the data shows a lower fraction of strong BLAs.  We appear
to be missing a population of weak BLAs; however we leave a detailed
comparison of such properties for future work during the COS era.

Our review of $\HI$ finds we reproduce most key observables; only
subtle disagreements among our various models are noticeable, and then
only for strong lines above $N(\HI)=10^{14} \cms$.  These disagreements
may be related to (1) detailed differences between processing observed
and simulated spectra, and (2) the inability of simulations to resolve
absorbing substructures in halo gas.  The strongest winds make 20\% fewer
$N(\HI)=10^{14.0-17.0} \cms$ absorbers at all redshifts below $z=1$, while
added turbulence appreciably broadens only absorbers with $N(\HI)\geq
10^{14.0} \cms$.  Much remains to be learned about the observed low-$z$
$\lya$ forest by comparing simulations to future observations with COS,
especially for BLAs; we leave for future work examining the entire Lyman
series with identical reduction and analysis methods applied to observed
and simulated spectra.

\subsection{$\OVI$ Observables} \label{sec:observables}

We now consider three robust $\OVI$ observables with trends
corroborated by multiple groups.  We explain their physical
significance, and then explore how each of our models, all at the same
resolution, fits these three sets of data.  Lastly, we compare models
at different resolutions to examine the issue of numerical resolution
convergence.

\begin{figure*}
\includegraphics[scale=0.90]{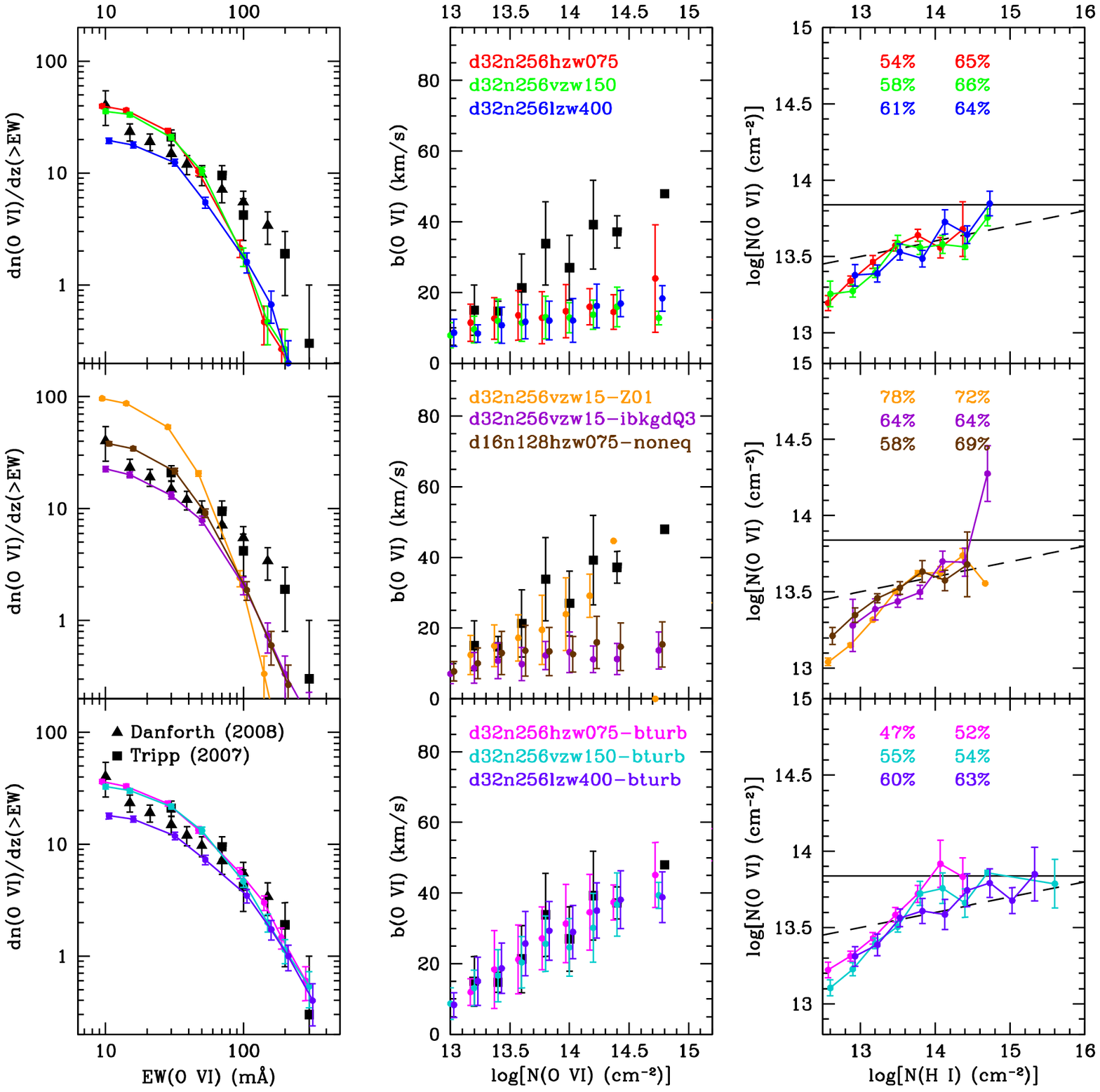}
\caption[]{Nine models are fit to three $\OVI$ observables: the
cumulative $EW$ distribution (left) with data from T08 (squares) and
DS08 (triangles), $b(\OVI)$ as a function of $N(\OVI)$ (center) over
$z=0.15-0.5$ from T08 components, and the $N(\OVI)$ as a function of
$N(\HI)$ (right) with a fit from \citet{dan05} (dashed line) and the
relation from T08 (where $N(\OVI)=10^{13.84} \cms$, solid line) along
with two percentages indicating the fraction of well-aligned $\OVI$
absorbers with $N(\OVI)=10^{13.5-14.0}$ (left) and
$N(\OVI)=10^{14.0-15.0}$ (right); the corresponding observed alignment
fractions are 57 and 43\% respectively.  The top panels explore how
different wind models fit.  The middle panels explore other model
variations such as a quasar-only ionization background scaled to the
$\HI$ ionization rate, a uniform $Z=0.1 \Zsolar$ metallicity
distribution, and non-equilibrium ionization ratios from \citet{gna07}
for $\OVI$ at CIE densities.  The bottom panels show the addition of
sub-SPH turbulence for all three wind models, which we suggest are
small-scale velocities below our resolution limit in our simulations,
yet provides the best fit to the data by broadening lines and
increasing the $EW$ of the strongest absorbers.  Error bars show 1
$\sigma$ errors in our sample on the left and right panels, while in
the center panel they indicate 1 $\sigma$ dispersions to indicate the
range of $b$-parameters.  Slight offsets in the x-axis values are for
the ease of visibility.}
\label{fig:OVI}
\end{figure*}

The cumulative $EW(\OVI)$ component distribution (1032 \AA~line) below
$z\leq 0.5$, shown in the left panels of Figure \ref{fig:OVI}, is the most
commonly published low-$z$ $\OVI$ observable.  It is relatively
independent of spectral resolution and noise, so long as a given
equivalent width systems is identifiable.  T08 performed a blind
$\OVI$ search (i.e. search for the $\OVI$ doublet alone without
requiring corresponding $\HI$), explicitly differentiating between
components and systems.  DS08 identify $\OVI$ associated with $\HI$
absorbers, and estimate they miss $\sim20\%$ of $\OVI$ absorbers this
way.  Their larger sample has the advantage of reaching down to
10~m\AA.  This is especially important, because \citet{tum05a} suggest
that a turnover in this distribution indicates the extent to which
metals are distributed to lower overdensities.  The turnover in the
DS08 data set is not as pronounced as in the smaller \citet{dan05}
dataset at 30 m\AA.  We will argue that the declining numbers are
weakly indicative of a metallicity gradient with density.  Figure
\ref{fig:fullslice} illustrates the decline in $\OVI$ at weak column
densities; fewer regions exist where $N(\OVI)\la 10^{13} \cms$ (green
regions in the bottom right) compared to $\HI$ (bottom left), which is
not observed to turnover at $N(\HI)\la 10^{13} \cms$
\citep[e.g.][]{pen04}.

Our second observable is $b(\OVI)$ as a function of $N(\OVI)$.  In the
literature, this plot is often used to diagnose the physical state of
absorbers, as in \citet{hec02}.  That paper models $\OVI$ absorbers
as radiatively cooling flows passing through the coronal temperature
regime, finding a linear correlation between $N(\OVI)$ and $b(\OVI)$
corresponding to the cooling flow velocity, and independent of density
and metallicity.  They also show non-IGM $\OVI$ absorbers ($\OVI$
around local group galaxies and starbursts) follow this relation very
closely, and \citet{leh06} finds that IGM absorbers agree with the
same trend.  However, this latter paper fails to match the predicted
$N(\NeVIII)/N(\OVI)$ ratios, and suggests the IGM lines may instead
be photo-ionized.  \citet{dan06} also do not see $\OVI$ absorbers
following the \citet{hec02} trend, but note that their lines may suffer
from blending and exaggerated $b$-parameters.

We use only the highest resolution (STIS) data of intervening absorbers
from T08, and plot these 55 absorbers spanning $z=0.13-0.50$ in 0.2
dex bins in the central panels of Figure~\ref{fig:OVI}, with 1 $\sigma$
dispersions.  We consider $b(\OVI)$ as a function of $N(\OVI)$ (rather
than vice-versa) because we will show that our absorbers are primarily
photo-ionized with the underlying relation being density-$N(\OVI)$.
A clear upward trend is obvious in this data sample with lines
$N(\OVI)>10^{14} \cms$ averaging 35.6 $\kms$.  These lines cannot be
explained by thermal broadening alone, because the thermal $b$-parameter
is only 17.7 $\kms$ for the optimal $\OVI$ CIE temperature of $\approx
300,000$~K.  Hence there is extra broadening in these lines; possibilities
include instrumental broadening, Hubble flow broadening, and turbulent
broadening, all of which we will explore.

The final observable is $N(\OVI)$ as a function of $N(\HI)$ for
well-aligned components, shown in the right panels of Figure
\ref{fig:OVI}.  \citet{dan05} fit this relation with a $N(\HI)^{0.1}$
power law, i.e. $N(\OVI)$ rises very mildly with $N(\HI)$.  They thus
argue that these two ions trace different IGM phases, with $\OVI$
tracing collisionally ionized gas, and consider this the multi-phase
ratio.  T08 also finds very little variation of $N(\OVI)$ with
$N(\HI)$, but show that single-phase photo-ionized models can
reproduce much of this trend if metallicities are allowed to vary.
They find their aligned absorbers are well-described by the median of
their robust $\OVI$ sample, $N(\OVI)=10^{13.84} \cms$, shown as a
solid line.  They further argue that well-aligned $\OVI$ with $\HI$ is
most likely photo-ionized; this conclusion is also reached by
\citet{tho08a}, who argue that slightly increasing $\OVI$ strength
with greater $N(\HI)$ implies a declining ionization parameter with
increasing gas column density.  The specifics of the physics revealed
by this trend are explored further in \S\ref{sec:HIalign}.

We define ``well-aligned'' $\OVI$ and $\HI$ absorption as having
$\delta v\equiv |v(\HI)-v(\OVI)| \leq 8 \kms$, which corresponds to a
Hubble-broadened spatial extent of $103\pm 3$ kpc given the
uncertainty in cosmological parameters from the WMAP5 results.  We
compare our models to the T08 dataset by applying the exact same
criteria in searching for the closest $\HI$ component to every $\OVI$
component.  We only consider absorbers at $N(\OVI)\geq 10^{13.5}
\cms$, which is essentially a complete sample in $S/N\geq 10$ spectra.
For the same reason, we require $N(\HI)\geq 10^{12.9}$, so that an
observed high-$S/N$ sight line does not have a greater alignment
fraction due to more detected weak $\HI$ absorbers compared to a
low-$S/N$ sight line.  A single $\HI$ component is allowed to be
associated with multiple $\OVI$ absorbers; the converse is {\it not}
applicable when we discuss the fraction of $\HI$ absorbers aligned
with $\OVI$ later in \ref{sec:env}.  T08 applies an alignment criteria
that considers instrumental resolution (see their \S2.3.3), but we use
a simple limit of 8 $\kms$ because almost every component is
determined to an error below this limit.  We use the T08 dataset over
the \citet{tho08a} dataset because the former is larger and spans a
larger range of $N(\OVI)$.  We make an exception for one strong system
at $z=0.20266$ in PKS~0312-770, since T08 does not fit $\HI$ owing due
to large uncertainties, and we instead use the parameters found by
\citet{tho08b}.  This is an important system, because it is one of the
strongest $\OVI$ absorbers with the strongest aligned $\HI$.

In the upper left area of the right-hand panels in
Figure~\ref{fig:OVI} we show two percentages, corresponding to the
fraction of well-aligned $\OVI$ components in two
intermediate-strength $N(\OVI)$ bins: $10^{13.5-14.0}$ and
$10^{14.0-15.0} \cms$.  For comparison, the T08 dataset has 57\% and
43\% of strong absorbers well-aligned in the low and high bins
respectively by our criteria.  T08 considers 37\% of their $\OVI$ {\it
systems} aligned, however this is a very different consideration where
such systems have a single $\OVI$ component well-aligned with $\HI$.
We do not consider simple versus complex systems as defined by T08,
because the definitions are not easily applicable to an automated
procedure as we run on our simulations; comparing simulated and real
systems is a challenge left for a future paper where the data
reduction and analysis are replicated as closely as possible.

In the following subsections, we step through each of our models, and
assess how the variations in input physics impact observable
properties.  We emphasize the observables that differentiate amongst
models in the discussion below.  We save the interpretation of
well-aligned absorbers for a detailed examination of ionization models
in \S\ref{sec:HIalign}

\subsubsection{Wind Velocity Strength} \label{sec:modelvar_wind}

The top panels of Figure \ref{fig:OVI} show a comparison between our
three wind models, with our default {\tt specexbin} parameters.  For
all three wind models, there is a clear dearth of high-$EW$ lines, and
the observed line widths are progressively larger than predicted to
higher $N(\OVI)$.  Hence wind strength does not affect the high-$EW$
components nor their $b$-parameters, which is an indication that the
strongest lines are saturated and depend little on their actual
metallicity as long as it is above a certain threshold.  Since varying
wind strength cannot explain these data, we require another
explanation, which we will explore in the coming sections.
\citet{cen06b} also find their simulations with and without superwinds
make no difference at $EW\geq 100$~m\AA; however our SPH code versus
their grid-based code treats the distribution of metals on large
scales very differently.  In our case, \gad~simulations without
outflows barely produce any IGM absorbers (Springel \& Hernquist 2003;
OD06), thus we do not explore this case at low-$z$.

\begin{figure}
\includegraphics[scale=0.80]{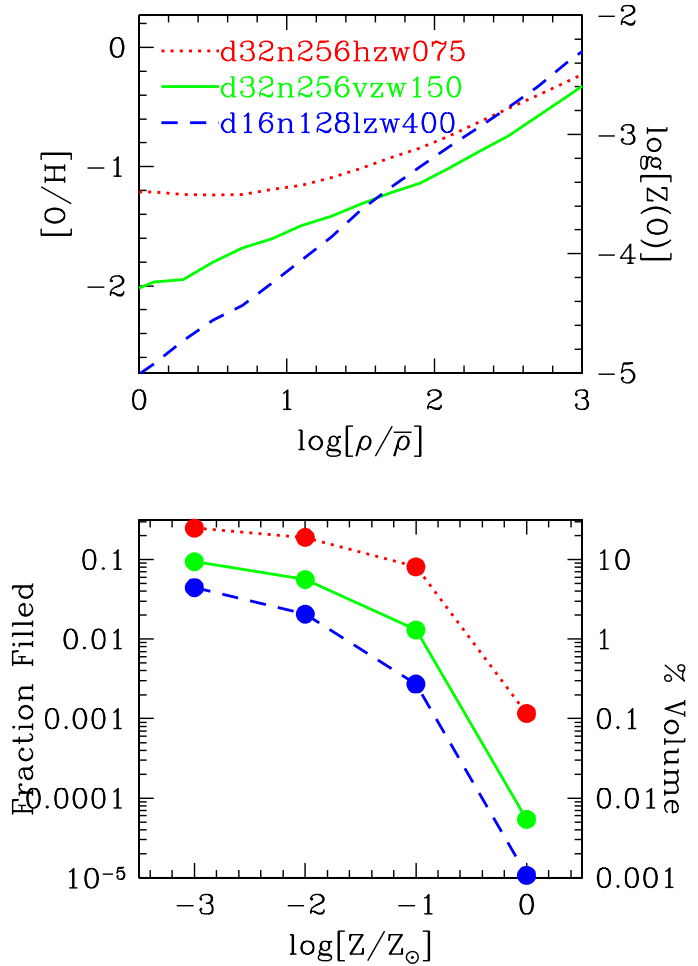}
\caption[]{The average metallicity per density for the three wind
models in the top panel show that while all wind strengths form a
metallicity-density gradient, stronger winds flatten this
relationship.  The volume filling factors at various limiting
metallicities in the bottom panel indicate wind strength plays a major
role.  At $Z=0.1 \Zsolar$, a factor of 27 reduction in volume filling
factor across the three wind models is reflected in observations only
by 50\% fewer weak components. }
\label{fig:rho_Z}
\end{figure}

Varying the strength of the winds has the largest observable
consequence for the number of low-$EW$ $\OVI$ components.  Comparing
{\it vzw} and {\it lzw} best illustrates the suggestion of
\citet{tum05a} that the turnover seen at low $EW$ is a consequence
of declining IGM enrichment further from galaxies, and can be used
to constrain the extent of the dispersal of metals.  \citet{fan01}
find their simulations generate a turnover in the $EW$ distribution
at $z=0$ due to a metallicity-density gradient.  Indeed, all our
simulations produce metallicity-density gradient as we display in
the top panel of Figure~\ref{fig:rho_Z}; OD06 also showed strong
gradients at higher redshift.  While the steeper gradient of the
{\it lzw} model yields an observational signature, the milder
flattening of the {\it hzw} below overdensities of 10 has no
observable effect.

$\OVI$ does not distinguish the density-metallicity gradient well;
however, a key point is that our $\OVI$ absorbers arise from a fairly
small volume filling factor of metals, and this is strongly governed
by $\vw$.  By $z=0.25$, our three simulations have volume filling
factors where $Z\geq 0.01 \Zsolar$ (determined from oxygen) of 1.9\%,
5.5\%, and 18.0\% as $\vw$ increases ({\it lzw}$\rightarrow${\it
vzw}$\rightarrow${\it hzw}).  We determine these filling factors from
lines of sight in physical space (i.e. no velocity smearing) using
{\tt specexbin}, plotting them in the bottom panel of
Figure~\ref{fig:rho_Z} for various metallicities.  $\OVI$ generally
arises from regions with $Z\geq 0.1 \Zsolar$, where volume filling
factors are only 0.3\%, 1.3\%, and 8.1\% with increasing $\vw$.  The
{\it lzw} model differentiates between the amount of volume filled
with half as many absorbers at $EW=15$ m\AA~and below, but the {\it
vzw} and {\it hzw} models are indistinguishable despite even larger
disparities in the volume filled.  This disparity grows even further
at the $\Zsolar$ limit; the {\it hzw} model enriches 18$\times$ more
volume than the {\it vzw} model, which fills only 1/15,000th of a
cosmological volume!  These relatively small filling factors in our
simulations are a consequence of the fact that, as we showed in \citet{dav07}
and OD08, winds at late times enrich an increasingly smaller comoving
volume, while many of the metals launched at early times reaccrete
onto galaxies.  Hence metals increasingly reside in compact
high-overdensity regions at lower redshifts.

It is remarkable that an order of magnitude change in the volume
filling factor at the fiducial $\OVI$ metallicity is observationally
indistinguishable.  It is not mysterious however, because the
mass-weighted oxygen abundance (only 7\% less in {\it vzw} compared to
{\it hzw}) determines most trends in the observables.  The Figure
\ref{fig:fullslice}~slice of the d32n256vzw150 simulation illustrates
that the volume enriched (upper right) does not correlate to where
observable $\OVI$ is (bottom right; compare blue, pink, \& white
regions).

The alignment of $\HI$ and $\OVI$ appears to provide no indication of
the metallicity distribution either.  The percentage of aligned
components is statistically indistinguishable in the two $N(\OVI)$
bins in the upper right panel of Figure \ref{fig:OVI}~as well,
which could be considered surprising given how differently oxygen is
distributed.  We take this as an indication that the peculiar
velocities are more important when considering aligned components
rather than spatial distribution.  It is disappointing that one cannot
constrain the metallicity-density gradient or the volume filling
factor from $EW(\OVI)$, despite $\OVI$ being an ideal tracer of
photo-ionized low-density metals \citep{dav98}.


\subsubsection{Uniform $Z=0.1 \Zsolar$ Distribution}

In the middle row panels of Figure \ref{fig:OVI} we explore variations
of post-run physics.  Applying a uniform metal distribution (orange
lines in the middle panels) goes in the opposite direction of how our
models need to be modified to fit the $EW$ distribution, since there are
now more metals in low-density regions and fewer in high-density
regions compared to our self-consistent enrichment models, which have
metallicity-density gradients shown in Figure \ref{fig:rho_Z}.
Conversely, it illustrates the possibility of explaining large
$b$-parameters with spatially extended structures undergoing Hubble
expansion.  Stronger components do show greater spatial broadening,
with $b\sim 30 \kms$ at $N(\OVI)\geq 10^{14.1} \cms$ corresponding to
spatial scales of $\sim400$ kpc.  With 100\% volume filling factor of
$Z\geq 0.1 \Zsolar$, the number of $\geq 10$ m\AA~components only
triples over the d32n256vzw150 wind-distributed metals, which has a
volume filling factor $80\times$ less; again, this shows that low-$EW$
lines are a weak tracer of volume filling factor.  Lines with
$EW\geq100$ m\AA~are statistically indistinguishable suggesting that
these thin lines are already saturated with only 0.1 $\Zsolar$.

The slope of $N(\HI)-N(\OVI)$ relation steepens due to more weak
$\OVI$ absorbers, because there is an excess of such components near
the $\OVI$ detection limit. This model does show a 17\% higher
incidence of well-aligned components with $N(\OVI)=10^{13.5-14.0}
\cms$ (78\%) than any other model, which is a statistically meaningful
difference.  Metals are smoothly distributed over the same
fluctuations that create $\HI$ absorption leading to this higher
fraction.  The fact that the T08 dataset finds a lower fraction (57\%)
is an indication that metals are distributed differently than baryons
traced by $\HI$.  In summary, applying a uniform metallicity does not
appear to reconcile the models and observations.

\subsubsection{Ionization Background}

The middle row of Figure \ref{fig:OVI} illustrates the impact of
varying the ionizing background.  37\% fewer $EW\leq 30$ m\AA~$\OVI$
lines create a more obvious turnover when the quasar-only background
is tripled ({\it ibkgd-Q3}), while leaving the $\lya$ forest
statistics nearly unchanged.  The peak of photo-ionized $\OVI$
ionization fractions moves to higher overdensities due to the higher
ionization parameter at 8.4 Rydbergs, resulting in $\OVI$ not tracing
lower overdensities ($\sim 10$) nearly as much, and leading to an
underestimate of weak absorbers relative to the data.  If future
datasets from COS show a greater downturn in the $EW$ distribution,
then perhaps this harder ionization field is more appropriate.  Little
is know about the ionization field at 8.4 Rydbergs, however the
quasar+galaxy \citet{haa01} background works extremely well for both
the $\HI$ and $\OVI$ observables we consider.

The {\it ibkgd-Q1} background leaves the $\OVI$ statistics unaltered,
but creates significantly more $\HI$ absorption, in conflict with
observations.  We do not plot this in Figure \ref{fig:OVI} as the only
change is the $\HI-\OVI$ relation moves rightward by nearly 0.5 dex due
to greater $\HI$ optical depths at a given overdensity.




\subsubsection{Non-Equilibrium Ionization}

Lastly, the middle row panels of Figure \ref{fig:OVI} illustrate the
impact of non-equilibrium cooling.  \citet{gna07} calculate
non-equilibrium ionization states and cooling efficiencies for rapidly
cooling shock-heated gas without any ionization background.  Since
recombination lags cooling, the temperature range over which
collisionally ionized $\OVI$ can exist extends below $T=10^5$ K.  When
we apply only their metallicity-dependent ionization fractions (their
isochroic case) to where $\OVI$ is primarily collisionally ionized
(i.e. $n_{h}>10^{-4.1} \cmc$ at $z=0.25$), we find a slight but
immeasurable increase in $\OVI$ absorption from complex systems
tracing halo gas.  Collisionally ionized $\OVI$ is primarily found
near galactic halos, which have a small volume filling factor at
low-$z$ (see \S\ref{sec:collion}).

\citet{cen06b} added non-equilibrium ionization directly to their
simulations, and find an appreciable reduction in $\OVI$ tracing
lower overdensities, because it is here that CIE timescales are
longer than shock-heating timescales; $\OVI$ remains in CIE longer
in the equilibrium case.  Similar two-temperature behavior was shown
by \citet{yos05} to have observational consequences in WHIM and ICM
gas when applied to their cosmological simulations.  Hence it is
possible that including non-equilibrium ionization self-consistently
during the simulation may yield a larger difference.

As an initial attempt towards this, we run the d16n128hzw075-noneq
simulation with non-equilibrium metal-line cooling from \citet{gna07}
{\it during} the run.  However, this creates no appreciable difference
in $\OVI$ either.  The broader metal-line cooling bumps actually
appear to make it slightly easier for baryons to cool into galaxies,
and there is a slight increase in $\Omega_*/\Omega_b$ at $z=0$ (6.3\%
versus 5.7\% from our d16n128hzw075 test simulation).  However, $\OVI$
observables are insensitive to such a small change in star formation,
and the brown curves in the middle row panels of Figure \ref{fig:OVI}~are
indistinguishable from the assumption of CIE.

Density-dependent non-equilibrium ionization effects with the
inclusion of an ionization background could affect photo-ionized
$\OVI$, especially given the fact that these metals get into the
diffuse IGM via shocked outflows.  Recombination timescales exceed the
Hubble time in such regions, and the two-temperature behavior of ions
and electrons may have unforeseen consequences, possibly moving oxygen
seen as $\OVI$ to higher ionization states.  It remains to be seen how
big of an effect this is, and non-equilibrium effects on photo-ionized
$\OVI$ will be necessary to explore.

Overall, non-equilibrium ionization effects may have some impact on
weak $\OVI$ systems, but is unlikely to provide an appreciable change
to $b$-parameters as required to match the $N(\OVI)-b(\OVI)$ relation.
We plan to implement full non-equilibrium oxygen network during our
simulation run to study this possibility further, but it seems
unlikely to yield a large difference.

\subsubsection{Sub-SPH Turbulence}

We now consider the by-hand addition of sub-resolution turbulence,
given by $b_{turb}$ as described in Equations~\ref{eqn:bturb1}
and \ref{eqn:bturb2}.  In this case, T08's observed $N(\OVI)-b(\OVI)$
relation in the bottom center panel of Figure~\ref{fig:OVI} is reproduced
by construction.  However, an added consequence of the broader lines
is that some of the lines that were saturated in the non-turbulent
case now lie on the extended linear portion of the curve of growth.
Their equivalent widths grow dramatically, and these models provide
a good fit to the cumulative $EW$ distribution at 50 m\AA~and above.
Our favored model, the d32n256vzw150-bturb, matches the T08 $EW$
distribution as well as the~10 m\AA\ data of DS08.  All the wind models
shown in the bottom panels behave in the same way relative to each other
as discussed in \S\ref{sec:modelvar_wind}.

A major caveat is that, as we show in \S\ref{sec:resolution}, the
overall abundance of $\OVI$ absorbers is not completely converged with
numerical resolution, in the sense that higher resolution runs tend to
show more absorbers at high $EW$.  Hence the good agreement with
d32n256vzw150-bturb should be taken with some caution, and our wind
model may need further refinement.  The robust result here is that
only the addition of turbulence is able to change the {\it shape} of
the $EW$ distribution to resemble that observed.

\begin{figure*}
\includegraphics[scale=0.90]{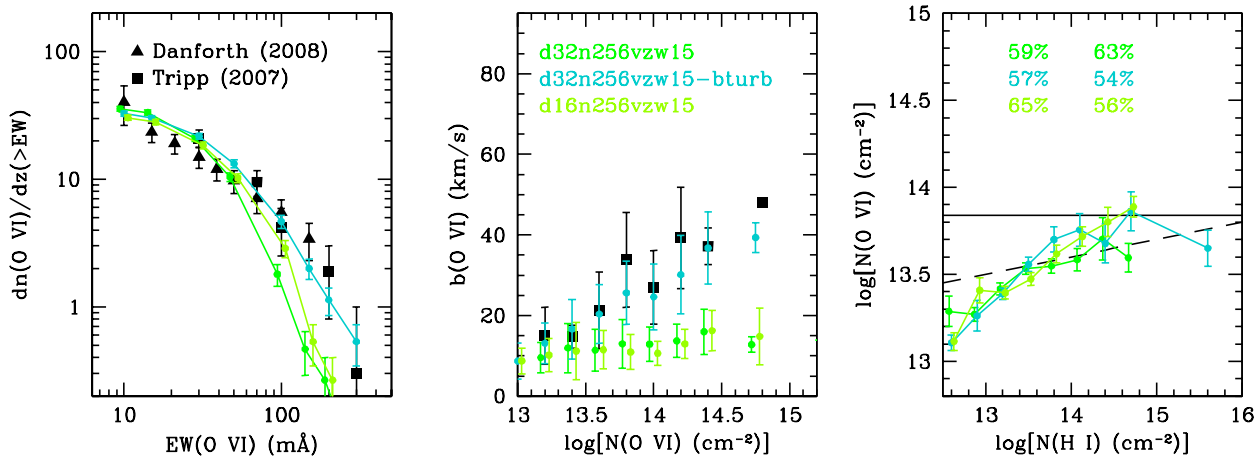}
\caption[]{Three models with {\it vzw} winds are plotted against
the same three $\OVI$ observables in Figure \ref{fig:OVI}.  The
d32n256vzw150 and d32n256vzw150-bturb are plotted on the same panels
to directly show how turbulence affects observational signatures,
while the d16n256vzw150 model explores the effect of increasing mass
resolution by 8$\times$.  }
\label{fig:OVIres}
\end{figure*}

We focus on the differences between turbulent and non-turbulent cases
in Figure \ref{fig:OVIres}, where we have included the d32n256vzw150
and d32n256vzw150-bturb models along with the higher resolution
d16n256vzw150 model, which we discuss below.  Turbulence blends
components, primarily weaker ones in complex systems, which would be
identified individually without turbulence; this is evidenced by 21\%
fewer 10-30 m\AA~components\footnote{The cumulative $EW$ distribution
only shows a 7\% difference summing all absorbers $\geq 10$ m\AA.},
and could be only a minor factor in explaining the turnover at
low-$EW$.  When considering $\Omega(\OVI)$ by summing all components
between $10^{12.5}-10^{15.0}$ using equation 6 of DS08, we find
$\Omega(\OVI)=3.8\times 10^{-7}$ with turbulence\footnote{This value
compares favorably with DS08's value of $4.1\pm0.5\times 10^{-7}$
above a similar limit ($30$ m\AA), but we do not wish to draw too much
into this comparison, since our simulation resolution limits may be
hiding more $\OVI$, see \S\ref{sec:resolution}.}, only 5\% more than
without; turbulence should not affect column densities much, because
the same amount of $\OVI$ remains in a sight line.

Both simulations have the same amount of $\OVI$ along a line of sight,
but AutoVP will often fit components with and without turbulence
dramatically differently.  An example is an absorber at $z=0.19$ where
the turbulently broadened line is fitted with
$N(\OVI)=10^{14.73}\cms$, $b=38\kms$, and $EW=305$ m\AA, versus
$N(\OVI)=10^{14.33}\cms$, $b=9.4\kms$, and $EW=90$ m\AA\ in the
non-turbulent case.  Without turbulence, much of the $\OVI$ remains
hidden in the saturated line center.  T08 finds no lines with at least
$10^{14.3}\cms$ having $b$-parameters under 33 $\kms$ using STIS.
This is the typical absorber that greatly enhances the high end of the
turbulent $EW$ distribution, increasing it by 2.8$\times$ despite
having the same underlying physical distribution of $\OVI$.  along the
sight line.

Every single one of our models follows shows increasing $N(\OVI)$ with
$N(\HI)$, which as \citet{tho08a} argue indicates a decreasing
ionization parameter for stronger $\OVI$, and suggests a photo-ionized
origin.  The aligned absorber percentages in the lower bin are only
2\% apart with and without turbulence, and agree statistically with
the T08 percentages.  There are more aligned absorbers at
$N(\HI)>10^{14.5} \cms$, because of the blending of separate $\HI$
components in very dense regions due to added turbulence.

It is worthwhile to ask whether it is fair to add such a ``fudge factor''
in our models just to fit the data.  In its favor, it does separately
alter the shape of the $EW$ distribution as desired, and there seem to
be no other effects capable of doing so.  Still, it is an unsatisfyingly
ad hoc addition, and in \S\ref{sec:turbulence} we attempt to further
explore the physics responsible for $b_{turb}$, and understand their
implications for small-scale motions in the IGM.

\subsubsection{Resolution Convergence} \label{sec:resolution}

We use the d16n256vzw150 simulation only to test resolution
convergence, and not to explore the general properties of $\OVI$
absorbers, since it does not contain a statistical sample of galaxies
at $M^*$ nor a representative volume of low-$z$ environments.
Although the d16n256vzw150 simulation produces 40\% more star
formation, this is not evidenced in the low end of the $EW$
distribution in Figure \ref{fig:OVIres}~where there exist 26\% {\it
fewer} absorbers between 10-50 m\AA~than the d32n256vzw150 simulation.
The excessive late-time star formation in the d16n256vzw150 simulation
(as mentioned in \S\ref{sec:modelruns}) enriches mainly higher
overdensities probed by stronger $\OVI$ absorbers (50\% more 100
m\AA~$\OVI$ absorbers) and lower ionization metal species.  Although
the d16n256vzw150 simulation is resolving more galaxies down to $\sim
10^{8} \msolar$, these galaxies do not seem to be a significant
contributor to the IGM $\OVI$.  Instead, the difference for low-$EW$
absorbers appears to depend on the hydrodynamics of outflows at
different resolutions-- winds from the same mass galaxy at higher
resolution do not enrich as effectively low density regions where weak
$\OVI$ absorbers arise.  This could be a result of more resolved
substructure leading to greater hydrodynamic slowing at higher
resolutions.  We caution that the differences could also be partly
numerical artifacts inherent to SPH, such as excessive viscosity and
poor two-phase medium separation.  Furthermore, there is no evidence
we can resolve the sub-SPH turbulence any better at $\times 8$ greater
resolution as evidenced by the $b$-parameters showing nearly no change
between the two simulations; this turbulence must be added explicitly
at cosmological resolutions.

The $N(\HI)/N(\OVI)$ ratio in right panel of Figure~\ref{fig:OVIres}
again shows essentially no difference between the two resolutions.
There are more strong aligned $\OVI$ absorbers above
$N(\OVI)=10^{14.1} \cms$ due to greater enrichment of overdense
regions in the d16n256vzw150 box.

We do not explore our d64n256vzw150 box, because (1) this simulation
significantly underestimates SF at $z>2$ as it cannot resolve below
$M^*$ at these redshift, and (2) the mean interparticle spacing is
greater than the sizes of absorbers ($50-100$ kpc) we find in the next
section.  The $32 \hmpc$ $256^3$ is above the critical resolution
where we can resolve the production of metals in $\OVI$ absorbers and
the sizes we derive for these absorbers.

Overall the resolution convergence of $\OVI$ absorber properties,
while mostly statistically consistent in Figure \ref{fig:OVIres},
indicate some numerical issues within our simulations.  It appears
unlikely that modest increases in resolution will dramatically change
the basic conclusion that sub-SPH turbulence is needed to match $\OVI$
observations.  Running simulations with sufficient resolution to study
small-scale turbulence while properly modeling nonlinear structures to
$z=0$ is well beyond present computing abilities.  We further stress
the need to run galactic-scale simulations in the future to fully
resolve the absorbing structures within haloes and intragroup gas
where the strongest $\OVI$ absorbers appear arise in our simulations.

Finally, concerning box size convergence, it is true that the $32
\hmpc$s does not contain larger modes that could become non-linear by
$z=0$; however we will show that we are primarily exploring the
diffuse IGM around sub-$M^*$ galaxies, which are statistically
well-sampled in our simulation.  This is evidenced by the fact that
our $\OVI$ and $\HI$ statistics in the d16n128vzw150 box are
statistically consistent with those of the d32n256vzw150 box.

\section{Physical conditions of $\OVI$ absorbers} \label{sec:OVIphys}

From here on we focus on our favored model with sub-resolution
turbulence added, d32n256vzw150-bturb, and examine the physical
nature of $\OVI$ absorbers.  We first examine the density, temperature,
metallicity, and absorber size of IGM $\OVI$ components, finding
that most observed $\OVI$ systems are consistent with being
photo-ionized.  Shorter cooling times of a clumpy metal distribution
are argued as the driver for $\OVI$ to reach such temperatures.  We
then discuss the correspondence of $\HI$ with $\OVI$, using the
$N(\HI)-N(\OVI)$ relation to explore various photo-ionization,
collisional ionization, and multi-phase ionization models.  Lastly,
we justify adding turbulent velocities on physical grounds, explaining
why it should not be surprising that turbulence provides a dominant
component to many IGM absorber profiles.  This will lead us to the
next section where we examine the location and origin of $\OVI$ 
relative to galaxies.

Physical parameters ascribed to absorbers such as density and
temperature can be weighted in multiple ways.  Possibilities include
$\HI$ and $\OVI$ absorption weighting, or just normal mass weighting.
We weight such quantities using the formula
\begin{equation}\label{eqn:weighting}
X(W) = \frac{\sum{W\times X}}{\sum{W}}
\end{equation}
where $X$ is the property and $W$ is the weight.  For example, if a
quantity is weighted by $\OVI$ absorption, $W=m\times Z_{\rm O}\times
f(\OVI)$, where $m$ is mass of the element (e.g. SPH particle),
$Z_{\rm O}$ is the mass fraction in oxygen, and $f(\OVI)\equiv
\OVI/{\rm O}$ (the $\OVI$ ionization fraction).  $X$ is often
temperature or density itself.  The difference between $\HI$ and
$\OVI$-weighted quantities is critical for understanding the
multi-phase nature of absorbers.


We examine 70 continuous lines of sight with $S/N=20$ between
$z=0.5\rightarrow 0$; we call this is our high quality sample.  1511
$\OVI$ components are identified above $N(\OVI)=10^{12.5}\cms$.  This
sample allows us to obtain reasonable statistics of all absorbers
between $N(\OVI)=10^{12.9-14.9} \cms$, which we will refer to as the
``observed range'' since nearly all observed IGM $\OVI$ absorbers lie
in this interval.

\subsection{A photo-ionized model for $\OVI$ absorbers} \label{sec:photomodel}

\begin{figure}
\includegraphics[scale=0.80]{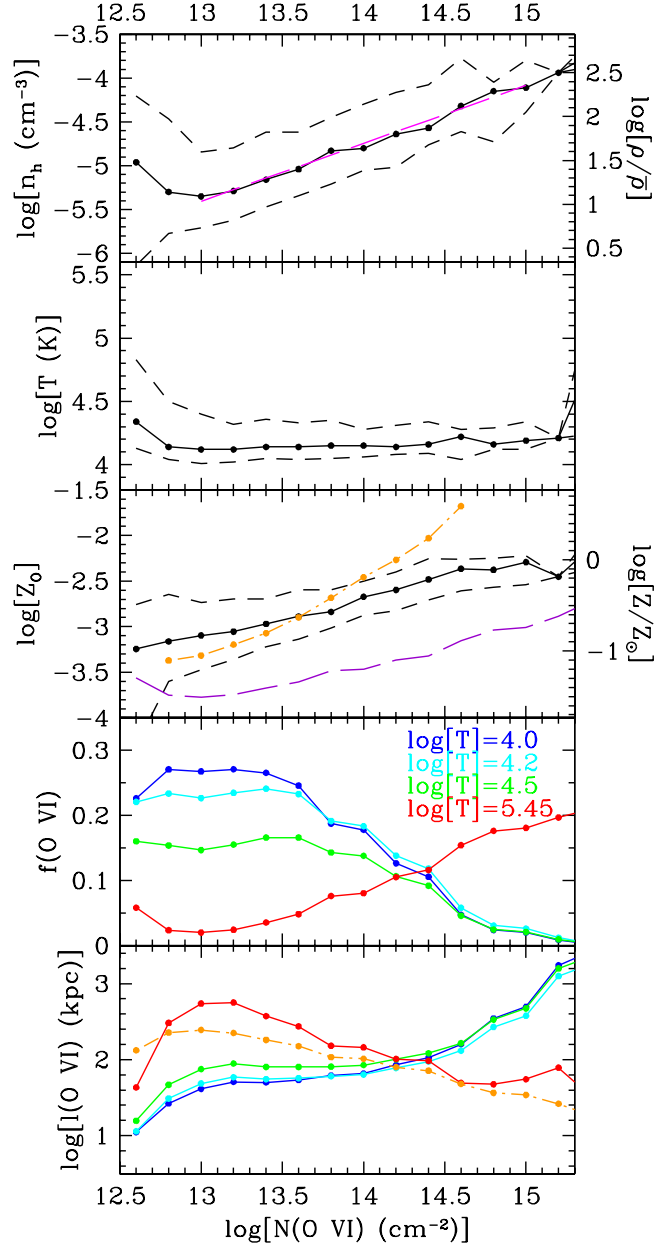}
\caption[]{Physical properties as functions of $\OVI$ column density
in the d32n256vzw150-bturb model are plotted a solid black lines with
points in the top 3 panels for all our simulated absorbers in 70
high-quality lines of sight below $z=0.5$.  Dashed lines correspond to
1$\sigma$ dispersions.  The long-dashed magenta line is the fit from
Equation \ref{eqn:nh_NOVI}.  We also plot the simulations-average
metallicity as a function density (converted to $N(\OVI)$ via Equation
\ref{eqn:nh_NOVI}, purple long-dashed line) and the metallicity one
would infer using aligned $N(\HI)-N(\OVI)$ absorbers as in Equation~\ref{eqn:met_est} (orange
dashed-dot line).  We consider $\OVI$ fractions from CLOUDY at three
photo-ionized temperatures (2nd to last panel), which suggest
photo-ionized $\OVI$ absorbers are typically 50-100 kpc long (last
panel).  $\OVI$ in CIE will produce more compact absorbers at higher
$N(\OVI)$ (red lines), which appear to trace gas near and within
halos.  The $\HI$-derived absorber length (orange dashed-dot line)
assuming $\OVI$ traces the same gas as $\HI$ appears to be a poor
approximation for strong absorbers, suggesting such aligned absorbers
are multi-density.}
\label{fig:OVIphys1}
\end{figure}

We examine the density, temperature, metallicity, and sizes of
$\OVI$ absorbers based on the observational quantity $N(\OVI)$.
Figure~\ref{fig:OVIphys1} (top two panels) plots the median
$\OVI$-weighted density and temperatures at the line centers.  Over
two decades of column density covering the observed range, $\OVI$
absorbers show a steady increase in $n_{\rm H}$, while the temperature
stays nearly constant at $T\sim 10^{4.15}$ K.  Also plotted in
dashed lines are the 16th and 84th percentile values corresponding
to the 1 $\sigma$ dispersions.  The scatter in density is significantly
larger, $\sigma=0.3-0.4$ dex, compared to the scatter in temperature,
$\la0.2$ dex over much of the observed range.  Our first main
conclusion is that the vast majority of $\OVI$ absorbers at all but
the very strongest column densities are photo-ionized, with
temperatures less than 30,000~K.

We already presented the fit to the density as a function of $N(\OVI)$
in Equation~\ref{eqn:nh_NOVI}, which fits the relation quite well
regardless of added turbulence (magenta dashed line).  The increase in
density alone can explain 66\% of the $\OVI$ column density increase,
with all else being equal.  The remaining increase must arise from
variations in metallicity, ionization fraction, and absorber length.
Metallicity is expected to explain some of the increase, because our
simulations show a consistent metallicity-density gradient, as shown
in Figure~\ref{fig:rho_Z}.  Fitting the points in the third panel of
Figure~\ref{fig:OVIphys1} yields the relation
\begin{equation}\label{eqn:ZOVI}
\logrm[Z_{\rm O}] = 0.45\times \logrm[N(\OVI)] - 8.96,
\end{equation}
for $N(\OVI)=10^{12.9-14.9}~\cms$.  Although density (66\%) and
metallicity (45\%) variations can account for all of the dependence of
$N(\OVI)$, we will show that ionization fraction and absorber length
also vary.

The metallicities reported here always exceed the fiducial value of
0.1~$\Zsolar$ often assumed when modeling low-$z$ $\OVI$ absorbers.
Our absorbers are generally found to lie between 0.15-1.0 $\Zsolar$,
which is consistent with metallicity determinations from aligned $\HI$
absorbers~\citep[; T08]{pro04,leh06,coo08}.  We contrast the
oxygen-weighted metallicity of $\OVI$ absorbers
(Equation~\ref{eqn:ZOVI}) to the mass-weighted oxygen
metallicity-density gradient averaged over the entire simulation box
at $z=0.25$:
\begin{equation}\label{eqn:Zbar}
\logrm[\bar{Z}_{\rm O}] = -4.26 + 0.36 \logrm[\rho/\bar{\rho}] + 0.075 (\logrm[\rho/\bar{\rho}])^2,
\end{equation}
for $\rho/\bar{\rho}=1-3000$, which is a fit to the green curve in the
top panel of Figure~\ref{fig:rho_Z}.  This shows a 34\% rise over the
same $N(\OVI)$ range, and is represented by the purple dashed line in
the middle panel of Figure~\ref{fig:OVIphys1}.  The metallicities of
$\OVI$ absorbers are on average $4-6\times$ higher than the general
IGM over the observed range.  This is a clear sign that $\OVI$
absorbers trace inhomogeneously distributed metals.  The low scatter
in the oxygen metallicity-density relation indicates that $\OVI$
absorbers rarely trace gas with average IGM metallicity.  Hence the
global metallicity-density relationship is a poor descriptor of the
enrichment level in metal-line absorption systems.


The fourth panel in Figure~\ref{fig:OVIphys1} shows the $\OVI$
ionization fraction, computed by assuming the density$-N(\OVI)$
relation shown in the top two panels, for various temperatures,
assuming a \citet{haa01} quasar-only background.  For a typical $\OVI$
temperature of $T=10^{4.2}$ K, $f(\OVI)$ changes significantly from
0.23 to 0.029 over the observed range.  It depends little on
temperature for lines above $10^{14} \cms$.  The peak ionization
fraction for collisional ionization ($T=10^{5.45}$) is around 0.2, but
it is only approached for the strongest lines.  This is another reason
why over most of the observed range, photo-ionization is the dominant
mechanism for $\OVI$ absorbers.

Thus far, the dependence of $N(\OVI)$ using photo-ionization is
partially explained by density (66\%) and metallicity (45\%), but
the declining ionization fractions above $n_{\rm H}=10^{-5.0} \cmc$
anti-correlate with $N(\OVI)$ (-45\%).  The remaining dependence
(34\%) therefore must be in the absorber size ($l(\OVI)$).  Absorber
size is calculated in the bottom panel of Figure~\ref{fig:OVIphys1}
by dividing the column density by the number density of $\OVI$:
\begin{equation}
l(\OVI) = \frac{N(\OVI)}{n_{\rm H}(\OVI)\; \frac{Z_{\rm O}}{f_{\rm H}} \; \frac{m_{\rm H}}{m_{\rm O}}\; f(\OVI)}~{\rm cm}
\end{equation}
where the $f_{\rm H}=0.76$ is the mass fraction in hydrogen, and $
\frac{m_{\rm H}}{m_{\rm O}}$ is the ratio of the atomic weights
(0.0625).  Absorbers remain between 50-100 kpc at $N(\OVI)<10^{14.0}
\cms$ where $f(\OVI)\approx 0.20-0.25$, and rise to many hundreds of
kpc at $N(\OVI)=10^{14.5} \cms$ assuming photo-ionization.  The large
required path length for the very strongest absorbers makes
photo-ionized absorbers rarer, and makes collisionally ionized systems
($T=10^{5.45}$) with smaller sizes (red curve) more viable.  We will
discuss such systems further in \S\ref{sec:collion}.


In summary, virtually all $\OVI$ IGM absorbers are photo-ionized.
Weak absorbers trace filaments with overdensities in the range of
$10-20$; intermediate absorbers trace overdensities $20-100$; and strong
absorbers trace gas within or near galactic halos.  We stress that some
collisionally ionized $\OVI$ absorbers do exist, particularly for the
strongest absorbers where a much shorter pathlength can yield sufficient
$N(\OVI)$, but these are better viewed as galactic halo rather than
IGM absorbers.  Strong $\OVI$ absorbers associated with the MW halo
indicate collisionally ionized oxygen, because otherwise their long
pathlengths would be unable to fit within the halo \citep{sem03}; our
absorber length analysis agrees with this assessment.  Since our sight
lines occasionally pass through such halos, we discuss such absorbers
in \S\ref{sec:collion} when we consider environment.  Finally, while we
have used the d32n256vzw150-bturb model, the trends identified here are
nearly identical without sub-SPH turbulence added, and are essentially
independent of wind model.  In other words, the photo-ionized nature of
IGM $\OVI$ absorbers is a highly robust prediction of our simulations.

\subsection{Cooling Times}\label{sec:cooling}

The inhomogeneous nature of the metal distribution is vital for the
photo-ionizational explanation of $\OVI$, because metal-line cooling
(which we implemented in \gad~in OD06) is much more effective when
metals are concentrated.  Consider the cooling time of gas in CIE
at $z=0.25$:
\begin{equation}\label{eqn:tcool}
\tau_{cool} = 4.12\times10^{-17} \frac{T}{\delta \Lambda(T,Z)}~{\rm yrs}
\end{equation}
where $\Lambda$ is the cooling rate (ergs s$^{-1}$
cm$^{3}$)\footnote{See \citet{gna07} for a derivation of the general
  form of this formula.}.  Let us consider the cooling time of a
typical $10^{13.0} \cms$ absorber (corresponding to overdensity,
$\delta=13$), assuming it cools from $3\times10^5$ K.  If we use
$\bar{Z}$ from Equation \ref{eqn:Zbar}~ (i.e. we assume the mean IGM
metallicity) and assume solar abundance ratios for all species, the
cooling time is 42 Gyr.  Conversely, using the metallicity of $\OVI$
absorbers from Equation \ref{eqn:ZOVI}~yields a cooling time of
10~Gyr.  Hence weak absorbers cannot cool significantly within a
Hubble time at the mean metallicity, but can do so at the mean $\OVI$
absorber metallicity.  Similarly, a strong system with
$N(\OVI)=10^{14.5} \cms$ ($\delta\sim100$) has cooling times for a
mean IGM metallicity and for $\OVI$ absorber metallicity of
$\tau_{cool}=2.9$ and 0.6 Gyr, respectively.  Of course, densities in
the early Universe were higher by $(1+z)^3$, and cooling times depend
inversely, but metallicities were lower, so these cooling times are
probably overestimates.  Still, for observed $\OVI$ absorbers the
cooling time is generally shorter than a Hubble time ($\tau_{Hubble}$)
when the $\OVI$ absorber metallicity is considered.

Weaker absorbers take longer to cool, but as we will show in
\S\ref{sec:age} weak absorbers trace some of the oldest metals
injected at high-$z$, while stronger absorbers trace more recent metal
injection.  This means that all but the very strongest $\OVI$
absorbers have had plenty of time to cool to photo-ionized temperatures
since their injection into the IGM.  Physically, this is why the
clumpy metal distribution where $Z\sim 5\bar{Z}$, causes most
$\OVI$ absorbers to be photo-ionized.

In practice, superwind feedback shock-heats metals to higher
temperature ($T>10^6$ K) in denser environs than those traced by
photo-ionized $\OVI$.  The result is a range of cooling times with
lower metallicities more likely to have $\tau_{cool}>\tau_{Hubble}$
and remain in the WHIM or hot IGM, and higher metallicities often
with $\tau_{cool}\ll \tau_{Hubble}$.  This latter case includes
metals traced by photo-ionized $\OVI$.  The strength of the feedback
is a sensitive determinant for which $(\rho,T)$ path an SPH particle
will follow, with stronger feedback putting many more metals into
the WHIM and hot IGM (OD08).  Our main point here is to show that
metals heated to WHIM temperatures at $\OVI$ absorber metallicities
will usually cool to photo-ionized temperatures by today.

\subsection{Alignment with $\HI$}

We have thus far discussed $\OVI$ absorbers independently of $\HI$.
Any model for $\OVI$ absorbers should also reproduce the general
properties of {\it aligned} $\HI$ absorbers, in addition to reproducing
global $\lya$ forest properties~\citep[e.g.][]{dav99, dav01b}.
We start by emphasizing that a vast majority of $\OVI$ absorbers
are aligned at some level with $\HI$.  However, from the $\HI$- and
$\OVI$-weighted density and temperature of well-aligned absorbers,
we will demonstrate that aligned absorbers often arise in different
phases of gas.  We introduce the idea of multi-phase photo-ionized
models with different density and temperatures for the $\OVI$ and
$\HI$ components.  We then estimate $\HI$-derived absorber length,
showing these can differ from the $\OVI$ lengths often because they
are tracing different phases of gas.  Lastly, we fit multi-phase
ionization models to the $N(\HI)-N(\OVI)$ plane, to understand what
types of absorbers populate various regions.  This can aid in
interpretation of data sets such as T08 and upcoming data from COS.

\begin{figure}
\includegraphics[scale=0.80]{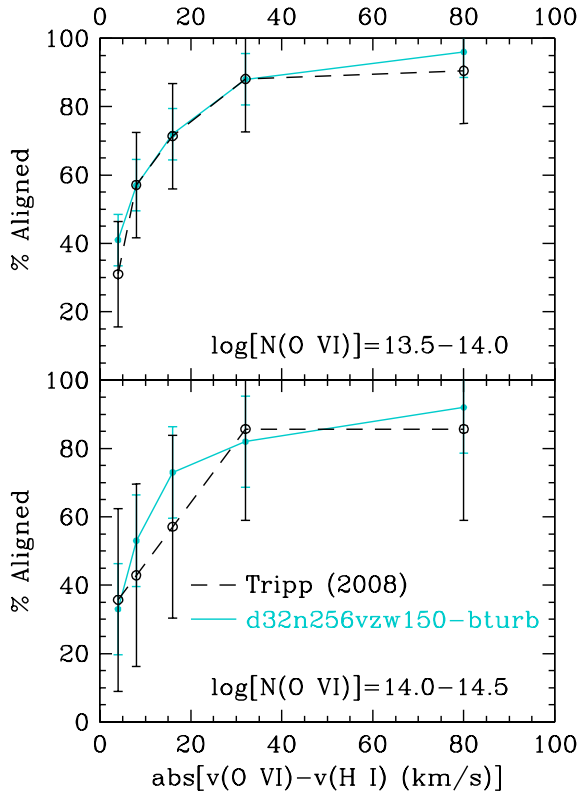}
\caption[]{The alignment fraction of $\OVI$ absorbers with $\HI$ at
various separations in two bins covering intermediate absorbers
($N(\OVI)=10^{13.5-14.0}$ and $10^{14.0-14.5} \cms$).  The T08 data is
compared to the $S/N=10$ sight lines of the d32n256vzw150-bturb
simulation, showing nearly perfect agreement in the lower bin (top)
and adequate agreement in the upper bin (bottom).  The high aligned
absorber fractions are an indication of photo-ionized $\OVI$.  }
\label{fig:OVI_align}
\end{figure}

The alignment fraction as a function of velocity separation for
this model, as for most of our models, agrees well with observations,
as we show in Figure \ref{fig:OVI_align} for intermediate absorbers
split into two bins.  Well-aligned absorbers ($\delta v\leq 8\kms$
or $\leq 100$ kpc in a Hubble flow) comprise 57\% of all $\OVI$
absorbers in the d32n256vzw150-bturb simulation when considering
intermediate absorbers, in statistical agreement with T08 (54\%)
(see \S\ref{sec:observables}).  We use the $S/N=10$ lines of sight
in this case, because it is more similar to the data $S/N$; the
high quality sample has a 62\% alignment fraction, since more
components are identified.  The agreement in the $N(\OVI)=10^{13.5-14.0}
\cms$ bin is nearly perfect.  There are fewer aligned absorbers in
the data for the higher column density bin ($N(\OVI)=10^{14.0-14.5}
\cms$), but this is not a statistically significant difference since
the observed sample has only 14 absorbers.

Misaligned absorbers are an important population, being a signature of
collisionally ionized $\OVI$ as we discuss in \S\ref{sec:collion}.
Our discussion in that section suggests strong absorbers have a
greater chance of being collisionally ionized; however we argue in the
following subsections that the high number of aligned absorbers in
both the data and simulations is a strong indication of photo-ionized
$\OVI$.

\subsubsection{Density and Temperature} \label{sec:HIalign}

\begin{figure*}
\includegraphics[scale=0.80]{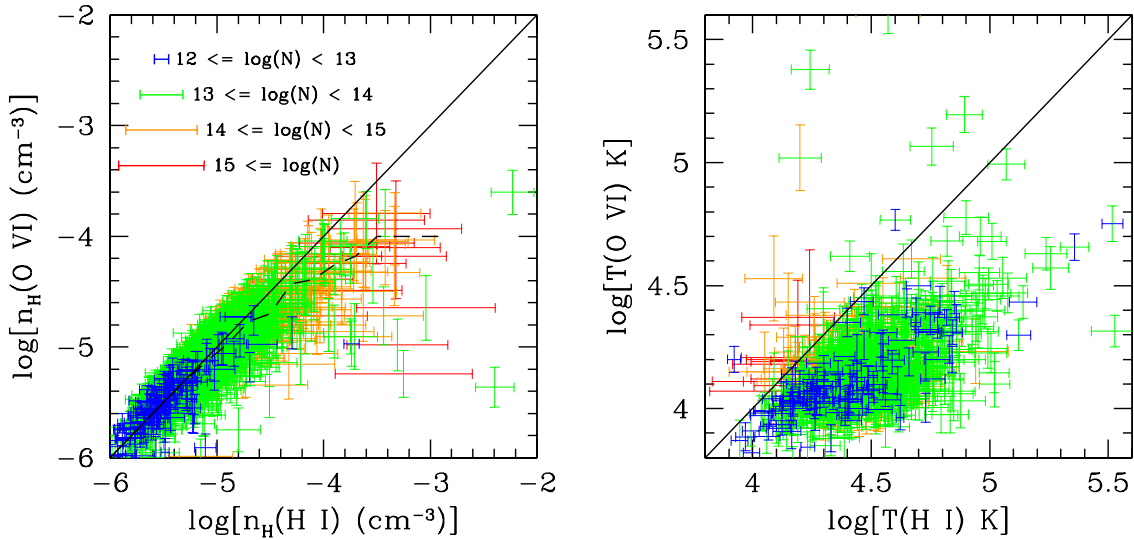}
\caption[]{The multi-phase nature of well-aligned absorbers is shown
by plotting $\HI$ and $\OVI$-weighted temperatures and densities
against each other (left \& right panels respectively) from 70 sight
lines.  The column density of absorbers are indicated by both the size
and color of the error bar with $N(\HI)$ in the x direction and
$N(\OVI)$ in the y direction.  Densities diverge above $n_h=10^{-4.8}
\cmc$ where $\OVI$ more often traces gas outside halos while $\HI$
traces gas within.  The dashed line indicates the relation we use
between the two densities.  Temperatures are generally cooler for
metal-enriched gas indicating metal-line cooling plays an important
factor in gas, which indicates an inhomogeneous distribution of
metals.}
\label{fig:align_rho_T}
\end{figure*}

While we have argued that the vast majority of $\OVI$ absorbers are
aligned at some level \citep[e.g.][]{tho08a}, do such aligned
absorbers trace the same underlying gas?  To answer this we plot the
$\HI$ and $\OVI$-weighted densities and temperatures for well-aligned
absorbers against each other in Figure \ref{fig:align_rho_T}.  $\HI$
and $\OVI$ trace the same density gas below $n_{\rm H}\approx
10^{-4.8} \cmc$, but above this the absorbers tend to become
multi-phase in density.  The declining ionization fraction of
photo-ionized $\OVI$ makes it a poor tracer of gas inside halos at
overdensities above a couple hundred.  These rare instances are less
represented in well-aligned $\OVI$ absorbers due to the large peculiar
velocities of various components.  However, the spatial extent of
regions at overdensities $\sim 500$ traced by strong $\HI$ is small
($<100$ kpc), and gas at lower overdensities tracing strong $\OVI$
absorption is often close by and therefore appears aligned.

The weighted temperatures show significantly less alignment, and
reveal something that may be counter-intuitive -- metal-enriched gas
is cooler than unenriched gas.  It is counter-intuitive in the sense
that to reach the IGM via superwind feedback, gas is driven at high
velocities and therefore must shock heat.  However, stronger
metal-line cooling outweighs this by low-$z$.  The $\HI$ temperatures
span a wider range, just falling short of the $T=10^5$ K limit, which
enters the realm of BLAs ($b(\HI) = 40 \kms$).  The two temperatures
show that even in aligned absorbers at similar densities, $\OVI$ and
$\HI$ are not tracing the same gas.  Again, we suggest that $\OVI$ is
arising from a clumpy distribution that picks out enriched regions of
the more smoothly varying $\lya$ forest.  Several rare instances of
$\OVI$ in CIE is aligned with cooler gas with $\HI$ absorption.  


\subsubsection{Absorber Size}

\begin{figure}
\includegraphics[scale=0.80]{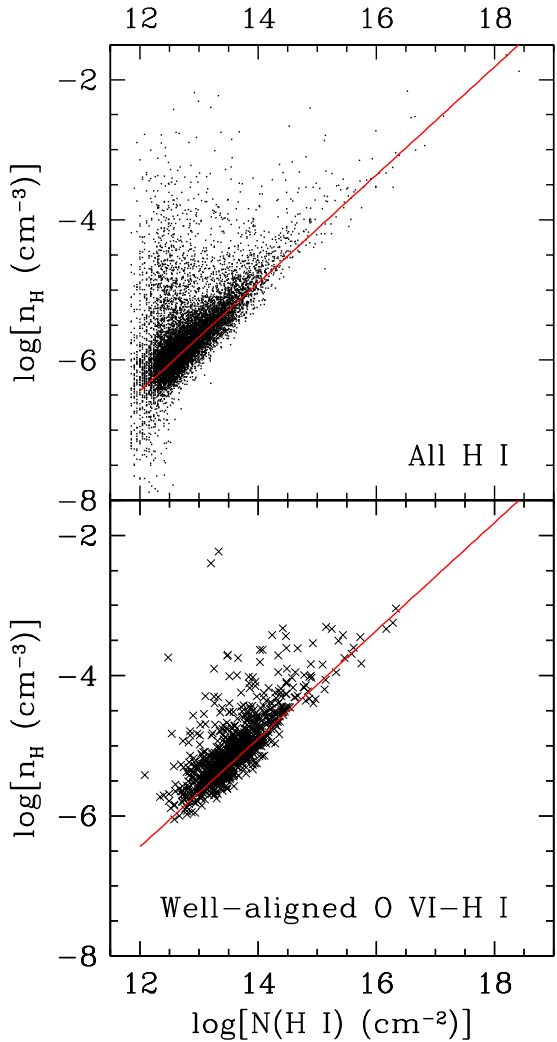}
\caption[]{The $\HI$ column density is related to $n_{\rm H}$ over 7
decades as shown by the fit to the median density between
$N(\HI)=10^{12.0-14.5} \cms$.  $\HI$ absorbers aligned with $\OVI$
follow this relation as well, however in both cases scatter exists as
weak absorbers sometimes trace high densities in complex systems.  The
$n_{\rm H}-N(\HI)$ relation is used to determine the $\HI$-derived
absorber size.  }
\label{fig:nh_NHI}
\end{figure}

Before introducing ionization models predicting line ratios, we must
consider the $\HI$-weighted absorber size, which is different than
$l(\OVI)$.  We show the $\HI$-weighted density for all $\HI$
absorbers, even those without $\OVI$, in the top panel of Figure
\ref{fig:nh_NHI}.  A fit to the median of these absorbers between
$N(\HI)=10^{12.0-14.5} \cms$ gives the relation
\begin{equation}\label{eqn:nh_NHI}
\log[n_{\rm H}~(\cmc)] = -15.68+0.77\; \log[N(\HI)]
\end{equation}
at $\langle z\rangle=0.25$, and well describes the lower density
boundary of $\HI$ absorbers over 7 orders of magnitude.  This
relationship is also derived by \citet{dav99}, finding $\log(n_{\rm
H}) = -15.13 + 0.7\; \log[N(\HI)]$ at $z=0.25$, and analytically by
\citet{sch01} by assuming the characteristic size of an absorber is
well-described by the IGM Jeans length, $\log(n_{\rm H}) = -14.68 +
2/3\; \log[N(\HI)]$.  Our current simulations~\citep[which are much
improved over][]{dav01b} find an overall slightly steeper dependence
and a higher normalized $n_{\rm H}$.  The upward scatter in the top
panel of Figure \ref{fig:nh_NHI} are the absorbers likely to be part
of complex $\HI$ systems; similar scatter is seen by \citet{dav01b}.
The well-aligned absorbers plotted in the bottom of Figure
\ref{fig:nh_NHI} are consistent with equation~\ref{eqn:nh_NHI} (red
line), showing that they generally are not a biased population of
$\HI$ absorbers, except that stronger $\HI$ is more likely to have
aligned $\OVI$\footnote{See the end of \S\ref{sec:env} for the
predicted fraction of $\HI$ absorbers with $\OVI$.}.

We take Equation~\ref{eqn:nh_NHI} as the relation between
$N(\HI)-n_{\rm H}$ for all aligned absorbers, despite the scatter, and
use the $\HI$ fraction for gas at $T=10^{4.3}$ K photo-ionized by a
quasar-dominated ionization background, $\logrm[f(\HI)] =
0.17+0.99\; \logrm(n_{\rm H})$, to get the neutral hydrogen
column density relationship:
\begin{equation}\label{eqn:NH_nh}
\logrm[N({\rm H})~(\cms)] = 20.19 + 0.31\; \log(n_{\rm H}).
\end{equation}
Dividing $N({\rm H})$ by $n_{\rm H}$ yields the absorber length,
\begin{equation}\label{eqn:lh1_nh}
\logrm[l(\HI)~({\rm cm})] = 20.19 - 0.69 \; \log(n_{\rm H}).
\end{equation}
We plot $l(\HI)$, using the assumption that $n_{\rm H}(\OVI)= n_{\rm
  H}(\HI)$ as an orange dash-dot line in Figure \ref{fig:OVIphys1}~to
contrast with the $\OVI$-derived absorber size, $50-100$ kpc at
$N(\OVI)<10^{14.5}\cms$.  Below $N(\OVI)=10^{14.0} \cms$, it seems
fair to assume the same densities since $n_{\rm H}<10^{-4.5} \cmc$,
and the resulting $\OVI$ absorber lengths fit within the $\HI$
absorber lengths.  Above $10^{14.0} \cms$, $l(\HI)$ becomes shorter
than $l(\OVI)$ for several reasons.  First and perhaps most important,
the densities where the two species arise diverge above $n_{\rm
  H}=10^{-4.8} \cmc$, with $\OVI$ arising from lower density gas
distributed over larger lengths.  Second, stronger absorbers are less
likely to be aligned due to larger peculiar velocities, making this
comparison less meaningful.  Third, strong absorbers are more likely
to have metals at CIE temperatures, raising the ionization fraction
and lowering $l(\OVI)$.  Finally, the successive scatter in each step
above to get $l(\HI)$ adds up when considering the spread in
temperatures.

As an interesting aside, we investigate the ``$N({\rm H})$ conspiracy''
for low-$z$ metal line systems presented by \citet{pro04}.  They find
six $\OVI$ absorbers in the PKS 0405-123 spectrum averaging $N({\rm
H}) = 10^{18.7} \cms$ with a scatter of only 0.3 dex.  In fact, this is
straightforwardly understood in the context of our photo-ionized model for
$\OVI$ absorbers.  The $N({\rm H})$ conspiracy comes about because $\OVI$
arises mostly from photo-ionized overdensities ($\sim10-200$ at $z=0.25$),
which leads to a similarly small range, $N({\rm H})=10^{18.50-18.91} \cms$
(using eq.~\ref{eqn:NH_nh}), in our model.  This range is in excellent
agreement with \citet{pro04}, providing further evidence that $\OVI$
in our simulations is tracing the cosmic web as observed.  $N({\rm H})$
column densities below $10^{18.5} \cms$ generally indicate photo-ionized
absorbers according to \citet{ric06}.

A common method to estimate oxygen metallicities of aligned absorbers
from observations is
\begin{equation}\label{eqn:met_est}
Z_{\rm O} = \frac{N(\OVI)}{N(\HI)}\times \frac{f(\HI)}{f(\OVI)} \times \frac{m_{\rm O}}{m_{\rm H}} \times f_{\rm H}.
\end{equation}
Unfortunately, this may significantly mis-estimate the true
metallicity.  To illustrate this, we plot this ``observed
metallicity'' as the orange dash-dotted line in the central panel of
Figure \ref{fig:OVIphys1}.  At weak $\OVI$ column densities, the
metallicity is underestimated, while at above $N(\HI)>10^{14.0} \cms$
it overestimates the actual metallicity.

In summary, the $\HI$ properties of well-aligned $\OVI$ absorbers
are fairly similar to general $\HI$ absorber properties.  $\HI$ and
$\OVI$ absorber sizes can be substantially different in stronger
absorbers, because even aligned $\HI$ and $\OVI$ arise in different
phases of gas.  Finally, $N({\rm H})$ remaining relatively invariant
over a large range of $N(\OVI)$ is expected if well-aligned absorbers
arise from typical $\HI$ absorbers.

\subsubsection{Ionization Models} \label{sec:ionmodels}

\begin{figure*}
\includegraphics[scale=0.80]{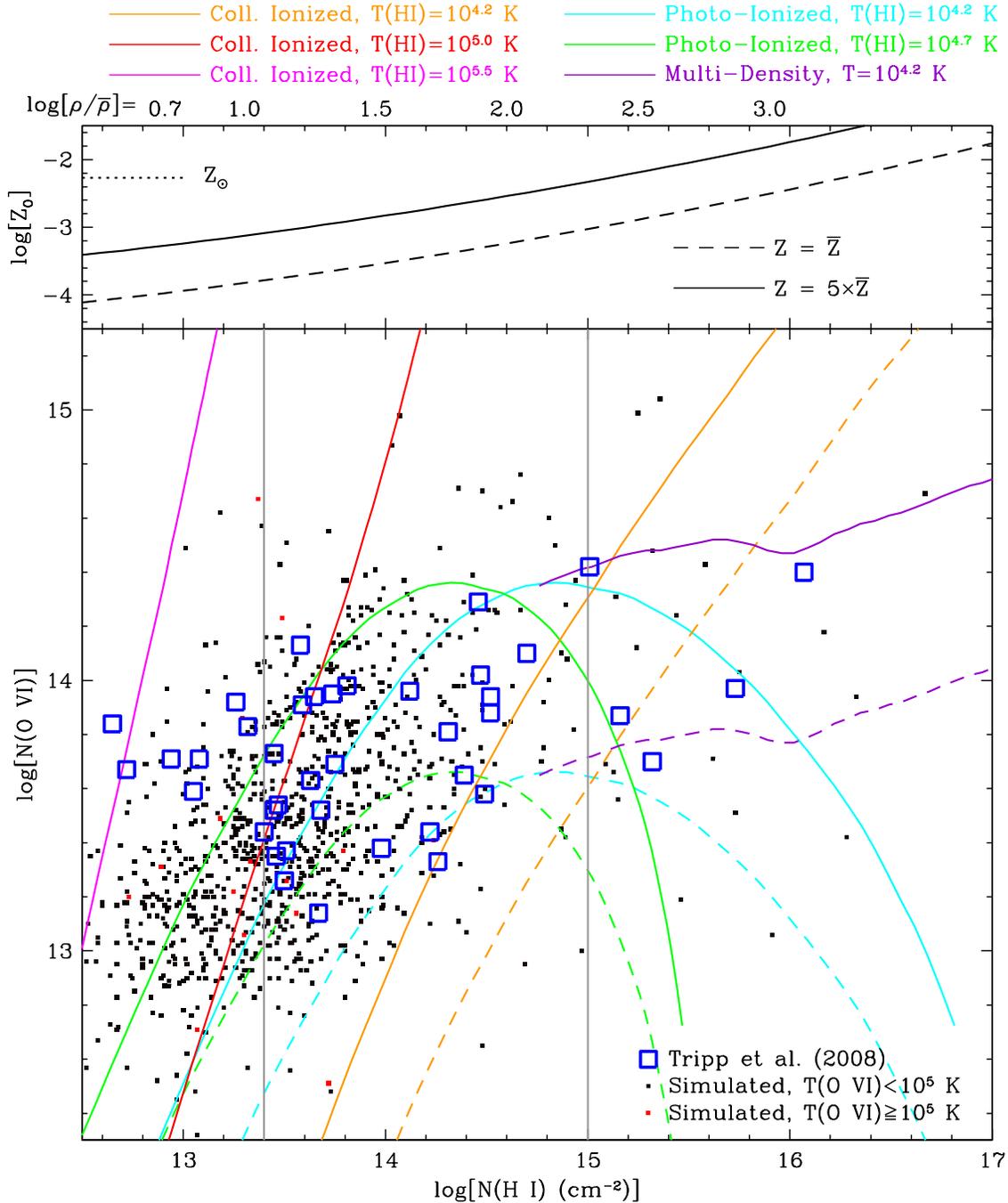}
\caption[]{The $N(\HI)-N(\OVI)$ plane shows the aligned absorbers from
our high-quality lines of sight (black points, $T(\OVI)<10^5$ K; red
points, $T(\OVI)\geq 10^5$ K) along with absorbers observed by T08
(open blue squares).  Several ionization models (colored lines) are
shown using either the metallicity-density relationship, $\bar{Z}$, in
our $z=0.25$ d32n256vzw150 simulation (dashed lines) or 5$\times
\bar{Z}$ (solid lines) since metals in $\OVI$ absorbers appear to be
significantly above the cosmic average.  The top panel shows this
relationship translated to $N(\HI)$ using Equation \ref{eqn:nh_NHI}
and the relation $n_{\rm H} = 3.63\times10^{-7}\rho/\bar{\rho} \cmc$,
which applies at $z=0.25$.  Collisionally ionized $\OVI$
($T(\OVI)=10^{5.45}$ K) aligned with $\HI$ at various temperatures
(orange, red, \& magenta) can explain most absorbers, but the
absorbers are vastly photo-ionized and better explained by cyan and
green ionization models ($T(\OVI)=10^{4.2}$ K).  The purple ionization
models are multi-density with $\OVI$ arising from lower densities than
$\HI$ when $n_{\rm H}\geq 10^{-4.8} \cmc$.  }
\label{fig:h1_o6}
\end{figure*}

Having the length scale of aligned absorbers, along with densities and
temperatures, we now have everything necessary to fit ionization
models to $\OVI-\HI$ line ratios.  This is portrayed in Figure
\ref{fig:h1_o6} using the $N(\HI)-N(\OVI)$ multi-phase plane
considered by \citet{dan05} and T08.  Note that these authors plot
$N(\HI)$ versus $N(\HI)/N(\OVI)$; we prefer to plot the column
densities against each other directly and not the ratio, because it is
simpler and makes trends clearer.  Small points represent simulated
aligned absorbers.  We plot colored lines for various two-temperature,
multi-phase models where oxygen is either photo-ionized
($T(\OVI)=10^{4.2}$ K) or in CIE ($T(\OVI)=10^{5.45}$ K), and $\HI$ is
one of several different temperatures.  Our models use the $z=0.25$
metallicity-density relation (Equation \ref{eqn:Zbar}) shown on top
(dashed lines), but we also consider metallicities 5$\times \bar{Z}$
(solid lines) as more relevant given the $\OVI$ absorber metallicities from
\S\ref{sec:photomodel}.  It is difficult to create photo-ionized
$\OVI$ absorbers with column densities much greater than $10^{14.3}
\cms$ using the 5$\times \bar{Z}$ relation, because $f(\OVI)$ declines
as $N(\HI)$ rises creating a maximum in $N(\OVI)$.

The CIE models (orange, red, \& magenta) follow the behavior
highlighted by \citet{dan05} where $N(\OVI)$ is independent of
$N(\HI)$, which sweeps across a broad swath of this plot.  In our
case, the $\HI$ temperature ranging from $T=10^{4.2}\rightarrow
10^{5.5}$ K is the variation that allows this scenario to explain most
of the T08 well-aligned absorbers ($\delta v \leq 8\kms$ separation,
blue squares) while $\OVI$ is always collisionally ionized.  This
scenario seems somewhat implausible at face value.  Meanwhile, the
photo-ionized models with two choices of $T(\HI)$ (cyan and green)
bracket the majority of $\HI$-derived temperatures in Figure
\ref{fig:align_rho_T}, and also can explain most, {\it but not all},
of the T08 absorbers.  Lastly, we introduce a multi-density model
assuming $T=10^{4.2}$~K (purple) for densities above $n_{\rm
H}=10^{-4.5} \cmc$ where we assume different absorber lengths for
$\OVI$ and $\HI$ by plugging in the divergent density relationship
from Figure \ref{fig:align_rho_T} (dashed line) into Equation
\ref{eqn:lh1_nh}.  This results in longer absorber pathlengths for
$\OVI$ than $\HI$.  Unless otherwise noted, we assume aligned
absorbers are described by photo-ionization.

We compare the number densities of $\OVI$ systems to aligned T08
absorbers in three bins denoted by the vertical gray lines in Figure
\ref{fig:align_rho_T}-- $N(\HI)<10^{13.4}$, $10^{13.4-15.0}$, and
$>10^{15.0-17.0} \cms$, and we overlay our high quality sample of
70 lines of sight (black dots, $T(\OVI)<10^5$ K; red dots, $T(\OVI)\geq
10^5$ K).  Our sample covers $\Delta z=35$ with $\sim 90\%$
completeness at $N(\OVI)=10^{13.0} \cms$.  We directly compare with
the T08 dataset only for absorbers of at least intermediate strength,
for which they have $\Delta z=2.62$ (their effective pathlength at
$EW=30$m\AA).

Beginning with the strongest $\HI$ bin first, T08 finds 5 such
absorbers ($1.9~\Delta z^{-1}$), which is nearly 2 $\sigma$ more than
we find (14, $0.4~\Delta z^{-1}$).  These absorbers correspond to gas
in and around halos, and rarely fall below $N(\OVI)=10^{13.6} \cms$ in
either set.  Two of T08's absorbers and many of our absorbers are
above the solid $T(\HI)=10^{4.2}$ K photo-ionized model (cyan) which
seemingly rules out photo-ionization, because higher metallicities are
required.  We suggest that a subset of these absorbers are best
described by the multi-density photo-ionized model (purple), with
$\OVI$ arising from longer absorbers outside a halo, while $\HI$
traces denser halo gas.  The higher observed frequency of strong
absorbers in the data may indicate non-uniformity in the ionization
background, where halo substructure unresolved in our simulations
increases the number of absorbers with $N(\HI)>10^{15} \cms$, which
our simulations under-predict.

It is significant that few of our simulated absorbers and none of
T08's correspond to values lie below $\bar{Z}$ in either of the
single-phase models (green and cyan dashed lines) above
$N(\HI)=10^{15} \cms$, although these should be detectable in the
data; this behavior also exists in the \citet{tho08a} dataset.  The
$\HI$ often traces halo gas while the photo-ionized $\OVI$ arises from
a longer pathlength of gas at $\delta\sim 100$, increasing the chance
of intersecting with a clump of metals well above $\bar{Z}$.  Only
29\% of our strong $\HI$ absorbers have well-aligned $\OVI$, but 87\%
of them have $\OVI$ within 80 $\kms$ (see the end of \S\ref{sec:env}
for more statistics).  Our interpretation is there is almost always
$\OVI$ with strong $\HI$, but the $\OVI$ is much more extended and
peculiar velocities dominate, resulting in well-aligned strong $\HI$
absorbers often being chance alignments.

We find a vast majority of aligned absorbers in the intermediate bin
($8.5~\Delta z^{-1}$), within the Poisson errors of T08 ($7.3~\Delta
z^{-1}$).  Most of our absorbers including those down to our detection
limit fall between $1-10\bar{Z}$, which should happen according to our
photo-ionized $\OVI$ model.  These absorbers tracing $\delta=10-100$
are more likely to be associated with hotter $\HI$, for which the
$T(\HI)=10^{4.7}$ K photo-ionized model may be more relevant.  This
model encompasses a cluster of T08 absorbers with
$N(\HI)\approx10^{13.4-13.6} \cms$, which is near the the peak
concentration of our simulated absorbers.  Sub-$\bar{Z}$ absorbers
should theoretically exist in this bin, although it is rare for
metallicities traced by $\OVI$ to fall below $\bar{Z}$.  Part of this
discrepancy may be explained by smaller $l(\OVI)$ ($\sim 50$ kpc)
aligned with greater $l(\HI)$ ($\sim 100$ kpc), underestimating the
true metallicity.  COS should help confirm or exclude the presence of
low-metallicity $\OVI$ absorbers.

The lowest strength aligned absorbers are slightly under-predicted by
our simulation ($1.9~\Delta z^{-1}$) relative to the T08 dataset
($3.1~\Delta z^{-1}$, a 1.1 $\sigma$ difference).  Four of 7 T08 weak
aligned absorbers are in simple systems, and bear a resemblance to
their proximate absorbers with $N(\HI)<10^{13.0} \cms$.  The $\OVI$
column densities are at least $\times 3$ greater than the $\HI$ in the
data in every case, which is true only a quarter of the time for the
simulated absorbers.  We suggest the possibility of a non-uniform
ionization background at the Lyman limit due to the proximity of AGN
could result in weaker $\HI$ for an absorber, while $N(\OVI)$ remains
unchanged, because the 8.4 Rydberg ionization field is more uniform as
these photons have a much longer mean free path.  This could push a
number of aligned absorbers from the middle $\HI$ bin into the lowest
bin in Figure \ref{fig:h1_o6}.

We find 5.1 absorbers per $\Delta z$ where $N(\HI)<10^{13.4} \cms$ and
$EW(\OVI)=10-30$~m\AA, whereas there are none observed by T08.  COS
should be able to access such absorbers, which should indicate the
enrichment level of the IGM at overdensities of 5-10.

Considering strong $\OVI$ absorbers ($N(\OVI)\geq 10^{14.5} \cms$)
with aligned $\HI$, we find 14 ($0.4~\Delta z^{-1}$) while T08 finds
none.  These absorbers are predominantly photo-ionized, but excluded
by our ionization models, because the metallicity would have to be at
least $10 \bar{Z}$ in most cases, which rarely occurs.  Most of these
absorbers fall off the $N(\HI)-n_{\rm H}$ relation from Equation
\ref{eqn:nh_NHI} indicating they they are part of a complex system
with significant peculiar velocities, and possibly chance alignments.
Finding no such absorbers in the data, we may be overestimating the
$\OVI$ columns of such strong absorbers as we discuss in
\S\ref{sec:collion}.

To summarize, an in-depth examination on the $N(\HI)-N(\OVI)$ plane shows
we can explain some trends in the data with our photo-ionized multi-phase
explanation, most importantly that $\OVI$ appears to rise quite gradually
with increasing $\HI$~\citep[cf. Figure 1 of][]{dan05}.  Our absorbers
however appear more clustered at $N(\HI)=10^{13.5-14.5} \cms$ compared
with the T08 dataset, and we believe a varying ionization background
near 1 Rydberg is the best way to distribute over a wider range of $\HI$
column density.  We encourage the use of {\it Hubble}/COS to greatly
expand the aligned $\HI-\OVI$ weak absorber statistics, as these are one
of the most straight-forward and effective ways to constrain ionization
conditions of metals and the enrichment levels of filamentary structures
($\delta\sim 10$).

\subsection{Turbulence in the IGM}\label{sec:turbulence}

By forwarding the d32n256vzw150-bturb model, we are claiming that there
is some form of turbulence in the low-$z$ IGM well below the resolution
of our simulation that is responsible for broad $\OVI$ lines.  We arrive
at this claim by exploring every other broadening mechanism (spatial,
temperature, and instrumental), and finally settling on sub-SPH motions
as the only viable explanation.  We use ``turbulence" as a blanket term
to cover velocity shear, bulk motions, shock disturbances, and other
random motions unresolvable in our simulations.  Our main point is
that such motions increasingly dominate the line profiles for stronger
low-$z$ $\OVI$ absorbers.  In this section we connect (circumstantially,
at least) turbulence to the outflows that enrich and stir the IGM.  We begin by
showing that examples of turbulence in the IGM and halo gas are common,
and then use simple physical relations from \citet{kol41} turbulence to
broadly motivate our prescription of sub-SPH turbulence.

Turbulence or some other non-thermal broadening is commonly invoked
to explain $\OVI$ line widths in the IGM \citep[e.g.][]{dan06}.
Aligned $\HI-\OVI$ absorbers allow a constraint on how much of a
$b$-parameter is thermal and non-thermal.  Both T08 and \citet{tho08b}
agree that such systems almost always have $T<10^5$ K, with non-thermal
broadening significant if not dominant.  Even in our models without
turbulence, the same exercise reveals a significant non-thermal
broadening, a finding also confirmed in the simulations of
\citet{ric06}.  However, such alignments do not {\it necessarily}
imply turbulent motions, and instead may only indicate that metal
absorbers arise from different gas than $\HI$.

A more relevant measurement may be aligned $\HI-\HeII$ components,
which do not rely on the metal distribution, and have been shown by
\citet{zhe04} to indicate the IGM is dominated by turbulence at
$z=2-3$.  \citet{fec07} downplay the amount of turbulence in the IGM,
finding that only 45\% of such aligned components favor turbulent
broadening.  Our invocation of turbulence requires the association
with outflows and metals; therefore we disfavor turbulence in the
low-density, diffuse IGM, which is likely unenriched.  Furthermore,
\citet{dav01b} show that the low-$z$ $\lya$ forest is well described
by thermal line widths alone, and we reaffirm this point by showing in
\S\ref{sec:lya} that lines with $N(\HI)< 10^{14} \cms$ are unaffected
by our addition of turbulence.

A more direct way to measure turbulence is to look for velocity
differences on the smallest scales possible, as \citet{rau01} did in
lensed quasars.  By observing adjacent $\CIV$ profiles in paired lines
of sights, they find $\delta v\sim 5 \kms$ on scales of 300 pc at
$\langle z\rangle\sim 2.7$.  They apply the Kolmogorov steady-state
assumption whereby kinetic energy, $\frac{1}{2}\langle
v_{KE}^2\rangle$, injected at a rate $\epsilon_0$ cascades through
increasingly small eddies at $\epsilon\approx\epsilon_0$ until
viscosity transforms this energy to heat.  The dimensionless nature of
Kolmogorov turbulence allows the application of a simple scaling
between size, $l$, and rms velocity, $v_{\rm rms}$, such that
$\bar{v}_{\rm rms}^2\approx(\epsilon l)^{2/3}$.  Therefore, similar to
\citet{rau01}, we use the energy transfer rate $\epsilon\approx v_{\rm
rms}^3/l$ to parametrize turbulence.  For $\OVI$ absorbers we assume
$\bar{v}_{\rm rms}\sim b_{turb}$ and the transverse $l\sim l(\OVI)$,
such that
\begin{equation}\label{eqn:epsilon}
\epsilon\sim \frac{b_{turb}^3}{l(\OVI)}.  
\end{equation}
This application of transverse $\delta v$'s to $b$-parameters makes
more sense for $\OVI$, because $\OVI$ has been shown to be distributed
on scales much larger than a kpc whereas lower ionization species
(e.g. $\CIV$, $\CIII$, $\SiIV$) show sub-kpc structure indicating
small cloudlets \citep{lop07}.  Therefore, a single broad $\OVI$
profile may comprise many small metal-enriched cloudlets traced by low
ionization species of the sort suggested by \citet{sim06} and
\citet{sch07}.  In this scenario $\OVI$ traces more extended metals at
lower densities where turbulence on scales of 50-100 kpc dominates the
line width, while small, dense cloudlets traced by low ionization
species, possibly in sub-kpc clouds, show less broadening due to the
Kolmogorov scaling.  The result is the type of low-$z$ system commonly
observed, where low ionization species have multiple thinner profiles
along with one large broad $\OVI$ profile \citep[e.g.][]{tho08b},
which we suggest may be primarily photo-ionized.  This is only a
hypothesis, and should be tested in galactic-resolution simulations
where turbulence and the formation of dense clumps can be better
resolved.

\begin{figure}
\includegraphics[scale=0.90]{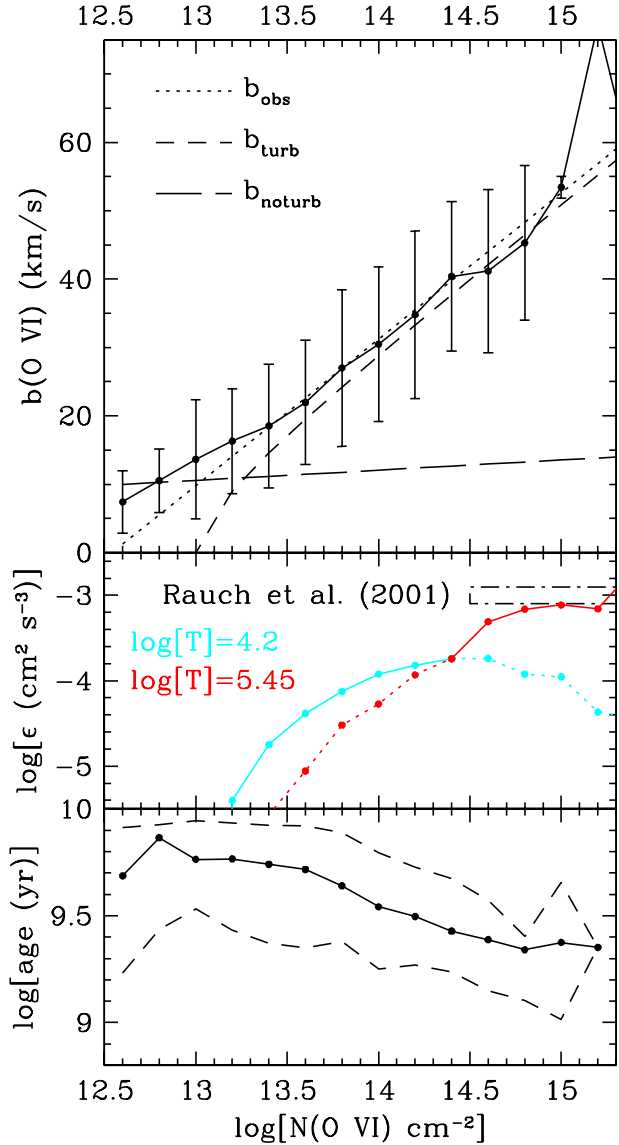}
\caption[]{The $b$-parameters from the d32n256vzw150-bturb model are
plotted as data points with 1 $\sigma$ dispersions.  $b_{obs}$ is a
linear fit to the T08 $b(\OVI)-N(\OVI)$ relation, and $b_{noturb}$ is
the fit to the d32n256vzw150 model.  From these fits, the relation for
$b_{turb}$ is calculated in Equation~\ref{eqn:bturb1}.  If we consider
the photo-ionized and collisionally ionized absorber sizes in the
bottom panel of Figure~\ref{fig:OVIphys1}, we can calculate the energy
dissipation rate $\epsilon$ (center panel), which indicates lower
rates than observed at high-$z$ by \citet{rau01} for photo-ionized
absorbers tracing the diffuse IGM.  The median ages are plotted
(bottom panel) to show that more turbulence may be associated with
metals ejected more recently.  }
\label{fig:OVIphys2}
\end{figure}

\citet{rau01} calculates $\epsilon\sim10^{-3} \cmsst$ for high-$z$
$\CIV$ absorbers, and uses the dissipation timescale,
\begin{equation} \label{eqn:tdiss}
\tau_{diss}\sim \frac{<v_{\rm KE}^2>}{2 \epsilon}
\end{equation}
to find that the turbulence should dissipate in $\approx 10^8$ yrs.
The average velocity of the kinetic energy input, $\langle v_{\rm KE}
\rangle$, is speculated to arise from some sort of galactic-scale
feedback.  This short timescale, along with $\CIV$ arising from
overdensities enriched recently by SF galaxies (OD06), indicates a
more violent environment than the overdensities $\OVI$ absorbers trace
at low-$z$; therefore we consider this energy transfer rate an upper
limit for low-$z$ $\OVI$.  We calculate and plot $\epsilon$ as a function of
$N(\OVI)$ in the central panel of Figure \ref{fig:OVIphys2} for both
the photo-ionized and CIE cases by assuming $b_{turb}$ from the top
panel and $l(\OVI)$ from the bottom panel of
Figure~\ref{fig:OVIphys1}.

If we use photo-ionized absorber lengths below $10^{14.5} \cms$
($l(\OVI)\sim 50-100$ kpc), we find energy transfer rates at most
$\sim2\times 10^{-4} \cmsst$ dropping rapidly at lower $N(\OVI)$.
$l(\OVI)$ grows rapidly above $10^{14.5} \cms$ for photo-ionized absorbers
reducing $\epsilon$, but the absorber length from $\OVI$ in CIE compresses
such absorbers to be $\ll 100$ kpc, so that these absorbers (often
tracing halo gas) are able to fit within a halo.  $\epsilon$ grows
to match the \citet{rau01} value for the strongest $\OVI$ absorbers.
Such absorbers we will argue in \S\ref{sec:collion}, may be analogous to
the $\OVI$ absorbers associated with the MW halo and thick disk, which
must be collisionally ionized \citep{sem03,fox04}.  Considering the thick
disk absorbers with 60 $\kms$ corresponding to a structure on at most
several kpc \citep{sav03}, we subtract off the temperature broadening
($17.7 \kms$ at $T=10^{5.45}$ K), the instrumental broadening ($\sim20
\kms$ for FUSE), and safely assume no spatial broadening to arrive at
$b_{turb}\sim54 \kms$.  If we assume pure Kolmogorov turbulence and
$l(\OVI)=3$ kpc, we find a much higher transfer rate, $1.7\times10^{-2}
\cmsst$.  Similar size assumptions for the \citet{sem03} absorbers, which
are more difficult to estimate due to their less constrained distance,
give $b_{turb}\sim30 \kms$ and $\epsilon= 3 \times10^{-3} \cmsst$.
These values are higher than in the IGM at $z>2$ and within MW molecular
clouds, $\epsilon\sim 4\times10^{-4} \cmsst$ \citep{lar81}, although more
similar to $\HII$ regions $\epsilon\sim 2.5\times 10^{-2} \cmsst$ for a
100 pc-sized region \citep{ode91}.  The latter do not show a Kolmogorov
spectrum, but may be more related to outflows driven by supernovae,
which may include IVCs and HVCs in the MW halo.  The key point here is
that the turbulence we invoke for most $\OVI$ systems is a small fraction
of the energy transfer rate seen at high-$z$ and smaller still compared
to turbulence associated with active regions within our Galaxy.


Do our absorbers follow a Kolmogorov spectrum?  We should see evidence
of an increasing $b_{noturb}$ in the higher resolution d16n256vzw150
simulation compared to d32n256vzw150 if it is.  Our added turbulence,
$b_{turb}$ from Equation \ref{eqn:bturbform}, should be reduced by
a factor of $\frac{1}{2}^{1/3}\sim 0.8$ when halving the resolved
length scales according to Equation \ref{eqn:epsilon} if $\epsilon$
is constant; however we see no evidence of an increasing $b_{noturb}$
with resolution.  Therefore, either turbulence begins at smaller scales
with a larger $\epsilon$, or it does not follow a Kolmogorov spectrum.
There may be a precedent for the former as simulations of SNe-driven
ISM turbulence by \citet{jou08} find that energy injection occurs over
a broad range of scales.  The turbulent energy injection of outflows
may very well occur on scales less than the resolution of our simulations
($\sim 20-50$ kpc for halo gas) meaning that $\epsilon$ is larger.

Why would turbulence decline toward lower overdensity?  As a possible
answer, we consider the ages of individual SPH particles defined by
the last time they were launched in a wind.  We define the average
``age" of an absorber as an $\OVI$-weighted age of its contributing SPH
particles (equation~\ref{eqn:weighting}).  Plotting the median age of
$\OVI$ absorbers with 1$\sigma$ dispersions (bottom panel of Figure
\ref{fig:OVIphys2}) shows that younger absorbers have higher column
densities, which in turn have more turbulent broadening in our model.
The average age of metals in a $10^{13.0} \cms$ is about 6 Gyr while for a
$10^{14.5} \cms$, the age is 2.5 Gyr.  Note that this is an {\it average}
age, and higher column density absorbers may be composed of a number of
SPH particles with a range of ages, a significant fraction of which are
below 1~Gyr.  Our turbulence model is therefore consistent with the idea
that turbulence is injected by outflows, and then dissipated on a timescale
comparable to (or perhaps somewhat shorter than) a Hubble time.

Finally, we stress the need for high-resolution simulations to trace
outflows as they move from the galactic ISM into the IGM.  The largest
contributor of ISM turbulence in star-forming galaxies appears to be
supernovae \citep{mac04}, with $>90\%$ of the kinetic energy from
SNe-driven turbulence contained shortward of 200 pc~\citep{jou06}.
Although cosmological simulations are far from capable of resolving such
scales, an implementation of a sub-grid model of turbulent pressure
appears conceivable \citep{jou08}.  Meanwhile, the simulations of
\citet{fuj08} track the formation of shell fragments by Rayleigh-Taylor
instability thought to be responsible for $\NaI$ absorption tracing
wind feedback around star-forming galaxies \citep[e.g.][]{mar05a}.
The differential velocities produce $\NaI$ line widths of $320\pm120
\kms$, which we would describe ascribe as turbulent broadening using
our broad definition.  We hypothesize that the hot medium driving the
shell fragments that escape into the IGM evolve into the low density
metals traced by $\OVI$ with denser cloudlets traced by low ionization
species (e.g. $\CII$, $\CIII$, \& $\SiII$).  Extending galactic-scale
simulations, including treatment for small-scale turbulence ($<200$ pc),
to follow winds into the IGM under an ionizing background should help
show how turbulence affects the line profiles of various species.

To summarize, ``painting on'' turbulence post-simulation should rightfully
be considered controversial.  However, it also likely to be unrealistic
that gas on 20-100 kpc scales pumped by galactic outflows has no internal
motions on smaller scales.  The energy dissipation rate of the required
turbulence at low-$z$ is just a fraction of that observed for high-$z$
$\CIV$, which traces more recent SF activity.  Our discussion here
emphasizes that IGM metal-line absorption profiles can be dominated by
small-scale velocities, analogous to the line profiles of Galactic $\HII$
regions, molecular clouds, IVCs, and HVCs.  In fact, the diffuse $\lya$
forest may be a rare instance of something relatively unaffected by
turbulent velocities.  Resolution appears to resolve better some of
the small scale velocities in the very strongest absorbers, but if
the d16n256vzw150 simulation is any guide, the resolution required
to accurately model such small-scale motions is likely much higher.
Lastly, we have added turbulence as a function of density, which is
only a first-order approximation for the more relevant properties of
environment and ages of metals; an interesting future study (particularly
in comparison to COS data) would be to see how exactly turbulence
originates and what parameters best describe it.

\section{Origin and Environment of $\OVI$} \label{sec:origin}

The primarily photo-ionized nature for $\OVI$ absorbers suggests
this ion traces metals in the warm diffuse IGM.  We consider here the
origin of various absorbers by determining when their metals were last
injected by winds.  The age-density anti-correlation alluded to in the
previous section naturally segues into a discussion of environment,
where we consider an absorber's relation to galaxies responsible for
enriching the IGM.  We then discuss collisionally ionized $\OVI$, noting
that it appears primarily within halos of $\sim M^*$ galaxies which we
suggest is analogous to $\OVI$ associated with the MW galactic halo.
Finally, we show simulated COS observations of $\OVI$ absorption systems,
demonstrating how such observations can provide insights about environment
and the evolutionary state of IGM gas.

\begin{figure*}
\includegraphics[scale=0.50]{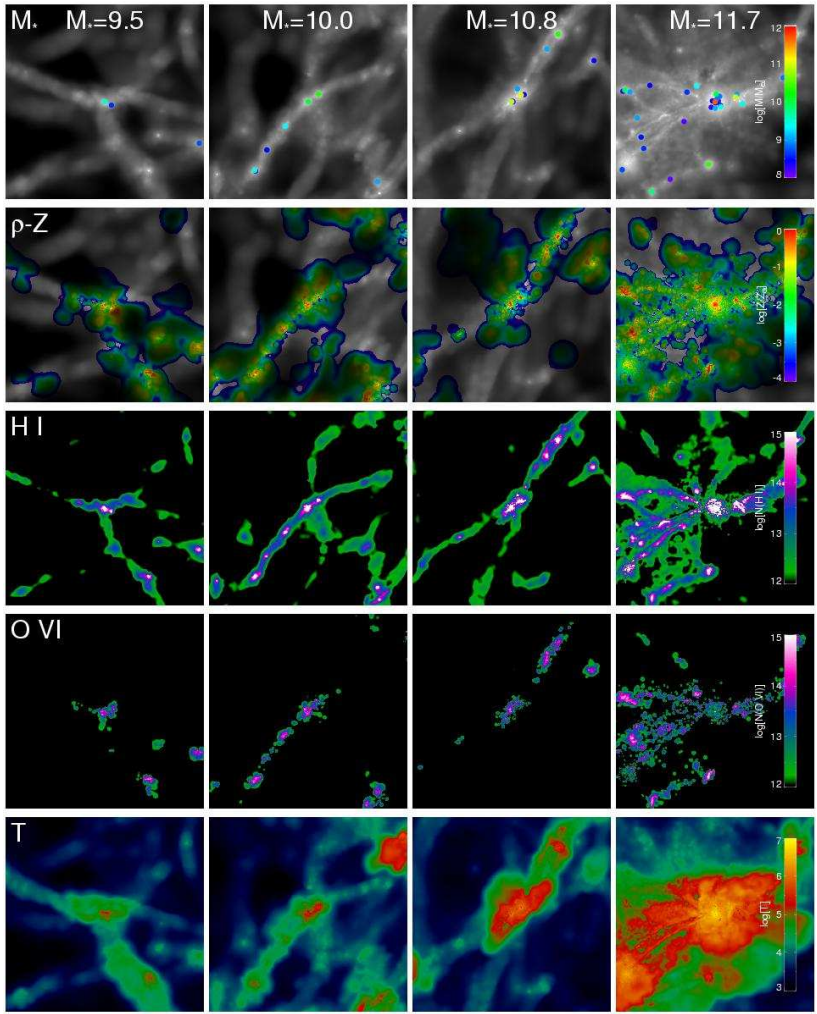}
\caption[]{The IGM environments in $8\times 8 \hmpc \times 25 \kms$
  slices around 4 different-sized galaxies at $z=0.25$.  {\it Top
  row:} Galaxy locations within the slice are indicated by colored
  dots corresponding to the galaxy stellar mass; the greyscale
  indicates gas overdensity.  {\it Second row:} Colors indicate the
  enrichment level of the gas with the greyscale.  {\it Third row:}
  $\HI$ column densities.  {\it Fourth row:} $\OVI$ column densities.
  {\it Bottom row:} Average gas temperature of the IGM.  Four
  different galaxy environments are displayed ranging from sub-$L^*$
  (left 2 columns), an $L^*$ galaxy analogous to the Local Group
  (third column), and a massive group environment (last column).  The
  strength and extent of $\OVI$ does not necessarily correlate with
  the galaxy mass, unlike $\HI$.  $\OVI$ is mostly photo-ionized in
  these simulations tracing overdensities of 10-200.  Often
  photo-ionized $\OVI$ is co-spatial with the WHIM and hot halo gas
  (regions of red and yellow).}
\label{fig:4galslices}
\end{figure*}

Figure \ref{fig:4galslices} illustrates various physical and observational
properties of the IGM around four selected galaxies ranging over two
orders of magnitude in stellar mass.  These figures show zoomed-in views
of some of the trends seen in Figure \ref{fig:fullslice}.  Metals follow
galaxies, and $\OVI$ in particular traces filamentary structures.
Despite a large range in stellar mass in the left three columns,
the extent of $\OVI$ is not qualitatively different.  Furthermore,
the stronger $\OVI$ in the group environment on the right panels are
found around galaxies on the outskirts of this group and not around the
$M_*=10^{11.7} \msolar$ in the center.  Much of this has to do with the
majority of $\OVI$ in our simulations tracing the diffuse, photo-ionized
IGM rather than hot halo gas.  In this section we will concentrate on the
locations of galaxies (top row) compared to $\OVI$ absorbers (4th row).


To quantify trends in origin and environment, we have modified our
{\tt specexbin} spectral generation code to assign the stellar mass
($M_{*}$) and distance ($r_{gal}$) to the galaxy identified via Spline
Kernel Interpolative Denmax\footnote{SKID;
{\tt http://www-hpcc.astro.washington.edu/tools/skid.html}} with the
greatest dynamical influence for each SPH particle.  Note that the
d32n256vzw150 simulation resolves galaxies down to $\approx
10^{9}\msolar$.  We define a {\it neighboring galaxy} to each SPH
particle as the one with the smallest fractional virial distance to
the particle (i.e. the minimum $r_{gal}/r_{vir}$, where
$r_{vir}\propto M_{dyn}^{1/3}$).  To calculate the dynamical mass
($M_{dyn}$) from the stellar mass we use the relation from
\citet{dav09}, which shows the fraction of baryons in stellar mass for
a given halo mass as calculated from this simulation and the
d64n256vzw150 simulation\footnote{See the relation in the left middle
panel of this paper's Figure 1.}.  {\tt specexbin} tracks age as well
as the originating galaxy mass and launch $\vw$ for enriched outflow
particles, but we limit ourselves to considering an absorber's age
only, saving an in-depth analysis of the originating galaxies of
absorbers for future work.  The neighboring $M_{*}$ and $r_{gal}$ we
refer to here often have no relation to that of the particle's
originating galaxy, especially considering the processes in galaxy
evolution and structural formation have altered the environments of
$\OVI$ absorbers, which often trace metals released into the IGM at a
much earlier time; this contrasts to the closely related
galaxy-absorber connection found in simulations with the same wind
model in \citet{opp09} at $z=6$ where winds appear to be clearly
outflowing from their parent galaxies.  Unlike the normal version of
{\tt specexbin}, we only use one snapshot at $z=0.25$ over the range
of $z=0.5\rightarrow 0$, although the varying ionization fields and
Hubble expansions are applied from the redshift along the line of
sight; this explains why ages never go above 10 Gyr.

\begin{figure*}
\includegraphics[scale=0.66]{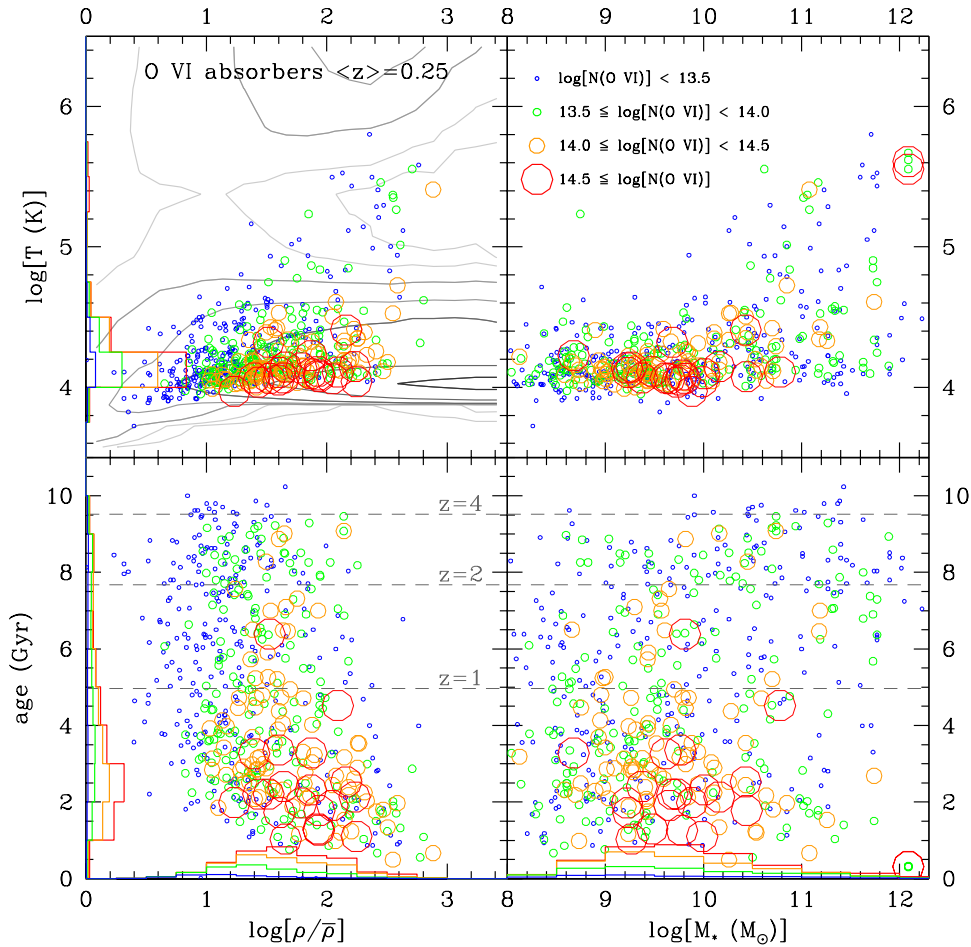}
\caption[]{$\OVI$ absorbers from the high-quality d32n256vzw150-bturb
  sample are plotted in four phase space planes considering two
  physical parameters ($\OVI$-weighted density and temperature), the
  age of gas tracing $\OVI$ absorbers (i.e. the time since the wind
  launch), and the stellar mass of the neighboring galaxy (i.e. the
  galaxy with the least fractional virial distance).  Logarithmic
  contours at 0.5 dex steps in the $\rho-T$ phase space correspond to
  the metallicity-weighted density (lighter contours are less).
  $\OVI$ photo-ionized absorbers trace metals up to an overdensity
  ($\delta$) of 300 where $f(\OVI)$ drops to almost zero, and $\OVI$
  at higher density must be collisionally ionized.  Few of these
  latter absorbers are found, because metals have a ``zone of
  avoidance,'' here due to efficient metal-line cooling through this
  region.  Absorbers can be followed around the four panels by eye to
  see how these parameters are related for various $\OVI$ strengths.
  A density-age anti-correlation exists with $\OVI$ in CIE associated
  with recent activity, often in close proximity with $M^*$ galaxies.
  Absorbers below $N(\OVI)=10^{14} \cms$ do not correlate generally
  with environment showing little dependence on the neighboring
  $M_{gal}$.  Such absorbers trace photo-ionized metals released in
  winds often at $z>1$.  Histograms along the side show the summed
  $\Sigma N(\OVI)$, with each color corresponding to the column
  density range indicated by the key in the upper right panel.  Less
  than 10\% of $\Sigma N(\OVI)$ is collisionally ionized.  }
\label{fig:4params}
\end{figure*}

We plot $\OVI$-weighted densities, temperatures, ages, and neighboring
galaxy masses at absorber line centers in Figure \ref{fig:4params} at
$z<0.5$ in 30 high-quality lines of sight from our d32n256vzw150-bturb
model.  The $\rho-T$ phase space panel (top left) reiterates our
finding that most $\OVI$ absorbers are photo-ionized near $T=10^4$ K
with an easily noticeable density-$N(\OVI)$ correlation.  The
over-plotted logarithmic metallicity-weighted contours show low-$z$
metals have either cooled to $T\sim 10^{4}$ K, where $\OVI$ is an
ideal tracer of these diffuse IGM metals, or are much hotter, $T\ga
10^6$ K, where cooling times are long.  In between is the metal ``zone
of avoidance'' where the rare $\OVI$ CIE absorber traces rapidly
cooling halo gas.  We provide an in-depth analysis of this region in
\S\ref{sec:collion}.

\subsection{Ages of $\OVI$ Absorbers} \label{sec:age}

The ages of $\OVI$ absorbers in the bottom left panel of Figure
\ref{fig:4params} generally anti-correlate with density.  Absorbers
below $N(\OVI)=10^{14} \cms$ (more often in simple systems) usually
correspond to metals in filaments often injected during the high-$z$
epoch of intense cosmic star formation.  Strong absorbers trace metals
on the outskirts of halos injected at $z<1$, while some of the
strongest absorbers near $N(\OVI)=10^{15} \cms$ trace metals cycling
in halo fountains within the virial radius as described by OD08 on
timescales $\la 2$~Gyr.  Histograms of $\sum{N(\OVI)}$ are drawn for
each property with the colored histograms stacked upon each other
corresponding to the column density range; intermediate absorbers
(green \& orange) hold the majority of $\OVI$ in the IGM.
A median age of 3~Gyrs is found with a spread of 0.6~dex.  Interestingly
$\CIV$ at $z=2$ shows a similar age spread and median ($\sim1$ Gyr) in
terms of a fraction of a Hubble time~\citep{opp07}, which is not
surprising considering that $\CIV$ at this redshift traces similar
overdensities as low-$z$ $\OVI$ (e.g. Simcoe et al.  2004; OD06).

The trend of increasing age with decreasing overdensity of metals results
from the outside-in pattern of IGM enrichment.  OD08 shows that winds
launched at $z=6\rightarrow 0.5$ all travel similar distances (60-100
proper kpc), thereby allowing a galaxy to enrich a larger comoving volume
at early times.  The increasing mass-metallicity relationship of galaxies
as the Universe evolves \citep[cf.][]{erb06,tre04} naturally results
in a metallicity-density IGM gradient as later galaxies enrich gas at
higher overdensities filling less comoving volume with more metal-rich gas.
Therefore the outer regions traced by the weakest $\OVI$ absorbers
are tracing the earliest epochs of IGM enrichment while remaining
relatively unaffected by later enrichment.  $\OVI$ is by far the best
UV transition able to trace such regions due to its high oscillator
strength, oxygen being the most copious metal, and its highly ionized
state ideal for tracing low densities where there are many ionizing
photons per ion\footnote{It would be fascinating to see if nucleosynthetic
yields indicating enrichment by low-metallicity or even Pop III stars are
associated with weaker $\OVI$ lines, although this would be difficult
as no other species has a similar transition in the UV.  High-$S/N$
observations of aligned $\CIV$ using COS could possibly probe similar
low overdensities with the trend of derived [O/C] with column density
possibly indicating evolving nucleosynthetic yields.}.  Conversely,
the overdense inner regions trace gas recycled multiply in winds often
not escaping from galactic halos, as described in OD08.  These regions
are relatively rare in a volume-averaged measurement as an absorption
line spectrum, but may harbor the strongest lines with $\OVI$ in CIE.

\subsection{Galaxy Environments of $\OVI$ Absorbers} \label{sec:env}

The strong correlation between density and $N(\OVI)$ suggests a
possible correlation between $M_{*}$ and $N(\OVI)$, considering that
we find more overdense environments associated with more massive
galaxies at low-$z$ in our simulations (OD08).  The right
panels of Figure \ref{fig:4params} do show such a trend, but only very
weakly; there is a large range of absorber strengths for a given
$M_{gal}$.  This is most easily noticed as the large area covered by
$\OVI$ absorbers in the $M_{*}$-age space.

\begin{figure}
\includegraphics[scale=0.90]{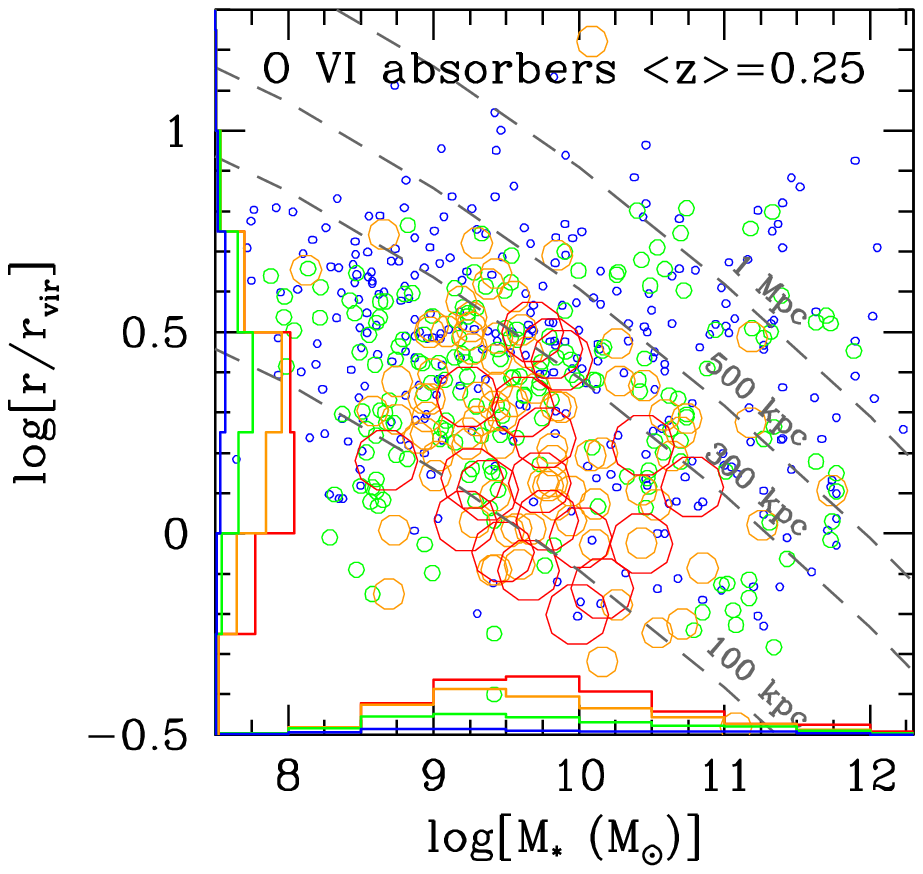}
\caption[]{The distance to the neighboring galaxy, shown as a multiple
  of the virial radius, is plotted against that galaxy stellar mass
  for all $\OVI$ absorbers.  The color and size key for $\OVI$ column
  density in Figure \ref{fig:4params} applies here.  Dashed lines show
  the physical distance assuming $z=0.25$, and the colored histograms
  correspond to the number of absorbers in each $\OVI$ column density
  bin.  Stronger $\OVI$ absorber are more likely to be dynamically
  associated with more massive galaxies, with the strong absorbers
  almost always within $2~r_{vir}$.  The range of physical distances
  seem similar to observed impact parameters.  We obtain the virial
  radius by using the relation from \citet{nav97} for a dark matter
  halo; we convert from stellar to dynamical mass using the relation
  in the left center panel of Figure 1 of \citet{dav09}.}
\label{fig:dgal_mgal}
\end{figure}

Another key variable we need to consider is distance from a
neighboring galaxy, which we plot as a multiple of the virial radius,
$r_{gal}/r_{vir}$, in Figure \ref{fig:dgal_mgal}.  Most $\OVI$
absorbers in the observed range reside several virial radii from the
neighboring galaxy, suggesting they are dynamically unassociated with
this galaxy.  Histograms along the side for each column density bin
indicate weaker absorbers lie at a greater virial distance from the
nearest galaxy, which is on average less massive.  The stellar mass
associated with $\OVI$ absorbers is $M_{gal}\sim10^{9.5-10}
\msolar\sim 0.03-0.1 M^*$.  These same absorbers are among the oldest,
suggesting merely incidental association with their present day
neighboring galaxy.  The absorbers at $N(\OVI)<10^{14} \cms$
associated with $M_{gal}>10^{11} \msolar$ are similarly incidental.
This is best illustrated by the chaotic structure of $\OVI$ absorbers
in the right panels of Figure \ref{fig:4galslices}.  \citet{gan08}
finds a similar association of their weaker absorbers ($EW<50$ m\AA)
with $\sim 0.1 L^*$ galaxies in the \citet{cen06b} simulation.

Our simulation results here are worth comparing to observations,
however we note that we consider only galaxy stellar masses, not
luminosities, and our detection limits are complete down to
$M_{gal}=10^9 \msolar$, or $\sim0.01 M^*$, which is better than any
survey currently obtains, except very locally.  Surveys by
\citet{pro06} and \citet{coo08} of the galaxy-absorber connection find
a large variety of environments for $\OVI$ absorbers, which could
possibly support our claim that absorbers in the observed range are
usually not directly associated with their neighboring galaxy.  The
majority of our absorbers are between 100-300~kpc from their nearest
galactic neighbor, which compares favorably with observations finding
the nearest projected neighbor at $\sim 100-200$ kpc away from $\OVI$
\citep{tri01, tum05b, tri06}, and the median distances of 200-270 kpc
from 0.1 $M^*$ galaxies derived by \citet{sto06}.  

\citet{sto06} stresses the need for deeper surveys below $0.1 L^*$ to
find the galaxies responsible for the bulk of the IGM enrichment at
low-$z$.  Indeed, 66\% of our intermediate absorbers neighbor $<0.1
M^*$ galaxies, which is even more impressive considering our neighbor
weighting scheme favors association with more massive galaxies.
However, the ages of such absorbers are almost always greater than 2
Gyr, often much more, indicating that such association is perhaps
incidental.  The scenario may be related to $\CIV$ absorbers at
$z\sim2-3$ showing the same clustering at several comoving Mpc as
Lyman-break galaxies \citep{ade03, ade05}, while \citet{por05} find
that this clustering arises even if the metals $\CIV$ traces were
injected at very high-$z$.  The same scenario appears to be in play
for $\OVI$ absorbers, which often have little dynamical association
with their neighboring galaxy and instead are injected at a much
earlier epoch.


If $\OVI$ is not linked to its immediate environment, then perhaps
this can explain \citet{pro06} finding 6~$\HI$ metal-free systems with
qualitatively similar environmental characteristics as their 6 systems
with $\OVI$ along the PKS 0405-123 sight line; the $\OVI$ does not
care about the immediate surroundings and therefore traces a wide
range of environments.  This supports the inhomogeneous nature of IGM
metal enrichment, where $\OVI$ absorbers trace metals above the mean
metallicity.  We also find a significant fraction of our $\HI$
absorbers without aligned $\OVI$ (compare third and fourth rows of
Figure \ref{fig:4galslices}).

\begin{figure}
\includegraphics[scale=0.80]{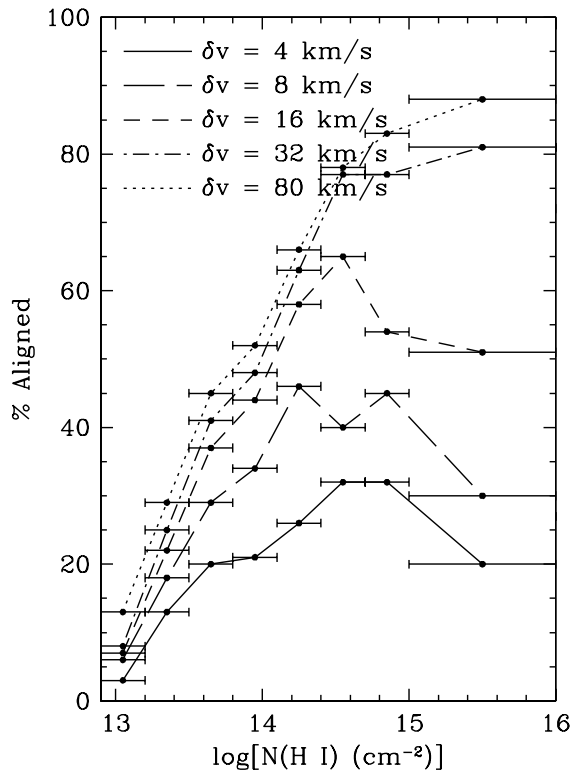}
\caption[]{The alignment fraction of $\HI$ absorbers with $\OVI$ with
at least $EW=30$ m\AA~at various $\delta v$'s using the
d32n256vzw150-bturb model.  Bins from $N(\HI)=10^{12.9-15.0} \cms$ are
0.3 dex wide with the last bin spanning $N(\HI)=10^{15.0-16.0}\cms$.
Bins above $N(\HI)= 10^{14.4} \cms$ are much more likely to be aligned
with $\OVI$ at some level.  The strongest $\HI$ absorbers are
associated 80\% of the time with $\OVI$, but are less likely to be
well-aligned, which we suggest is a signature of $\OVI$ near or in
halos more often collisionally ionized.  }
\label{fig:HI_align}
\end{figure}

As an aside, we plot the fraction of $\HI$ absorbers aligned with
$\OVI$ as a function of column density at various $\delta v$
separations in Figure \ref{fig:HI_align}, as long as $\OVI$ has an
equivalent width $\ge 30$ m\AA.  Note that unlike $\OVI$ aligned with
$\HI$ considered earlier, only one $\OVI$ component can be aligned
with a single $\HI$ component.  We provide this plot as a prediction
for comparison with current and future data.  We predict about half of
$N(\HI)= 10^{14.0} \cms$ absorbers should have $\OVI$ within 80
$\kms$, however only 20\% have well-aligned $\OVI$.  At $N(\HI)=
10^{15-16} \cms$, $\OVI$ lies within 80 $\kms$ 88\% of the time, but
the simulations suggest such absorbers are slightly less likely to be
well-aligned than the bin below it-- a possible signature of $\OVI$ in
CIE as we discuss next.

\subsection{Collisionally Ionized $\OVI$} \label{sec:collion}

The most obvious difference between collisionally ionized (CI) and
photo-ionized (PI) $\OVI$ is environment.  The few strong absorbers at
CI temperatures in Figure~\ref{fig:4params} trace gas within or just
outside halos of $\sim M^*$ galaxies undergoing rapid metal-line
cooling in the metal ``zone of avoidance.''  Their relatively young
ages, $\sim 0-2$ Gyr, indicate association with recent enrichment.
The strongest $\OVI$ absorber in PKS 1405-123 ($N(\OVI)=10^{14.8}
\cms$ at $z=0.1671$) from \citet{pro06} fits this description almost
perfectly; the $\OVI$ appears to be CI, nearly $\Zsolar$, and lies 108
kpc from a 4.1 $L^*$ galaxy (i.e. $<r_{vir}$), which shows evidence
for a burst of star formation $\sim 1$ Gyr ago.  While \citet{pro06}
expresses surprise no other galaxies lie near this absorber, we find
this appropriate because our similar simulated absorbers depend
predominantly on the environment associated with the parent halo; this
is a clear contrast with PI absorbers in the diffuse IGM.  

The strongest IGM $\OVI$ absorber yet observed at $z<0.5$, the
$N(\OVI)=10^{14.95} \cms$ absorber in PKS 0312-770, shows multi-phase
behavior according to \citet{leh08}, where a PI Lyman limit system is
aligned with $\OVI$-bearing gas with a likely CI origin.  This work
finds an 0.7 $L^*$ galaxy at an impact parameter of 38 kpc, with a
morphology indicating a possible galaxy merger.  Indeed, \citet{leh08}
interpret this system as possibly representing outflow materials
cycling in a halo fountain as the impact parameter is consistent with
a halo origin for this system.

We propose such CI $\OVI$ absorbers are primarily related to those
observed in IVCs and HVCs \citep[e.g.][]{sav03,sem03,fox06} associated
with the MW halo or possibly the intragroup medium of the Local Group.
This is consistent with the idea that CI $\OVI$ is the interface
between a hot, tenuous halo medium at $T\geq 10^{6}$ and
pressure-confined dense clumps of $\sim5\times 10^6 \msolar$ proposed
by \citet{mal04}, and supported by the observations of the
\citet{leh08} system.  The observed HVC $N(\OVI)-b(\OVI)$ trend is also
consistent with the \citet{hec02} relation resulting from radiatively
cooling flows of hot gases passing through the coronal temperature
regime, yielding CI $\OVI$.

One problem is that our absorbers are too strong; we find 14 very
strong absorbers ($N(\OVI)\geq 10^{15} \cms$) over $\Delta z=35$ while
not one such absorber is found in any quasar sight lines or associated
with the MW halo.  Such absorbers rarely occur and usually require CI
$\OVI$, because the pathlengths for PI $\OVI$ are too long.  Our
simulations are very under-resolved to adequately model such dense
clumps using $m_{\rm SPH}\sim 3\times10^7 \msolar$.  Halo $\OVI$ in
CIE is likely to exist in many smaller structures creating weaker
lines, which also may explain their wide $b$-parameters as one
component is made up of many smaller parcels of gas at random,
turbulent velocities.  Hundreds or even thousands of SPH particles per
$\sim5\times 10^6 \msolar$ cloud are likely needed to model the large
range of densities in pressure equilibrium.

Another reason for excessively strong CI $\OVI$ absorbers is possibly
overestimated metallicities, assuming they are similar to MW HVCs.
HVCs in general show metallicities of 0.1-0.3 $\Zsolar$
\citep{tri03,sem04b,fox05}, with rare exceptions of super-solar HVCs
near the galactic disk \citep[e.g.][]{zec08}; conversely, our halo
absorbers average $\Zsolar$.  Our winds are confined in $M^*$ halos
recycling in halo fountains in a cycle less than a Gyr, which may be
too fast.  Again, resolution may be an issue as our single SPH
particles fall straight through hot galactic halos, when they
physically need to be pressure-confined.  Two-phase boundaries in SPH
are difficult as a particle shares a similar smoothing length and
therefore density with its neighbors while it is at a very different
temperature.  Overestimated ram-pressure stripping reduces tangential
velocities adding to this numerical artifact at the resolutions we
explore (see \citet{ker08} \S4.3 for a discussion of this ``cold
drizzle'').  If slower recycling is more realistic, fewer instances of
metal enrichment occur, leading to lower metallicities while reducing
our (overly high) star formation rates in massive galaxies.  Accurate
hydrodynamical modeling of $\OVI$ halo absorbers requires an increase
of several orders of magnitude in resolution as well as
non-equilibrium cooling and ionization.  The non-equilibrium
ionization fractions of \citet{gna07} may alleviate the overestimate
of strong CI $\OVI$ absorbers by allowing lower $f(\OVI)$ at $T<10^5$
K making more weak $\OVI$ halo absorbers with a larger filling factor.

Two characteristics of CI $\OVI$ are their multi-phase nature and
their mis-alignment with $\HI$.  The multi-phase nature of $\OVI$
appears in observations showing coincident $\CIII$ and strong $\HI$ in
the strongest $\OVI$ absorbers \citep{pro04,dan06,coo08}.  Often our
strongest absorbers show evidence that $\OVI$ itself is multi-phase,
since such absorbers occupy temperatures in between PI and CI $\OVI$
(see the broad range of temperatures in the red histogram in the upper
left panel of Figure \ref{fig:4params}).  The CLOUDY tables we use
require that absorbers where $10^{4.5}<T<10^{5.3}$ K and $\delta>100$
have multi-phase $\OVI$ since $f(\OVI)$ values are minuscule in this
$\rho-T$ phase space.

The mis-alignment characteristic appears when we consider only
simulated $N(\OVI)$ above $10^{13.5} \cms$; such absorbers where
$T>10^{5}$ K are well-aligned ($\delta v \leq 8 \kms$) only 36\% of
the time compared to 55\% for lower temperature absorbers.  Very
mis-aligned absorbers ($\delta v > 80 \kms$) occur 21\% of the time
for CI absorbers and only 12\% for the cooler sample.  The fact that
only 7\% of the T08 absorbers are what we consider very mis-aligned,
including those without any $\HI$ at all, seems to support the case
for predominantly PI $\OVI$.

We note the analytical model of \citet{fur05} can explain all $\OVI$
as collisionally ionized when considering structural formation shocks
propagating out to a $8\times$ the virial radius.  It is not
straightforward to quantify how far our virial shocks propagate, but
qualitatively from simulation animations\footnote{See {\tt
http://luca.as.arizona.edu/\~{}oppen/IGM/}} it doesn't seem to
propagate much beyond $2~r_{\rm vir}$.  Moreover, virial shocks don't
carry metals as they propagate outwards.  Finally, virial shocks at
these distances are usually at low enough densities that the high
ionization parameter makes $\OVI$ in CIE impossible; any $\OVI$ must
be photo-ionized at lower temperatures.  While \citet{fur05} suggests
an even split between CI and PI $\OVI$ if the shocks go to
$4~r_{vir}$, our CI $\OVI$ is rarely found outside $2~r_{vir}$.  Like
them, we find strong absorbers should remain nearly unaffected by the
extent of virial shocks considering none of our absorbers over
$N(\OVI)=10^{14.7} \cms$ lie outside of $2~r_{vir}$.

The frequency of our $\OVI$ {\it systems} with $N(\OVI)\geq 10^{14.5}
\cms$ (i.e. summing all components within 100$\kms$) is 1.8 $\Delta
z^{-1}$.  We find a somewhat higher frequency of intersecting halos
$\geq 0.1 M^*$ halos ($1.4 \Delta z^{-1}$ for $>M^*$ and $1.8 \Delta
z^{-1}$ for $0.1-1 M^*$), however the frequency of a strong absorber
within a virial radius of such galaxies occurs about a tenth of the
time.  While we find CI $\OVI$ relates to the immediate environment of
a galaxy in a way PI $\OVI$ does not, not all strong absorbers are in
CIE and more often trace gas within two virial radii around sub-$M^*$
galaxies.  A fascinating survey with COS would be to explore sight
lines intersecting galactic halos for a range of galaxy masses and
types ($\leq 150$ kpc).  The incidence of $\OVI$ as a function of
impact parameter will provide a handle on the filling factor of
$\OVI$, which we suggest here is more often collisionally ionized
within halos.  A similar survey of $\CIV$ at $z<1$ by \citet{che01}
finds a sharp cutoff above $\sim 100$ kpc.  However, strong $\OVI$
associated with $\HI$ does not imply that $\OVI$ is within the halo,
as we have shown that the associated $\OVI$ often arises from lower
overdensities outside.  $\OVI$ may not have a similarly sharp cutoff
at a specific impact parameter as $\CIV$ if this is the case.

To summarize, our simulations preliminarily suggest that the vast
majority of $\OVI$ absorbers in quasar sight lines are likely
photo-ionized.  Our CI $\OVI$ fraction is less constrained; while
we find a small fraction ($<10\%$), this may increase if we could
resolve the structures responsible for CI $\OVI$, which we feel are
related to the HVCs in our Galaxy and necessary to explain the
strongest $\OVI$ systems.  However, if we consider that CI $\OVI$
is associated with stronger absorbers, as other simulations also
find \citep{cen01,fan01}, the steeper fit to the differential column
density distribution ($d^2n/dzdN(\OVI)\propto N(\OVI)^{-2.0}$, DS08)
compared to other species ($d^2n/dzdN(\HI)\propto N(\HI)^{-1.7}$,
$d^2n/dzdN(\CIII)\propto N(\CIII)^{-1.8}$, also DS08) suggests there
is not a large reservoir of CI $\OVI$ at high column densities.
$\HI$ and $\CIII$ have rising ionization fractions with higher
density at PI temperatures; this is a clear contrast to $f(\OVI)$,
which drops sharply at rising densities, possibly explaining part
of the difference in the power law fits.  Hence the CI versus PI
$\OVI$ argument is not completely settled, but it seems very likely
that most weak and intermediate $\OVI$ absorbers are photo-ionized,
and that collisionally ionized $\OVI$ is usually associated with
galactic halo gas.

\subsection{COS Simulated Observations and Environment} \label{sec:COS}

\begin{figure*}
\includegraphics[scale=0.70,angle=-90]{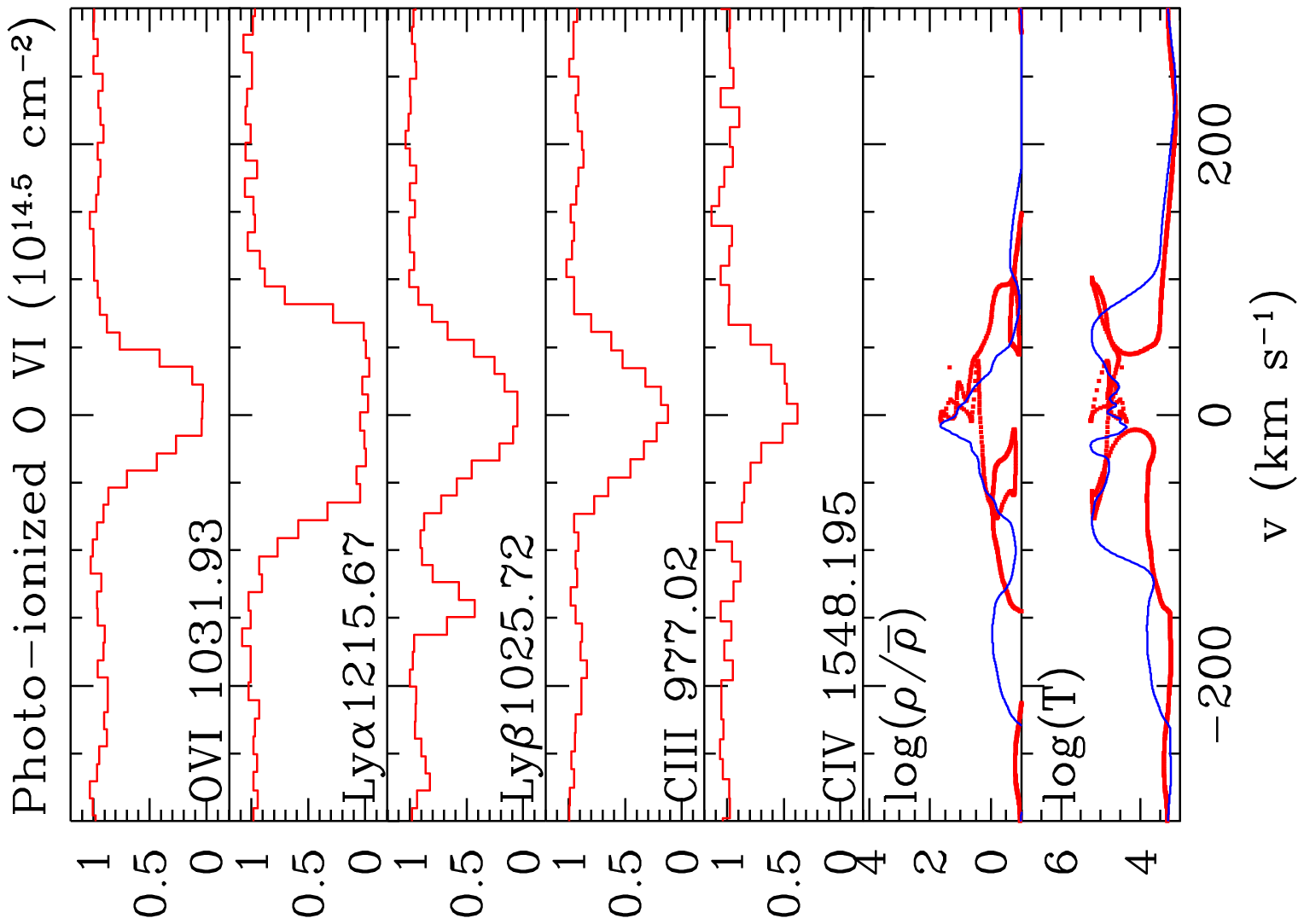}
\includegraphics[scale=0.70,angle=-90]{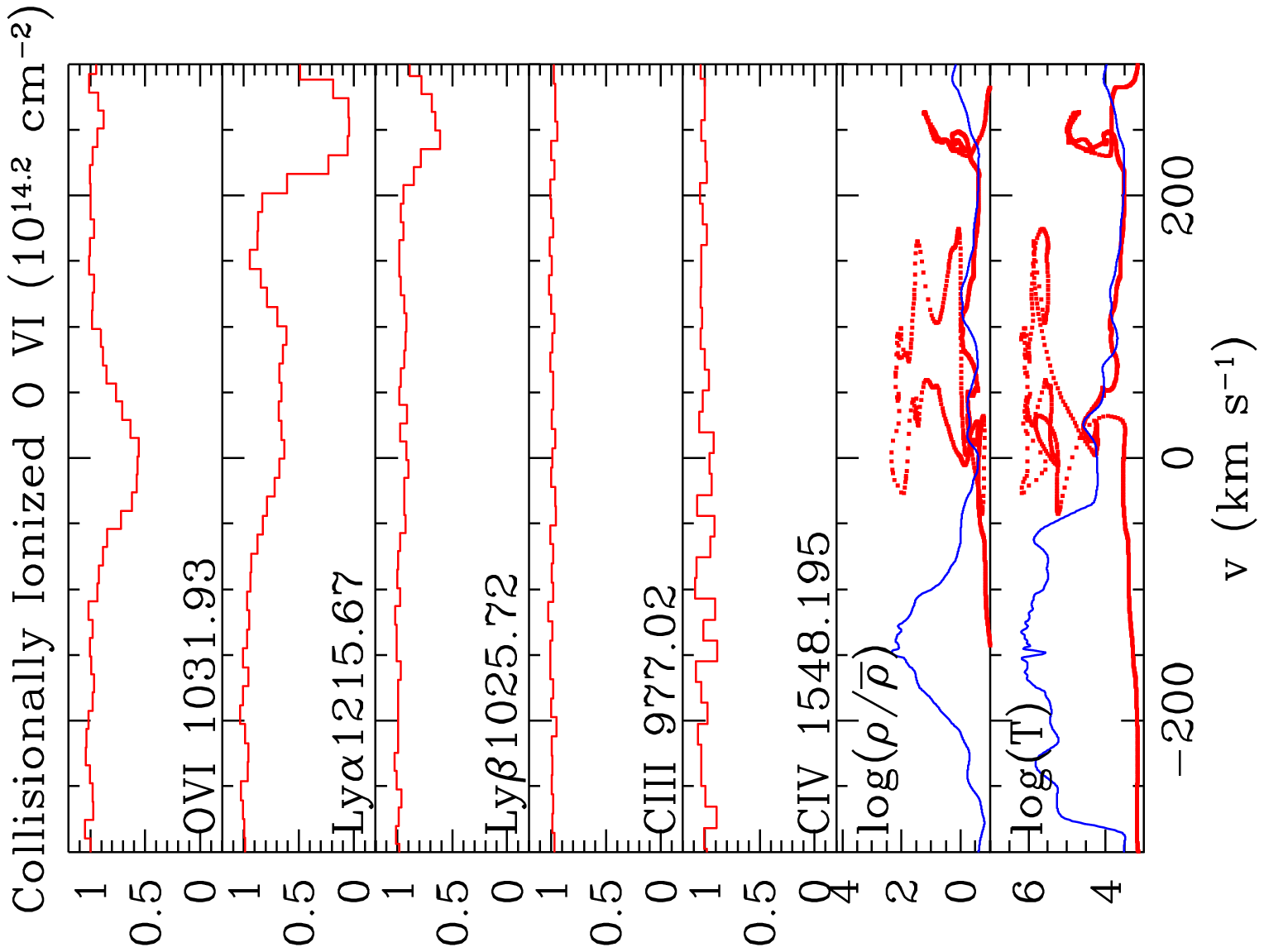}
\caption[]{Two metal-line systems simulated at prospective COS
resolution and $S/N$.  Absorption lines and physical parameters
plotted in red are in velocity space, while blue indicates the spatial
distribution of density and temperature in the bottom two panels
without peculiar velocities.  The left system at $N(\OVI)=10^{14.5}
\cms$ is comprised of photo-ionized gas well-aligned with $\HI$ at
$N(\HI)=10^{15.0} \cms$ tracing gas 195 kpc from a $\sim 0.1 M^*$
galaxy.  The right system is a $10^{14.2} \cms$ $\OVI$ absorber in CIE
lying 270 kpc from a super-$M^*$ galaxy showing a weak asymmetric
$\HI$ profile and no lower-ionization metals lines.  Both absorbers
have metallicities of $0.45 \Zsolar$.  Peculiar velocities dominate
over spatial distribution of metals and baryons as in the systems.
The two systems, comparable to actual observed systems, emphasize the
different environments from which $\OVI$ absorbers arise.  }
\label{fig:sysspec}
\end{figure*}

As a prelude to the types of observations enabled by {\it Hubble}/COS,
we consider $\OVI$ simulated absorption systems from a bright quasar
continuum generated using the COS Spectral Simulator\footnote{\tt
http://arlsrv.colorado.edu/cgi-bin/ion-p?page=ETC.ion} in Figure
\ref{fig:sysspec} where we assume 10,000 second integrations at all
bands.  Our point in this figure is to contrast the metal-line systems
of PI and CI $\OVI$ absorbers to see how environment can lead to
significant observational signatures.  The PI $\OVI$ absorber with
$b(\OVI)=35 \kms$ at $z=0.25$ is well-aligned ($\delta v=0.9 \kms$)
with a $\HI$ component where $N(\HI)=10^{15.0} \cms$ and $b(\HI)=47
\kms$ from $\lyb$.  This is a simple system with well-aligned $\CIII$
and $\CIV$ tracing gas 195 kpc from a $10^{9.8} \msolar$ galaxy
($4~r_{vir}$) with an average age of 4.4 Gyr and a metallicity of
$0.45 \Zsolar$.  We consider this absorber analogous to the
$z=0.20702$ absorber in HE 0226-4110 from \citet{tho08b}; their $\HI$
is slightly stronger and broken into two components although it
appears centered on the $\OVI$.

We show the density and temperature (mass-weighted) in the bottom
two panels in physical space (blue) and velocity space (red).
Peculiar velocities dominate over the spatial distribution even in
a PI absorber tracing the diffuse IGM.  At the line center different
parcels of gas overlap resulting in a multi-phase PI absorber
($T(\OVI)=10^{4.1}$ \& $T(\HI)=10^{4.3}$ K; $\delta(\OVI)=47$ \&
$\delta(\HI)=86$).  In this case, gas from surrounding voids accreting
into the local filament heats the extended gas extending $\sim 1$
Mpc around the galaxy to $T\sim 10^{5.2}$ K.  This gas is mostly
unenriched, despite being at CIE temperatures; any oxygen present
would be photo-ionized to a higher state at such low density anyway.
This suggests that there may be considerable WHIM gas that cannot
be traced by any metal lines, owing to the small filling factor of
metals in the IGM.

To demonstrate a typical CI $\OVI$ absorber, we show a $10^{14.2}
\cms$ component, which happens to be aligned with a weak BLA
($N(\HI)=10^{13.5} \cms$, $b(\HI)=50 \kms$).  In contrast to the type
of simple aligned absorbers T08 observes, which indicate
photo-ionization, the $\HI$ distribution here is multi-component,
extended, and weaker (another $10^{13.5} \cms$~$\HI$ component lies 80
$\kms$ away).  This system is multi-phase with $T(\OVI)=10^{5.01}$ K
indicating possibly some weak aligned PI $\OVI$, while
$T(\HI)=10^{4.16}$ K.  BLAs usually do {\it not} primarily trace gas at
temperatures above $10^{5}$ K, often because aligned $T=10^4$ K gas
dominates the absorption.  Lying 270 kpc from a massive $10^{11.5}
\msolar$ galaxy, this system traces gas with $Z=0.45\Zsolar$ in the
intragroup medium just outside of the halo.  Unlike most other CI
absorbers, which are often stronger, the metals traced are older, 6.6
Gyr. An observed analog may be the $z=0.1212$ system in H1821+643,
which indicates CI $\OVI$ \citep{tri01}, although the associated $\HI$
is somewhat stronger and broader in that system.

\begin{figure}
\includegraphics[scale=0.70,angle=-90]{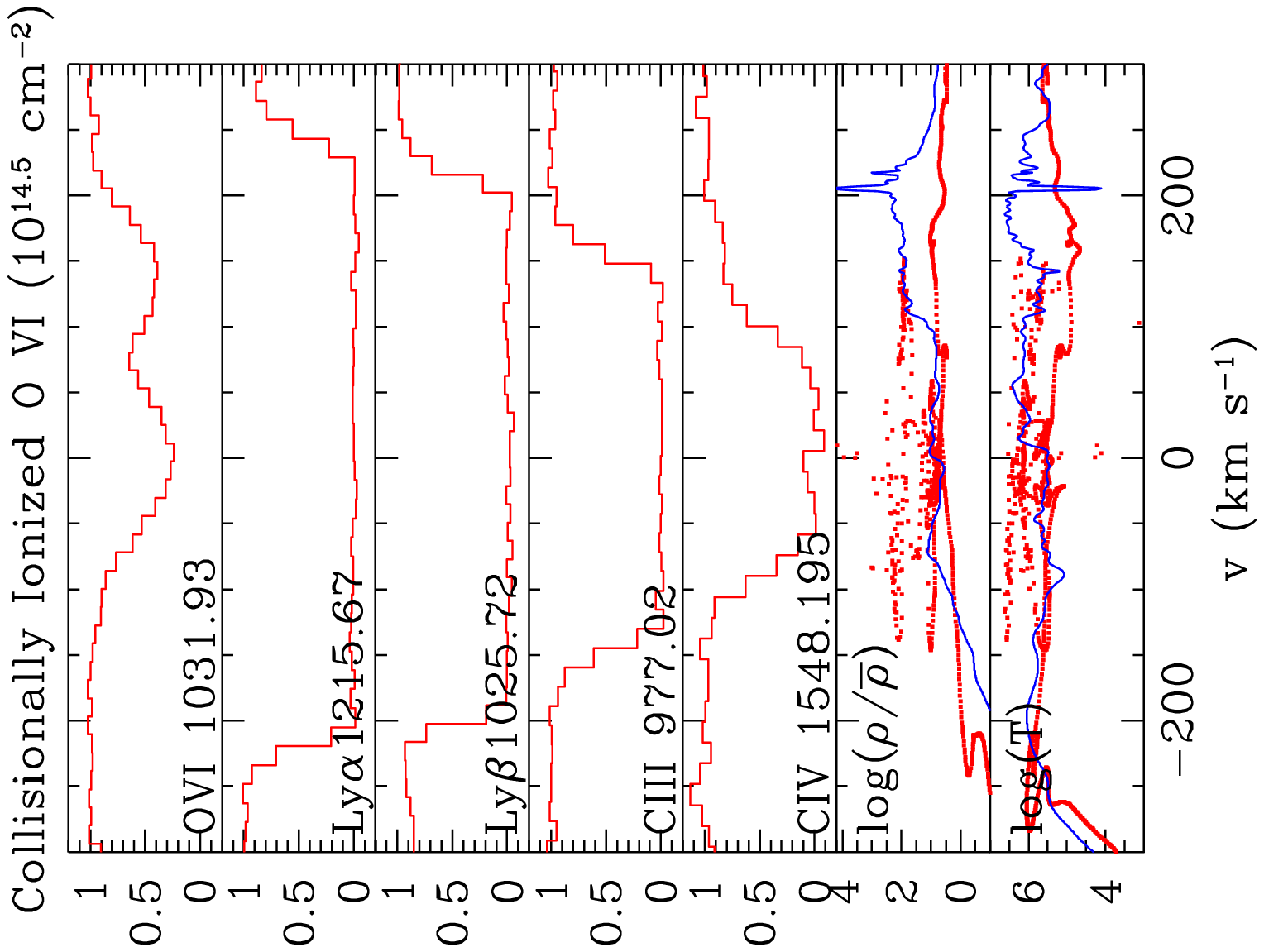}
\caption[]{Centered on a $10^{14.5} \cms$ $\OVI$ component in CIE,
  this multi-component system is a result of $\OVI$ in the intragroup
  medium 195 kpc from a $M^*$ galaxy aligned with a Lyman limit system
  $N(\HI)=10^{17.9} \cms$ arising from a chance alignment with the ISM
  of a nearby dwarf galaxy, $M_{gal}=10^{8.9} \msolar$, which is also
  responsible for the strong $\CIII$ and $\CIV$.  Overlapping gas in
  velocity space aligns $\HI$ gas at $10^{4}$ K with $\OVI$ gas at
  $10^5$ K.  }
\label{fig:sysspec2}
\end{figure}

As an example of extreme multi-phase behavior, we show another system in
Figure~\ref{fig:sysspec2}.  Although the CI absorber at $z=0.36$ is
the same column density and has nearly the same metallicity as the PI
absorber in Figure \ref{fig:sysspec}, almost everything else about it
is different.  The absorber is wider, $b(\OVI)=51 \kms$, and is paired
with a component 103 $\kms$ away ($N(\OVI) = 14.4 \cms$ \& $b(\OVI)=47
\kms$).  The main absorber is clearly in CIE, $T(\OVI)=10^{5.45}$ K at
$\delta(\OVI)=350$, and is slightly more than $1~r_{vir}$ (195 kpc)
away from a $10^{11.1} \msolar$ galaxy; this indicates the absorber
arises in the intragroup medium with ages of $\sim 2$ Gyr.  Peculiar
velocities of several hundred $\kms$ heavily confuse the spatial
distribution, creating one of the most extreme multi-phase absorbers
in our sample.  This is a Lyman limit system, $N(\HI)=10^{17.9} \cms$,
only because of a chance alignment with the ISM of a dwarf galaxy
($M_{gal}=10^{8.9} \msolar$) centered only 5 kpc from the line of
sight.  In our sample of $\Delta z=35$, this is the $\HI$ absorber
closest to any galaxy tracing the highest overdensity,
$\delta(\HI)=5\times 10^4$, and is completely coincidental.
Self-shielding may raise $N(\HI)$ substantially, but almost
any other line of sight passing through this region would intersect
the more extended intragroup $\OVI$ and probably be mis-aligned with a
much weaker BLA tracing halo or intragroup gas.  The lower ionization
metal species arise primarily from the dwarf galaxy ISM.  This
absorber is similar to the $z=0.2028$ absorber observed in PKS
0312-770 by \cite{leh08}, although in that case the Lyman limit system
is possibly co-spatial with the $\OVI$-bearing gas and not the ISM of a
separate galaxy.

The type of simulation analysis demonstrated here for these three
systems should provide a prelude to a future direction in using
simulations for metal-line absorber modeling.  Hydrodynamical
simulations are necessary to model the complex environment and
peculiar velocity field, the apparently inhomogeneous distribution of
metals, and eventually the ionization conditions resulting from
spatially and temporally non-uniform ionization sources.  The first
challenge is to be able to accurately simulate the complex low-$z$
systems observed by STIS and soon COS, and the second is to understand
the physical and evolutionary processes within the simulations with an
emphasis on environmental context.  

Our simulations appear to be able to generate such complex absorbers,
however we admit that we have to search over a pathlength of $\Delta
z=35$ to find just the right CI absorbers that reproduce the
properties of some of the most well-studied low-$z$ $\OVI$ systems
found among lines of sight covering a much smaller path length.
Accurate self-consistent hydrodynamical modeling, especially in CI
absorbers, appears to be a ways away, requiring several order of
magnitude of improved resolution, radiative transfer, and non-equilibrium
effects.  We are more confident about our ability to model the PI
absorbers tracing the diffuse enriched IGM, but we only display
one, as these are a fairly homogeneous population.

As for environment, we have only taken a first step in that direction
when considering the mass and distance to the neighboring galaxy
and the age of the absorber.  The characteristics of the originating
galaxy, the velocity it was launched at, the number of times a metal
has been recycled, and the spread in ages of metals within an
absorber are just a few of the things we have considered here.  Such
considerations are more informative for photo-ionized $\OVI$ because
these absorbers appear to be relatively unrelated to their neighboring
galaxy.

\section{Summary}


We examine the nature and origin of $\OVI$ absorbers in the low redshift
($z<0.5$) IGM using \gad~cosmological hydrodynamic simulations including
outflows, exploring a variety of wind models and input physics variations.
We determine the best-fitting model by comparing to a suite of observed
$\OVI$ and $\HI$ absorber statistics, including the cumulative equivalent
width distribution, $b(\OVI)$ as a function of $N(\OVI)$, and the
$N(\OVI)$ as a function of $N(\HI)$.

Our first main result is that {\it only a model where we explicitly add
sub-resolution turbulence is able to match all observations.}  The
observations motivating this conclusion are the progressively larger
line widths for stronger $\OVI$ systems, along with the high incidence
of large equivalent width absorbers.  We discuss this further later
on.

The second main result is that {\it the vast majority of $\OVI$ absorbers
are photo-ionized, with only a few strong systems being collisionally
ionized.}  This result does not depend on turbulence or any other input
physics variations.  The governing variable in photo-ionized $\OVI$
strength is density, which steadily increases from overdensities of
$\sim10$ for the weakest observed absorbers tracing metals in extended
filaments, to $\sim200$ for the strongest observed absorbers tracing
metals in or around galactic halos.  Increasing metallicity with
density also helps $\OVI$ column densities become stronger.

Our third main result is that {\it metals in the IGM are distributed
inhomogeneously, and that $\OVI$ absorbers preferentially arise
from over-enriched regions.}  The average absorber traces metals
5$\times$ above the mean metallicity-density relationship in our
simulation ($0.15-1.0 \Zsolar$).  Only 1.3\% of the IGM volume
enriched to greater than $0.1 \Zsolar$, with a filling factor of
only 0.3\% resulting from weaker winds working nearly as well.  We
also test a uniform distribution of IGM metals, finding that this
creates too many weak absorbers.  Our results imply that applying
the mean IGM metallicity-density gradient to gas everywhere in order
to study metal absorption is not appropriate; the {\it scatter} in
the global metallicity-density relationship critically influences
$\OVI$ absorbers.

The clumpy metal distribution is an important reason why most $\OVI$
is at photo-ionization temperatures, because it strongly enhances
metal-line cooling.  Oxygen is a powerful coolant and is (not
coincidentally) particularly efficient in the narrow temperature
regime where $\OVI$ has a collisional ionization maximum.  This
creates a ``zone of avoidance" for enriched IGM gas between
$10^{4.5}-10^{5.5}$ K.  Previous theoretical studies that did not
account for metal line cooling incorrectly predict more collisionally
ionized systems.  Non-equilibrium ionization allows collisionally
ionized $\OVI$ to extend to lower temperatures \citep{gna07}, but the
differences are only important for relatively rare high density
regions, and do little to alter the overall $\OVI$ statistics.  The
high fraction of aligned absorbers in the \citet{tho08a} and
\citet{tri08} datasets supports most $\OVI$ absorbers being
photo-ionized.  While observed complex, multi-phase systems with
mis-aligned components indicating CIE $\OVI$ may be under-represented
in our simulations, it is unlikely that would dominate typical $\OVI$
systems under any circumstance.  A key implication of our findings is
that {\it $\OVI$ is generally a poor tracer of the Warm-Hot
Intergalactic Medium (WHIM), and cannot be used to infer the WHIM
baryonic content.}

In depth study of the properties of $\OVI$ absorbers and their
associated $\HI$ reveals that multi-phase temperatures and densities
are often needed to explain their properties.  Metal-line cooling
allows inhomogeneously distributed $\OVI$ to cool slightly below
$\HI$-weighted temperatures, while stronger $\OVI$ absorbers often
trace lower overdensities than aligned strong $\HI$, which typically
traces gas inside halos.  Such a multi-density photo-ionized model
appears necessary to explain strong $\OVI$ absorbers aligned with
$N(\HI)>10^{15} \cms$ when considering absorbers on the
$N(\HI)-N(\OVI)$ plane.  The \citet{tri08} dataset finds more weak
$\HI$ absorbers aligned with $\OVI$ than our simulations, possibly
indicating local fluctuations in the meta-galactic photo-ionizing
background around 1~Rydberg.

We find good agreement with observations for the distribution of
velocity separations between $\OVI$ and $\HI$ absorbers.  The typical
length of a photo-ionized $\OVI$ absorber is $50-100$~kpc, although
the peculiar velocity field often dominates resulting in mis-aligned
$\HI$ components, especially those tracing higher densities.  Only
42\% of our strong absorbers $N(\OVI)\ge 10^{14} \cms$ are aligned
within $4\kms$ while 91\% are aligned within $80\kms$.  This indicates
a clumpy distribution of $\OVI$ that does not exactly trace the
smoothly varying gas in the $\lya$ forest, but still arises in the
same underlying large-scale structure.

A clear anti-correlation exists between the density traced by $\OVI$
and its age as determined by the last time the metals were launched in
a wind.  This trend arises in our simulations because of the
outside-in pattern of metal enrichment; outer, lower overdensities are
enriched first when the Universe is young and physical distances are
small, and inner, higher overdensities are progressively more enriched
by later winds extending a smaller comoving distance. This is a
consequence of our finding in \citet{opp08} that winds travel similar
physical distances relatively independent of redshift or galaxy mass.
While a majority of metals falls back into galaxies, our work here
shows that low-$z$ $\OVI$ absorbers below $N(\OVI)\sim 10^{14} \cms$
provide one of the best ways to observe the oldest metals that remain
in the IGM.  We suggest the fascinating possibility of deriving
nucleosynthetic yields tracing the earliest stars if another high
ionization species, such as $\CIV$, can be observed at low densities
with the increased throughput of COS or future facilities.

Finally, we study the galaxy environment around $\OVI$ absorbers.
Photo-ionized $\OVI$ usually has little to do with either the mass of
the neighboring galaxy, averaging $\sim 0.03-0.1 M^*$, or the
distance, averaging $\ga 2r_{vir}$, because of the significant
galactic and large-scale structural evolution occurring while these
metals reside in the IGM.  This helps explain why observed
photo-ionized absorbers appear to trace a variety of environments,
with a typical distance of 100-300 kpc to the nearest galaxy.  The
opposite is true of the small minority of collisionally ionized
absorbers.  These strong absorbers are $<2r_{vir}$ from $M^*$ galaxies
and have ages indicating activity within the last 2~Gyr, and are
perhaps analogous to $\OVI$ observed in HVCs and IVCs in the Milky Way
halo.  The relatively young age for strong $\OVI$ signifies that these
metals recycle back into galaxies multiple times in what can be
described as halo fountains~\citep{opp08}.

The most uncertain and controversial part of this paper is the claim
that significant sub-resolution turbulence is present in $\OVI$
absorbers.  We determine the amount of turbulence needed by directly
fitting the observed $N(\OVI)-b(\OVI)$ relation.  This increases
$b$-parameters to observed levels by construction, while
simultaneously increasing large-$EW$ absorbers to the observed
frequency.  We attempt to justify this ``magic bullet", as well as
compare the implied turbulence to other instances of observed IGM
turbulence.  We point out that various authors have argued for
clumpiness in IGM metals extending well below the mass and spatial
resolution of our simulations~\citep{sim06,sch07}.  If such clumpiness
exists, then surely those clumps must have some relative motion, so
assuming completely static velocities below the SPH smoothing scale of
$20-100$~kpc is unrealistic.  Our model states that strong absorbers
can be dominated by turbulent motions, just as the line profiles
associated with molecular clouds, $\HII$ regions, IVCs, and HVCs
associated with our Galaxy are.  The energy dissipation required using
a Kolmogorov spectrum is a fraction of the turbulence observed by
\citet{rau01} in high-$z$ $\CIV$ absorbers, which is encouraging
considering the IGM at late times should be calmer, although it is
unclear whether Kolmogorov theory accurately describes IGM turbulence.
We show that this trend is quantitatively consistent with the idea
that turbulence dissipates as metals reside longer in the IGM.  Our
turbulent scenario makes the case that metal-line absorbers are made
up of numerous cloudlets with lower ionization species tracing
high-density clouds making multiple thin profiles, and $\OVI$ tracing
more of the low-density regions in between creating a single broader
profile; such is often observed to be the structure of low-$z$ metal
line systems.  In short, we believe that small-scale turbulence is a
viable possibility, although more detailed modeling is required to
fully understand its implications.  Applying a volume renormalization
zoom technique to cosmological simulations, allowing the resampling of
individual haloes in galactic-scale simulations, provides a realistic,
near-term possibility for following the evolution of turbulence over a
Hubble time.

While it may be disappointing that $\OVI$ is a poor WHIM tracer and
provides only a weak handle on the IGM metal distribution, our results
open up some new and interesting possibilities for using $\OVI$
absorbers to understand cosmic metals and feedback.  For instance,
low-$z$ $\OVI$ may provide a fascinating opportunity to study some of
the oldest IGM metals in the Universe through weak absorbers.  The
strongest absorbers appear related to the recycling of gas between the
IGM and galaxies, providing a unique window into how galaxies get
their gas \citep[e.g.][]{ker05, ker08}.  Detecting multiple ions in
weaker absorbers will provide a good handle on physical conditions
owing to the photo-ionized nature of $\OVI$.  Our work here is only a
first step towards understanding the nature and origin of $\OVI$ in
the low-$z$ Universe using numerical simulations, which we hope to
extend further by considering other metal species in a greater
evolutionary context.  We stress the need to model in detail the
complex metal-line systems that will undoubtedly be observed by COS.
The computational models presented, while state of the art, still fall
well short of what is necessary, particularly for stronger
collisionally ionized systems.  We anticipate that future
observational and modeling improvements will shed new light on the
metal distribution at the present cosmic epoch with all its important
implications.  This work provides a first step in that direction.

\section*{Acknowledgments}  \label{sec: ack}

We thank Todd Tripp for inspirational conversations that helped spur
this investigation in the first place.  Valuable conversations with
Blair Savage, Nicolas Lehner, Christopher Thom, and Hsiao-Wen Chen
provided useful guidance.  Support for this work was provided by NASA
through grant number HST-AR-10946 from the SPACE TELESCOPE SCIENCE
INSTITUTE, which is operated by AURA, Inc. under NASA contract
NAS5-26555.  Support for this work, part of the Spitzer Space
Telescope Theoretical Research Program, was also provided by NASA
through a contract issued by the Jet Propulsion Laboratory, California
Institute of Technology under a contract with NASA. Support was also
provided by the National Science Foundation through grant number
DMS-0619881.

\label{lastpage}

\end{document}